\documentclass[11pt,onecolumn]{IEEEtran}
\pdfoutput=1
\usepackage[T1]{fontenc}
\usepackage{amsmath}
\usepackage{amssymb}
\usepackage{graphicx}
\usepackage{dsfont}
\usepackage{cite}
\usepackage{extraipa,setspace}
\usepackage{mathrsfs}
\usepackage{color}
\usepackage{enumerate}
\usepackage{stmaryrd, textcomp,chemarr,accents}
\usepackage{hyperref}

\newtheorem{theorem}{Theorem}

\newtheorem{lemma}{Lemma}

\newtheorem{definition}{Definition}

\newtheorem{remark}{Remark}

\newcommand{\step}[2]{\stackrel{\textnormal{#1}}{#2}}
\newcommand{\set}[1]{\ensuremath{\mathcal{#1}}} 
\newcommand{\mc}[1]{\ensuremath{\mathcal{#1}}}
\DeclareMathOperator{\Exp}{\ensuremath{\mathbb{E}}} 
\renewcommand{\Pr}{{\mathbb{P}}}
\newcommand{\mkv}{-} 
\newcommand{\qed}{\nobreak \ifvmode \relax \else
\ifdim\lastskip<1.5em \hskip-\lastskip
\hskip0.5em plus0em minus0.5em \fi \nobreak
\vrule height0.4em width0.5em depth0.1em\fi} 

\newcommand{\ve}{\ensuremath{\varepsilon}}
\newcommand{\U}{\ensuremath{\set{U}}} 
\newcommand{\V}{\ensuremath{\set{V}}} 
\newcommand{\X}{\ensuremath{\set{X}}} 
\newcommand{\E}{\ensuremath{\set{E}}} 
\newcommand{\Y}{\ensuremath{\set{Y}}} 
\newcommand{\eS}{\ensuremath{\set{S}}} 
\newcommand{\Z}{\ensuremath{\set{Z}}} 
\newcommand{\A}{\ensuremath{\set{A}}} 
\newcommand{\B}{\ensuremath{\set{B}}} 
\newcommand{\C}{\ensuremath{\set{C}}} 

\newcommand{\R}{{\mathscr{R}}} 
\newcommand{\RQB}{{\R^*_\mathsf{QB}}} 
\newcommand{\GK}{{\mathsf{GK}}} 

\newcommand{\supp}{\mathsf{S}} 
\newcommand{\mpv}{\mathfrak{m}} 
\newcommand*{\bord}{\multicolumn{1}{|}{}}

\allowdisplaybreaks[4] 

\title{Common Reconstructions in the Successive Refinement Problem with Receiver Side Information}
\author{Badri N. Vellambi,~\IEEEmembership{Senior Member,~IEEE} and Roy Timo,~\IEEEmembership{Member,~IEEE}
\thanks{B. N. Vellambi is with the New Jersey Institute of Technology, Newark, NJ 07102, USA, badri.n.vellambi@ieee.org.}
\thanks{R.~Timo was with the Institute for Communications Engineering at the Technical University of Munich. He is now with Ericsson Research, Stockholm, roy.timo@ericsson.com.} 
\thanks{Some of the material in this paper was presented at the 2013 IEEE Information Theory Workshop in Seville, Spain, and the 2014 IEEE International Symposium on Information Theory in Honolulu, USA, 2014. This work was supported by the Australian Research Council's Discovery Project DP120102123.}
}

\begin{document}

\maketitle

\begin{abstract}
We study a variant of the successive refinement problem with receiver side information where the receivers require identical reconstructions. We present general inner and outer bounds for the rate region for this variant and present a single-letter characterization of the admissible rate region for several classes of the joint distribution of the source and the side information. The characterization indicates that the side information can be fully used to reduce the communication rates via binning; however, the reconstruction functions can depend only on the G\'{a}cs-K\"{o}rner common randomness shared by the two receivers. Unlike existing (inner and outer) bounds to the rate region of the general successive refinement problem, the characterization of the admissible rate region derived for several settings of the variant studied requires only one auxiliary random variable. Using the derived characterization, we establish that the admissible rate region is not continuous in the underlying source source distribution even though the problem formulation does not involve zero-error or functional reconstruction constraints. 

\end{abstract}


\section{Introduction}\label{sec-intro}

\IEEEPARstart{T}{his} paper considers a \emph{common-receiver reconstructions (CRR)} variant of the \emph{successive refinement} problem with receiver side information where the source reconstructions at the receivers are required to be identical (almost always). An encoder is required to compress the output of a discrete memoryless source (DMS) into two messages:

\begin{itemize}
\item a \emph{common} message that is reliably delivered to both receivers, and
\item a \emph{private} message that is reliably delivered to one receiver.
\end{itemize}
Each receiver has some side information jointly correlated with the source (and the other receiver's side information), and is required to output source reconstruction that meets a certain fidelity requirement. The CRR condition requires that these reconstructions be identical to one another. 

The CRR problem described above can be viewed as an abstraction of a communication scenario that could arise when conveying data (e.g., meteorological or geological survey data, or an MRI scan) over a network for storage in separate data clusters storing (past) records of the data. The records, which serve as side information, could be an earlier survey data or a previous scan, depending on the specific application. The framework considered here is the source coding problem that arises when data is to be communicated over a  degraded broadcast channel to two receivers that have prior side information, and the three terminals (the transmitter and the two receivers) use a separate source-channel coding paradigm~\cite{Vellambi-Timo-AusCTW2014}.

The problem of characterizing the \emph{achievable rate-distortion region} of the general successive refinement  problem with receiver side information is open~\cite{Steinberg-Merhav-SRProblem,Tian-Diggavi-SRProblem,Timo-Chan-Grant}. The version of the successive refinement problem where the private message is absent, known as the Heegard-Berger problem, is also open~\cite{Timo-Chan-Grant,HeegardBerger-BHProblem, Kaspi-HBProblem}. However, complete characterization exists for specific settings of both successive refinement and Heegard-Berger problems. For example, the rate region of the successive refinement problem is known when the side information of the receiver that receives one message is a degraded version of side information of the other receiver~\cite{Steinberg-Merhav-SRProblem}. Similarly, the Heegard-Berger problem has been solved when the side information is degraded~\cite{HeegardBerger-BHProblem}, mismatched degraded~\cite{Watanabe-HB}, or conditionally less noisy~\cite{Timo-Oechtering-Wigger-HB}. Additionally, the HB problem has also been solved under list decoding constraints (closely related to logarithmic-loss distortion functions)~\cite{Chai-HB}, degraded message sets~\cite{Benammar-Zaidi-HB}, and many vector Gaussian formulations~\cite{Unal-Wagner-HB-Vector-Gaussian1,Unal-Wagner-HB-Vector-Gaussian2}. 

The common reconstruction variant of the Wyner-Ziv problem was first motivated and solved by Steinberg~\cite{SteinbergCR}. Common reconstructions in other problems were then considered in~\cite {HB-CR2013}. Benammar and Zaidi recently considered the HB problem under a three-way common reconstructions condition with degraded message sets~\cite{Benammar-Zaidi-HB,Benammar-Zaidi-HB-IZS}. In our previous work~\cite{Vellambi-Timo-ITW2013}, we characterized the rate region for several cases of the HB problem with the CRR requirement. In this work, we present single-letter inner and outer bounds for the rate region of the successive refinement problem with receiver side information and the CRR requirement (termed as the SR-CRR problem). For several specific cases of the underlying joint distribution between the source and the side information random variables (including those in our previous work~\cite{Vellambi-Timo-ISIT2014}), we prove that the inner and outer bounds match, and therefore yield a characterization of the rate region. 

The characterization indicates that while the receiver side information can be fully utilized for reducing the communication rate by means of binning, only the G\'{a}cs-K\"{o}rner common randomness between the  random variables (i.e., both auxiliary and side information) available to the two receivers can be used for generating the reconstructions. This feature is also seen in our characterization for the HB problem with the CRR requirement in~\cite{Vellambi-Timo-ITW2013}. This single-letter characterization for the rate region of the SR-CRR problem derived in this work  is \emph{unique} in the sense that it is the first rate region formulation where the G\'{a}cs-K\"{o}rner common randomness explicitly appears in the single-letter constraint corresponding to the receiver source reconstructions.  Unlike the best-known bounds for the successive refinement problem, the characterization of the SR-CRR rate region (when the source satisfies a certain support condition) requires only one auxiliary random variable that is decoded by both receivers. Thus, the CRR requirement obviates the need for a second auxiliary random variable to absorb the private message. 

The paper is organized as follows. Section~\ref{sec-N} introduces some basic notation; Section~\ref{Sec:Gacs-Korner} reviews the concept of G\'acs-K\"orner common randomness; and Section~\ref{sec-PD} formally defines the successive refinement problem with the common receiver reconstruction constraint. The characterization of the paper's main contributions  are summarized in Section~\ref{sec-NewResults}, including the single-letter characterization of the rate region, and the proof of the discontinuity of the characterization with the source distribution. The reader will be directed to the respective appendices for the proofs of the results contained in Section~\ref{sec-NewResults}. Finally, Section~\ref{Sec:Conclu} concludes this work.

\section{Notation, G\'acs-K\"orner Common Randomness, and Problem Setup}

\subsection{Notation}\label{sec-N}

Let $\mathbb{N} = \{1,2,\ldots\}$ denote the natural numbers. Uppercase letters (e.g., $S$, $U$, $V$) represent random variables (RVs), and the script versions (e.g., $\eS$, $\U$, $\V$) denote the corresponding alphabets. All alphabets in this work are assumed to be finite. Realizations of RVs are given by lowercase letters (e.g., $s$, $u$, $v$). For RVs $A,B,C$, we denote $A\mkv B\mkv C$ if they form a Markov chain. Given RVs $A$ and $B$, we let $A\equiv B $ if and only if
\begin{equation}
H(A|B)=H(B|A)=0.
\end{equation}
Given $k$ jointly correlated random variables $A_1,\ldots, A_l$, their support set is defined by 
\begin{align}
\notag
\supp(A_1,\ldots,A_k) 
&= \big\{(a_1,\ldots,a_k) \in \A_1 \times \cdots \times \A_k:
p_{A_1\cdots A_k}(a_1,\ldots,a_k)>0\big\}.
\end{align}
For a set $T$, and $a,b\in T$, we let 
\begin{align}
{\mathds{1}}\{a=b\} &:= \left\{\begin{array}{ll} 1, & a=b \\ 0, & a\neq b \end{array}\right.,
\end{align}
and let $\bar{\mathds{1}}\{a= b\} :=   1- \mathds{1}\{a= b\}$. Vectors are indicated by superscripts, and their components by subscripts; for example, $x^n:= (x_1,\ldots,x_n)$ and $x^{n\setminus i} := (x_1,\ldots,x_{i-1},x_{i+1},\ldots,x_n)$. We will use $\mpv$ to denote the least positive probability mass over the support of random variables; for example, given a joint probability mass function (henceforth, pmf) $p_{A_1\cdots A_k}$,
\begin{align}
\mpv_{A_1}
&= \min\big\{p_{A_1}(a) : a_1 \in \supp(A_1) \big\},\\
\mpv_{A_1\cdots A_k}
&= \min\big\{p_{A_1\cdots A_k}(a_1,\ldots,a_k):
(a_1,\ldots,a_k)\in\supp(A_1,\ldots, A_k)\big\},\\
\mpv_{A_1|A_2}
&= \min\big\{p_{A_1|A_2}(a_1|a_2):
(a_1,a_2)\in\supp(A_1, A_2)\big\}.
\end{align}
The probability of an event $E$ is denoted by $\Pr(E)$, and $\Exp$ denotes the expectation operator. Lastly, for $\ve>0$. the set of $\ve$-letter-typical sequences of length $n$ according to pmf $p_X$ is denoted by $T^n_{\ve}[p_X]$~\cite{Kramer-MUIF}.

\subsection{G\'acs-K\"orner Common Randomness}\label{Sec:Gacs-Korner}

Given two jointly correlated random variables $X$ and $Y$, the G\'{a}cs and K\"{o}rner's \emph{common randomness} between $X$ and $Y$~\cite{Gacs-KornerCI} is the random variable $Z$ with the largest entropy such that $H(Z|X)=H(Z|Y)=0$. This notion of common randomness will play a key role in this paper. To define this notion of common randomness we introduce the following terminology. Given $(X,Y) \sim p_{XY}$ on $\X\times \Y$, let $\mathbb{G}^{X,Y}[p_{XY}]$ denote the bipartite graph with left nodes $\X$,  right nodes $\Y$, and edges between $x\in\X$ and $y\in\Y$ if and only if $p_{XY}(x,y)>0$. Now define an equivalence relation on $\Y$ by $y_1\xrightleftharpoons{} y_2$ if and only if they are in the same connected component of  $\mathbb{G}^{X,Y}[p_{XY}]$. Finally, let 
\begin{equation}\label{Eqn:GKMapping1}
\GK^{X,Y}: \Y \rightarrow \Y
\end{equation}
be any mapping satisfying
\begin{equation}\label{Eqn:GKMapping}
\GK^{X,Y}(y_1)=\GK^{X,Y}(y_2) 
\quad
\text{iff} 
\quad 
y_1\xrightleftharpoons{}  y_2.
\end{equation}

\begin{figure}[t!]
\begin{center}
\includegraphics[width=0.45\textwidth]{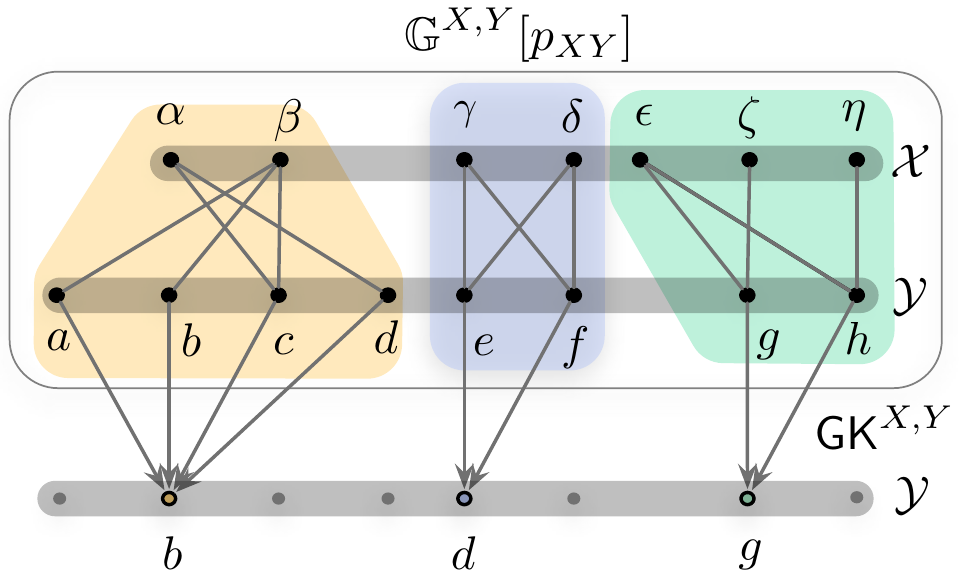}
\caption{Ilustration of $\mathbb{G}^{X,Y}[p_{XY}]$ and $\GK^{X,Y}$.}\label{FIGURE:1}
\end{center}
\end{figure}

Of course, there are multiple choices for the \emph{G\'acs-K\"orner} mapping in~\eqref{Eqn:GKMapping1}. However, all such choices are equivalent in the sense that if $\GK^{X,Y}_1$ and $\GK^{X,Y}_2$ satisfy~\eqref{Eqn:GKMapping} then 
\begin{equation}
\GK^{X,Y}_1(Y)\equiv \GK^{X,Y}_2(Y).
\end{equation}
As an illustration, let $\X=\{a,b,c,d,e,f,g,h\}$ and let $\Y=\{\alpha, \beta, \gamma, \delta, \epsilon, \zeta, \eta\}$. Consider the following pmf $p_{XY}$.
\begin{align}
p_{XY}=\frac{1}{20} \begin{array}{cc} 
\color{red} a\,\,\,\,\,\, b\,\,\,\,\,\, c\,\,\,\,\, d\,\,\,\,\,\,\,e\,\,\,\,\, f\,\,\,\,\, g\,\,\,\,\, h \color{black}& \\
\left[\begin{array}{ccccccccc}
0&0&1& 2&\bord\hspace{2.5mm}0&0&0&0\\
1&2&1& 0&\bord\hspace{2.5mm}0&0&0&0\\\cline{1-6}
0&0&0& 0&\bord\hspace{2.25mm}1&2&\bord\hspace{2.5mm}0 & 0\\ 
0&0&0& 0&\bord\hspace{2.25mm}2&1&\bord\hspace{2.5mm}0 & 0\\\cline{5-8}
0&0&0& 0&\hspace{0.45mm}0&0&\bord\hspace{2.5mm}2&1\\
0&0&0&0&\hspace{0.45mm}0&0&\bord\hspace{2.5mm}2&0\\
0&0&0&0&\hspace{0.45mm}0&0&\bord\hspace{2.5mm}0&2\\
\end{array}\right] & \color{red} \hspace{-2mm}\begin{array}{c} \alpha\\ \beta\\ \gamma\\ \delta\\ \epsilon\\ \zeta\\ \eta\end{array} \color{black}
\end{array}\label{eqn-Fig1-example}
\end{align}
Figure~\ref{FIGURE:1} illustrates the bipartite graph representation of $p_{XY}$, and depicts one possible choice for the G\'acs-K\"orner mapping $\GK^{X,Y}$ satisfying the requirement in \eqref{Eqn:GKMapping}. Notice that that $\mathbb{G}^{X,Y}[p_{XY}]$ contains three connected components, and hence $\GK^{X,Y}$ is a ternary RV taking values in $\{b,d,g\}$ if the chosen mapping is the one illustrated in the figure. Note that for this pmf, there are $4\times 2\times 2=16$ equivalent choices for the mapping.
 
From the definition above, two properties of the G\'{a}cs and K\"{o}rner's common randomness are evident.
\begin{itemize}
\item The G\'{a}cs and K\"{o}rner's common randomness between two random variables is \emph{symmetric}, i.e.,
\begin{equation}
\GK^{X,Y}(Y) \equiv \GK^{Y,X}(X).
\end{equation}
Note however that the above two G\'{a}cs and K\"{o}rner's common randomness variables take values over different alphabets even though each is a function of the other, i.e., $\GK^{X,Y}(Y)$ and $\GK^{Y,X}(X)$ are random variables over $\Y$ and $\X$, respectively.

\item Since the G\'{a}cs and K\"{o}rner's common randomness depends on the pmf $p_{XY}$ only through the bipartite graph, the G\'{a}cs and K\"{o}rner's common randomness between RVs computed using two pmfs $p_{XY}$ and $q_{XY}$ are identical if: (a) $\mathbb{G}^{X,Y}[p_{XY}]$ and $\mathbb{G}^{X,Y}[q_{XY}]$ are graph-isomorphic; and (b) the probabilities of the components of $p_{XY}$ and that of $q_{XY}$ are permutations of one another. To illustrate this, consider $p_{XY}$ of \eqref{eqn-Fig1-example} and the following pmf $q_{XY}$. 
\begin{align}
q_{XY}=\frac{1}{20} \begin{array}{cc} 
\color{red} \,a\,\,\,\,\,\, b\,\,\,\,\,\, c\,\,\,\,\,\, d\,\,\,\,\,e\,\,\,\,\, f\,\,\,\,\, g\,\,\,\,\, h \color{black}& \\
\left[\begin{array}{ccccccccc}
0&1&1& 1&\bord\hspace{2.5mm}0&0&0&0\\
1&0&1& 2&\bord\hspace{2.5mm}0&0&0&0\\\cline{1-6}
0&0&0& 0&\bord\hspace{2.25mm}2&3&\bord\hspace{2.5mm}0 & 0\\ 
0&0&0& 0&\bord\hspace{2.25mm}1&0&\bord\hspace{2.5mm}0 & 0\\\cline{5-8}
0&0&0& 0&\hspace{0.45mm}0&0&\bord\hspace{2.5mm}1&2\\
0&0&0&0&\hspace{0.45mm}0&0&\bord\hspace{2.5mm}3&0\\
0&0&0&0&\hspace{0.45mm}0&0&\bord\hspace{2.5mm}0&1\\
\end{array}\right] & \color{red} \hspace{-2mm}\begin{array}{c} \alpha\\ \beta\\ \gamma\\ \delta\\ \epsilon\\ \zeta\\ \eta\end{array} \color{black}
\end{array}
\end{align}
The G\'{a}cs and K\"{o}rner's common randomness depends between $X$ and $Y$ computed using either $p_{XY}$ or $q_{XY}$ yields a ternary random variable with the pmf $[ \frac{7}{20}\,\, \frac{3}{10}\,\, \frac{7}{20}]$.
\end{itemize}

In the remainder of this work, for a given pmf $p_{XY}$ over $\X \times \Y$, we will assume an arbitrary but fixed choice for the G\'{a}cs and K\"{o}rner's common randomness mapping that satisfies~\eqref{Eqn:GKMapping} without explicitly specifying this mapping. On account of notational ease, we will also drop the argument, and abbreviate it by
\begin{align}
\GK^{X,Y} := \GK^{X,Y}(Y). 
\end{align}
It is to be assumed that the argument is always the second random variable in the superscript, and consequently, the G\'{a}cs and K\"{o}rner's common randomness $\GK^{X,Y}$ is a random variable over the alphabet $\Y$ of the second variable. This is, quite simply, a only a \emph{notational bias}, since G\'{a}cs and K\"{o}rner's common randomness is indeed symmetric.

Before we proceed to formally state the problem investigated and the main results, we present two results that pertain solely to G\'acs-K\"orner common randomness that we will need in the main section of this work. Both results present decompositions of the G\'{a}cs-K\"{o}rner common randomness between two random variables when additional information about the support of the joint pmf of the two variables is known.

\begin{lemma}\label{Lem:Ancillary2}
Suppose that the support set of $(A_1,A_2,U,V)\sim q_{A_1A_2UV}$ satisfies 
\begin{align}
\supp(A_1,A_2,U,V) = \supp(A_1,A_2)\times \supp(U,V).
\end{align}
 Then,
\begin{equation}
\GK^{A_1U,A_2V}\equiv (\GK^{A_1,A_2},\GK^{U,V}).
\end{equation}
\end{lemma}
\begin{IEEEproof}
The proof can be found in Appendix~\ref{APP-A}.
\end{IEEEproof}

\begin{lemma}\label{lem-drop1var}
If $(X,Y,Z) \sim q_{XYZ}$ satisfies
\begin{equation}
\supp(X,Y,Z) = \supp(X,Y)\times \supp(Z),
\end{equation}
then 
\begin{equation}
\GK^{X,YZ} \equiv \GK^{X,Y}.
\end{equation}
\end{lemma}
\begin{IEEEproof}
Define a constant random variable $W$ over a singleton alphabet, say $\{w\}$ and let $q_W$ denote the degenerate pmf of $W$. Define $q_{XYZ} = q_W q_{XYZ}$. Then, one can see that 
\begin{align}
\supp(W,X,Y,Z) &= \supp(X,Y,Z) \times \supp(W) = \supp(X,Y)\times \supp(Z)\times\supp(W) \\
&=\supp(X,Y)\times \supp(Z,W).
\end{align}
Then, an application of Lemma~\ref{Lem:Ancillary2} yields
\begin{align}
\GK^{X,YZ} \equiv (\GK^{X,Y}, \GK(Z,W)) \step{(a)}\equiv\GK^{X,Y},
\end{align}
where (a) follows since $W$ is a constant RV.
\end{IEEEproof}

\subsection{Problem Setup --- Successive Refinement with the CRR Constraint}\label{sec-PD}

\begin{figure}[h!]
\begin{center}
\includegraphics[width=0.5\textwidth]{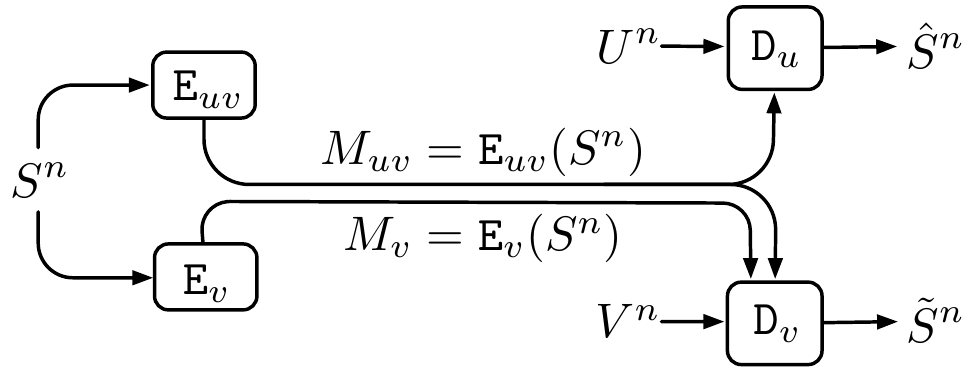}
\caption{Successive refinement with receiver side information and common receiver reconstructions (CRR).}\label{FIGURE:2}
\end{center}
\end{figure}

Let pmf $p_{SUV}$ on $\eS \times \U \times \V$, a reconstruction alphabet $\hat{\eS}$, and a bounded distortion function $d:\eS \times \hat{\eS} \rightarrow [0,\bar D]$ be given. We assume that $\bar D \in (0,\infty)$. A DMS $(S,U,V) \sim p_{SUV}$ emits an i.i.d. sequence 
\begin{equation}
(S^n,U^n,V^n) = (S_1,U_1,V_1), (S_2,U_2,V_2), \ldots,
(S_n,U_n,V_n).
\end{equation}
As illustrated in Fig.~\ref{FIGURE:2}, the engineering problem is to encode the \emph{source} $S^n$ into a common message $M_{uv}$ communicated to both receivers, and a private message $M_v$ communicated only the receiver having the side information $V^n$ so that the following three conditions are satisfied:
\begin{enumerate}
\item A receiver with access to the \emph{side information} $U^n$  and the common message $M_{uv}$ can output an estimate $\hat{S}^n$ of $S^n$ to within a prescribed average (per-letter) distortion $D$.
\item A receiver with access to the \emph{side information} $V^n$ and both (common and private) messages $M_{uv}$ and $M_v$ can output an estimate $\tilde{S}^n$ of $S^n$ to within a prescribed average (per-letter) distortion $D$.
\item The estimates $\hat{S}^n$ and $\tilde{S}^n$ (both defined on the set $\hat{\eS}^n$) are identical to one another almost always. 
\end{enumerate}
The aim then is to characterize the rates of the common and private messages that need to be communicated to achieve the above requirements. The following definition formally defines the problem.

\begin{definition}\label{defn-CRRRateregion}
Fix $D \geq 0$. We say that a rate pair $(r_{uv}, r_v) \in [0,\infty)^2$ is $D$-\emph{admissible} if for each $\ve>0$ there exist a sufficiently large blocklength $n\in\mathbb N$, and:
\begin{subequations}
\begin{enumerate}[(a)]
\item two encoders 
\begin{align}
\mathtt E_{uv}:\eS^n&\rightarrow \big\{1,2,\ldots,\big\lceil 2^{n(r_{uv}+\ve)}\big\rceil\big\},\\
\mathtt E_{v}:\eS^n&\rightarrow \big\{1,2,\ldots,\big\lceil 2^{n(r_{v}+\ve)}\big\rceil\big\};
\end{align}
\item and two decoders 
\begin{align}
\mathtt D_u&:\big\{1,2,\ldots,\big\lceil 2^{n(r_{uv}+\ve)}\big\rceil\big\}\times \U^n \rightarrow \hat{\eS}^n,\\
\mathtt D_v&:\big\{1,2,\ldots,\big\lceil 2^{n(r_{uv}+\ve)}\big\rceil\big\}
\times\big\{1,\ldots,\big\lceil 2^{n(r_{v}+\ve)}\big\rceil \big\}\times \V^n \rightarrow \hat{\eS}^n,
\end{align}
\end{enumerate}
\end{subequations}
such that the reconstructions
\begin{subequations}
\begin{align}
\hat{S}^n   &:=\mathtt  D_u\big(\mathtt E_{uv}(S^n),U^n\big),\\
\tilde{S}^n &:= \mathtt D_v\big(\mathtt E_{uv}(S^n), \mathtt E_v(S^n),V^n\big),
\end{align}
\end{subequations}
satisfy:
\begin{subequations}\label{eqn-RRDefn1}
\begin{align}
\frac{1}{n} \sum_{i=1}^n \Exp\big[d(S_i,\hat{S}_i)\big] &\leq D+\ve,\\
\frac{1}{n} \sum_{i=1}^n \Exp\big[d(S_i,\tilde{S}_i)\big]&\leq D+\ve, \\
\Pr\big[\hat S^n\neq \tilde S^n\big] &\leq \ve.
\end{align}
\end{subequations}
\end{definition}

\begin{definition}\label{Def:RateRegion}
The \emph{$D$-admissible rate region} $\mathsf{\R}(D)$ of the successive refinement problem with CRR is the set of all $D$-admissible rate pairs.
\end{definition}

The main problem of interest in this paper is to characterize the $D$-admissible rate region $\R(D)$. Define $\underline{D}$ by
\begin{equation}
\underline{D} :=\mathop{\min}_{\phi:\ \eS\rightarrow\hat\eS}\Exp d(S,\phi(S)).
\end{equation}
On one hand, if $0 \leq D < \underline{D}$,
then $\mathsf{\R}(D) = \emptyset$. On the other, if $D > \overline{D}$, then $\mathsf{\R}(D) = [0,\infty)^2$. For the interesting interval of non-trivial values of $D \in [\underline{D},\overline{D}]$, the rate region $\R(D)$ is a closed and convex subset of $[0,\infty)^2$. This nontrivial interval of values of $D$ will be the subject of our investigation for the remainder of the paper. 

\section{Main Results}\label{sec-NewResults}
In this section, we will present inner (achievability) and outer (converse) bounds for the $D$-admissible rate region $\R(D)$, and we show that these bounds are tight in a variety of nontrivial settings. We will characterize these bounds through the following three rate regions defined over three corresponding spaces of auxiliary random variable pmfs.

\subsection{Three single-letter rate regions and their properties}\label{Mainresults-rrdefns}
\begin{definition}\label{Def:PstarRegionk}
For $k \in \mathbb{N}$, let $\mc{P}^*_{D,k}$ denote the set of all pmfs $ q_{ASUV}$ defined on $\A \times \eS \times \U \times \V$ such that $(A,S,U,V)\sim q_{ASUV}$ satisfies the following conditions: 
\begin{enumerate}[(i)]
\item $q_{SUV}=p_{SUV}$; 
\item $|\A| \leq k$; 
\item $A\mkv S \mkv (U,V)$; and
\item there exists a function $f:\A\times\V\rightarrow\hat \eS$ for which
\begin{align}\label{eqn-Defn1eq3}
\Exp \big[d \big(S, f(\GK^{  AU,AV})\big)\big] \leq D.  
\end{align}
\end{enumerate}
\end{definition}

\begin{definition}\label{Def:RstarRegionk}
Let $\R_{k}^*(D)$ denote the set of all rate pairs $(r_{uv},r_v) \in [0,\infty)^2$ satisfying 
\begin{subequations}\label{Eqn:Rsk:Rates}
\begin{align}
r_{uv}        & \geq I(S;A|U)\label{eqn-*defnrate1}\\
r_v+r_{uv} & \geq I(S;A|V)\label{eqn-*defnrate2}
\end{align}
\end{subequations}
for some $(A,S,U,V)\sim q_{ASUV} \in \mc{P}^*_{D,k}$.
\end{definition}

\begin{definition}\label{Def:PddagRegionk}
For $k \in \mathbb{N}$, let $\mc{P}_{D,k}^\ddagger$ denote the set of all pmfs $q_{ABCSUV}$ defined on $\A \times \B \times \C \times \eS \times \U \times \V$ such that $(A,B,C,S,U,V) \sim q_{ABCSUV}$ satisfies the following constraints:
\begin{enumerate}[(i)]
\item $q_{SUV}=p_{SUV}$;
\item $\max\{|\A|,|\B|,|\C|\}\leq k$; 
\item $(A,B,C)\mkv S \mkv (U,V)$; and 
\item  there exists a function $f:\A\times\C\times\V\rightarrow\hat \eS$ for which
\begin{align}
{\textstyle\Exp \big[d (S, f\big(\GK^{  ABU,ACV}\big)\big]\leq D}.\label{eqn-Defn1eq1}
\end{align}
\end{enumerate}
\end{definition}

\begin{definition}\label{Def:RddagRegionk}
Let $\R_k^\ddagger(D)$ denote the set of all rate pairs $(r_{uv},r_v) \in [0,\infty)^2$ satisfying
\begin{subequations}
\begin{align}
r_{uv} &\geq I(S;A,B|U),\label{eqn-ddagdefnrate1}\\
r_v+r_{uv} &\geq I(S;B|A,C,U,V) +I(S;A,C|V) \label{eqn-ddagdefnrate2}
\end{align} 
\end{subequations}
for some $(A,B,C,S,U,V)\sim q_{ABCSUV} \in \mc{P}^\ddag_{D,k}$.
\end{definition}

\begin{definition}\label{Def:PdagRegionk}
For $k \in \mathbb{N}$, let 
\begin{equation}
\mc{P}_{D,k}^\dagger =  
\Big\{q_{ABCSUV} \in\mc{P}_{D,k}^\ddagger: B\mkv (A,S) \mkv C\Big\}.
\end{equation}
\end{definition}
\begin{definition}\label{Def:RdagRegionk}
Let $\R_{k}^{  \dagger}(D)$ denote the set of all rate pairs $(r_{uv},r_v)\in [0,\infty)^2$ satisfying 
\begin{subequations}\label{Eqn:DefRdagger}
\begin{align}
r_{uv} &\geq I(S;A,B|U)\label{eqn-dagdefnrate1}\\
r_v+r_{uv} &\geq \max\{I(S;A|U), I(S;A|V)\} + I(S;B|A,U) + I(S;C|A,V)\label{eqn-dagdefnrate2}
\end{align}
\end{subequations}
for some $(A,B,C,S,U,V)\sim q_{ABCSUV} \in \mc{P}^\dag_{D,k}$.
\end{definition}

We can establish the following preliminary inclusions between the three rate regions defined above.
\begin{lemma}\label{lem-*kddagkinclusion}
For any $k\in\mathbb{N}$,  $\R^*_{k}(D) \subseteq \R^\ddagger_k(D)$.
\end{lemma}
\begin{IEEEproof}
By simply choosing $B=C=A$ with $|\mc A|\leq k$, we cover all rate pairs that line in $\R^*_k(D)$, and hence, $\R^*_{k}(D)$ is a subset of $\R^\ddagger_k(D)$.
\end{IEEEproof}

\begin{lemma}\label{lem-dagkddagkinclusion}
For any $k\in\mathbb{N}$, $\R^\dag_{k}(D)\subseteq \R^\ddagger_k(D)$.
\end{lemma}
\begin{IEEEproof}
First, note that $\mathcal P_{D,k}^\dagger \subseteq \mathcal P_{D,k}^\ddagger$. So we are done if we show that the RHS of \eqref{eqn-ddagdefnrate2} is numerically smaller than that of \eqref{eqn-dagdefnrate2} for any pmf in $\mathcal P_{D,k}^\dagger$. To do that, pick  $p_{ABCSUV}\in \mathcal P_{D,k}^\dagger$ and consider the following argument.
\begin{align}
 I(S;B|A,C,U,V) +I(S;A,C|V) &= H(B|A,C,U,V)-H(B|A,C,S,U,V) + I(S;A,C|V)\notag\\
 &\step{(a)}=H(B|A,C,U,V)-H(B|A,C,S) + I(S;A,C|V)\notag\\
&\step{(b)}=H(B|A,C,U,V)-H(B|A,S) + I(S;A,C|V)\notag\\
&\step{(c)}\leq H(B|A,U)-H(B|A,S) + I(S;A,C|V)\label{eqn-Uintro}\\
& \step{(d)}= I(S; B|A,U) + I(S;A,C|U),\label{eqn-Uintro1}
\end{align} 
where (a) follows from the chain $(A,B,C)\mkv S \mkv (U,V)$; (b) follows from the chain $B\mkv (A,S) \mkv C$; (c) follows by dropping variables in the conditioning; and finally (d) follows by reintroducing $U$ in the second term of \eqref{eqn-Uintro} without affecting the numerically affecting the terms. Finally, the claim follows by noting that \eqref{eqn-Uintro1} is bounded below by the RHS of  \eqref{eqn-dagdefnrate2} thereby completing the proof of this claim.
\end{IEEEproof}

While the above two inclusions hold true for all DMSs $p_{SUV}$, we can establish stronger results if we know something more about $p_{SUV}$. In specific, if we know that the pmf $p_{SUV}$ satisfies the full-support condition of \eqref{eqn:Condition1}, then the following \emph{reverse} inclusion also holds albeit with some alphabet size readjustment. In other words, when the full-support condition is met, any rate pair that meets \eqref{eqn-ddagdefnrate1} and \eqref{eqn-ddagdefnrate2} (with auxiliary RVs $A,B$, and $C$) also meets \eqref{eqn-*defnrate1} and \eqref{eqn-*defnrate2} for a different auxiliary RV $A$ with an appropriately larger alphabet.

\begin{lemma}\label{thm-RegionEquivalence}
If the support of $(S,U,V) \sim p_{SUV}$ satisfies 
\begin{equation}\label{eqn:Condition1}
\supp(S,U,V)=\supp(S)\times\supp(U,V), 
\end{equation}
then
\begin{equation}
\R^{\ddagger}_{k}(D)
\subseteq
\R^{*}_{k^2}(D),
\quad 
\forall\ k \in \mathbb{N}.
\end{equation}
\end{lemma}
\begin{IEEEproof}
Lemma~\ref{thm-RegionEquivalence} is proved in Appendix~\ref{APP-B}.
\end{IEEEproof}

Observe that each of the three rate regions defined above (Definitions~\ref{Def:RstarRegionk}, \ref{Def:RddagRegionk}, and \ref{Def:RdagRegionk}) can potentially be enlarged by merely increasing $k$. In other words, we are guaranteed to have
\begin{align}
\R_1^\lambda(D)\subseteq \R_2^\lambda(D)\subseteq\R_3^\lambda(D)\subseteq\cdots, \qquad \lambda\in\{\star,\ddagger,\dagger\}.\label{eqn-sequenceofsets}
\end{align}

Since we do not impose restrictions on how we encode the source, we can allow the alphabet sizes of the auxiliary RVs to be finite but  arbitrarily large. Hence, it makes sense to introduce the following notation for the limiting rate regions allowing the alphabets of auxiliary random variables to be any finite set.
\begin{definition}
\begin{align}
\R^\lambda(D) := \bigcup_{k\in\mathbb{N}} \R^{\lambda}_{k}(D), \qquad \lambda\in\{\star,\ddagger,\dagger\}.
\end{align}
\end{definition}
However, from a computational point of view, it is preferable that the sequence of sets in \eqref{eqn-sequenceofsets} not grow indefinitely with $k$. The following result ensures that this, indeed, does not happen. It quantifies the bounds on the alphabet size of the auxiliary random variables beyond which there is no strict enlargement of these regions.

\begin{lemma}\label{thm-RegionEquivalence2}
For all integers $k \in \mathbb{N}$, we have the following.
\begin{subequations}
\begin{alignat}{3}
\R^*_{k}(D) &\subseteq \R^*(D)=\R^*_{|\eS|+2}(D) ,\label{eqn-rateregionalphabetbnd1}\\
\R^\ddagger_{k}(D)&\subseteq \R^\ddagger(D)=\R^\ddagger_{ |\eS|(|\eS|+6)|\hat\eS|^{|\U|}+4}(D),\label{eqn-rateregionalphabetbnd2}\\
\R^\dagger_{k}(D)&\subseteq \R^\dagger(D)=\R^\dagger_{ |\eS|(|\eS|+4)|\hat\eS|^{|\U|}+1}(D).\label{eqn-rateregionalphabetbnd3}
\end{alignat}
\end{subequations}
\end{lemma}
\begin{IEEEproof}
The proof for claim for  $\R^\ddagger(D)$ is presented in detail in Appendix~\ref{APP-C}. The proof for \eqref{eqn-rateregionalphabetbnd1} and \eqref{eqn-rateregionalphabetbnd3} is almost identical to that of \eqref{eqn-rateregionalphabetbnd2}, and the difference are highlighted in Remarks~\ref{rem-regionequiv2} and \ref{rem-regionequiv2b} in Appendix~\ref{APP-C}.
\end{IEEEproof}

We conclude this section with two properties of the above three regions.

\begin{lemma}\label{Lem:RstarConvex}
The regions $\R^*(D)$, $\R^\ddagger(D)$, and $\R^\dagger(D)$ are convex.
\end{lemma}

\begin{IEEEproof}
A proof of the claim for $\R^*(D)$ can be found in Appendix~\ref{APP-D}. The proofs of the convexity of  the other two regions are  identical, and are omitted.
\end{IEEEproof}

\begin{lemma}\label{Lem:RstarCont}
Let for each $i\in\mathbb{N}$, $D_i\in [\underline{D},\overline{D}]$ and $(r_{uv}^{(i)},r_{v}^{(i)})\in \R^*_{D_i,k}$ be given. Suppose that $\lim\limits_{i\rightarrow \infty} D_i = \mathsf D $, and $\lim\limits_{i\rightarrow \infty}  (r_{uv}^{(i)},r_{v}^{(i)}) = (\mathsf{r_{uv},r_v})$. Then, $(\mathsf{r_{uv},r_v}) \in  \R^*_{\mathsf D,k}$.
\end{lemma}

\begin{IEEEproof}
A proof of the claim for $\R^*(D)$ can be found in Appendix~\ref{APP-E}. The proofs for the other two regions are identical, and are omitted.
\end{IEEEproof}
By simply choosing  $D_i = D\in[\underline D, \overline D]$, $i\in \mathbb N$, we obtain the following result. 
\begin{remark}
For any $D\in[\underline D, \overline D]$, the regions $\R^*(D)$, $\R^\ddagger(D)$, and $\R^\dagger(D)$ are topologically closed.
\end{remark}
\subsection{A single-letter characterization for $\R(D)$}\label{Sec-SLChars}

In this section, we present our main results on the single-letter characterization of the $D$-admissible rate region. The first two present inner and outer bounds sandwiching the $D$-admissible rate region using the three limiting rate regions given in Definitions~\ref{Def:RstarRegionk}, \ref{Def:RddagRegionk}, and \ref{Def:RdagRegionk}.

\begin{theorem}\label{thm-achieve-thm-2} 
For any $D\in[\underline D, \overline D]$, the regions $\R^{*}(D)$ and $\R^{\dagger}(D)$ are inner bounds to the $D$-admissible rate region  $\R(D)$ of the successive refinement problem with the CRR constraint, i.e., 
\begin{equation}
\R^{*}(D)
\step{(a)}{\subseteq}
\R^\dagger(D)\step{(b)}\subseteq  \mathsf{\R}(D).
\end{equation}
\end{theorem}

\begin{IEEEproof}
The inclusion (a) follows from Lemmas~\ref{lem-*kddagkinclusion} and~\ref{thm-RegionEquivalence2} above, and a proof of the inclusion (b) can be found in Appendix~\ref{APP-F}.
\end{IEEEproof}

\begin{theorem}\label{thm-achieve-thm-2a} 
For any $D\in[\underline D, \overline D]$, the rate region $\R^{\ddagger}(D)$ is an outer bound to the $D$-admissible rate region  $\R(D)$ of the successive refinement problem with the CRR constraint, i.e., 
\begin{equation*}
\mathsf{\R}(D)
\step{(b)}\subseteq 
\R^\ddagger(D).
\end{equation*}
\end{theorem}

\begin{IEEEproof}
The proof of the inclusion in (a) can be found in Appendix~\ref{APP-G}.
\end{IEEEproof}

In the absence of the CRR constraint, Steinberg and Merhav's original solution to the \emph{physically-degraded side information} version of the successive refinement problem required three auxiliary random variables (later simplified to two by Tian and Diggavi~\cite{Tian-Diggavi-Aug-2007-A}) and two reconstruction functions. Benammar and Zaidi's solution to their formulation of the successive refinement problem with a common source reconstruction required two auxiliary random variables and a reconstruction function~\cite{Benammar-Zaidi-HB,Benammar-Zaidi-HB-IZS}. The following result, which is the main result in this work, establishes a single-letter characterization of the $D$-admissible rate region $\R(D)$ for several cases of side information. Unlike other characterizations, the rate region $\R(D)$ is completely described by a single auxiliary random variable, and a single reconstruction function whose argument is not the side information and the auxiliary random variable, but the G\'{a}cs-K\"{o}rner common randomness shared by the two receivers.

\begin{theorem}\label{thm-3}
If the DMS $(S,U,V) \sim p_{SUV}$ falls into one of the following cases,
\begin{itemize}
\item[A.] $\supp(S,U,V)=\supp(S)\times\supp(U,V)$,
\item[B.] $S\mkv V \mkv U$,
\item[C.] $S\mkv U\mkv V$ and $\supp(S,U)=\supp(S)\times\supp(U)$,
\item[D.] $H(S|U)=0$,
\item[E.] $S\mkv U\mkv V$ and $\supp(U,V)=\supp(U)\times\supp(V)$, or
\item[F.] $\min\{H(U|S),H(V|S)\}=0$,
\end{itemize}
then for any $D\in[\underline D, \overline D]$, the inner bound $\R^{*}(D)$ and the outer bound $\R^{\ddagger}(D)$ match, and 
\begin{equation*}
\mathsf{\R}(D) = {\R}^*(D) = \R^\ddagger (D).
\end{equation*}
\end{theorem}
\begin{IEEEproof}
Theorem~\ref{thm-3} is proved in Appendix~\ref{APP-I}.
\end{IEEEproof}
\begin{remark}
$\mathsf{\R}(D) = {\R}^*(D)$ when $H(S|V)=0$ or $H(U|V)=0$, but these are subsumed by Case B above. 
\end{remark}
Since in each of the cases in Theorem~\ref{thm-3}, the characterization is given by precisely one auxiliary random variable, it is only natural to wonder as to when a \emph{quantize-and-bin} strategy is optimal. In this strategy, the auxiliary random variable $A$ that the encoder encodes the source into is simply the reconstruction that the receivers require. The encoder upon identifying a suitable sequence of reconstruction symbols simply uses a binning strategy to reduce the rate of communication prior to forwarding the bin index to the receivers. Thus in this strategy, all three terminals (i.e., the transmitter included) are aware of the common reconstruction. To analyze cases in which the  quantize-and-bin approach is optimal, we define the corresponding rate region.

\begin{definition}
The \emph{quantize-and-bin} rate region $\RQB(D)$ is the union of all pairs $(r_{uv},r_v)\in[0,\infty)^2$ such that  
\begin{subequations}\label{Eqn:QandBin}
\begin{align}
r_{uv}\geq & I(S;\hat S|U),\\
r_v+r_{uv}\geq & I(S;\hat S | V),
\end{align}
\end{subequations}
where the union is taken over all reconstructions $\hat{S}$ on $\hat{\eS}$ with $\hat{S} \mkv S \mkv (U,V)$ and $\Exp[d(S,\hat S)] \leq D$.
\end{definition}

Clearly, by setting  $A=\hat S$ in the proof of Theorem~\ref{thm-achieve-thm-2} detailed in Appendix~\ref{APP-F}, we infer that $\RQB(D)\subseteq \R^*(D)$. Consequently, we can see that the quantize-and-bin region is always achievable. The following result establishes three conditions under which the quantize-and-bin strategy is not merely achievable, but optimal, i.e., $\RQB(D)=\R^*(D)$.

\begin{theorem}\label{thm-3-cor}
If the DMS $(S,U,V) \sim p_{SUV}$ falls into one of the following cases,
\begin{itemize}
\item[ A.] $\supp(S,U,V)=\supp(S)\times\supp(U,V)$ and $H(\GK^{U,V}(V)) = 0$, 
\item[   B.] $S\mkv U\mkv V$ and $\supp(U,V)=\supp(U)\times\supp(V)$, or
\item[   C.] $\min\{H(U|S),\ H(V|S)\} = 0$,
\end{itemize}
then, the quantize-and-bin strategy is optimal, i.e., 
\begin{equation}
\mathsf{\R}(D) = {\R}^*(D) = \R^\ddagger (D) = \RQB(D).
\end{equation} 
\end{theorem}

\begin{IEEEproof}
The proof can be found in Appendix~\ref{APP-J}.
\end{IEEEproof}

\subsection{A Binary Example}\label{Sec:BinaryExample}

\begin{figure}[h!]
\begin{center}
\includegraphics[width=0.4\textwidth]{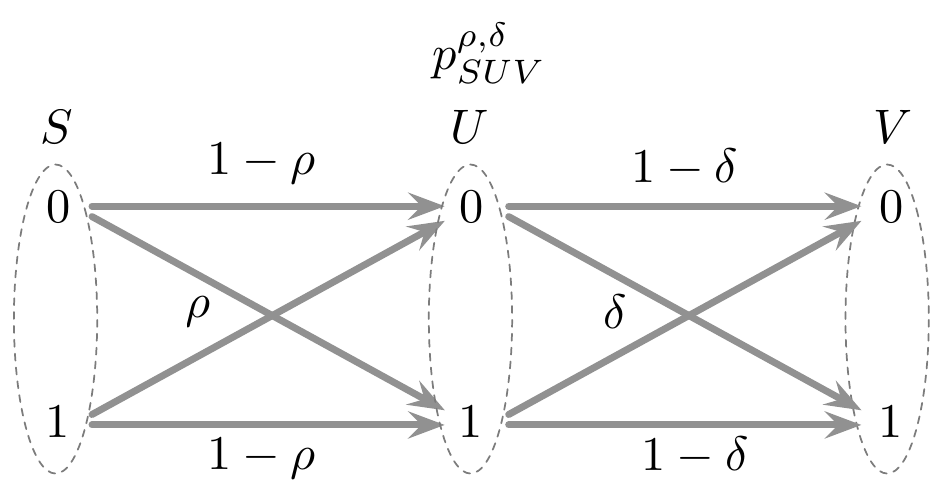}
\caption{A family of DMSs based on the binary symmetric channel.}
\label{FIGURE:3}
\end{center}
\end{figure}

In this section, we will present a binary example with $\eS = \hat{\eS} = \{0,1\}$ and with the reconstruction distortion measure $d$ being the binary Hamming distortion measure
\begin{equation}
d(s,\hat{s}) 
=
\left\{
\begin{array}{ll}
1 & \text{ if } \hat{s} \neq s\\
0 & \text{ otherwise}
\end{array}.
\right.
\end{equation}

As illustrated in Figure~\ref{FIGURE:3}, let $\big\{ (S,U,V) \sim p^{\rho,\delta}_{SUV}\big\}_{\rho,\delta \in [0,1]}$ denote a family of DMSs such that (a) $S$ is an equiprobable binary source; (b) $S\mkv U \mkv V$ form a Markov chain; and (c) the channels $p_{U|S}$, and $p_{V|U}$ are binary symmetric channels with crossover probabilities $\rho$ and $\delta$, respectively.  Note that for any $0<\rho,\delta<1$, the pmf $p^{\rho,\delta}_{SUV}$ satisfies the conditions of both Case A and Case B of Theorem~\ref{thm-3-cor} above. Hence, we see that the quantize-and-bin strategy is optimal, and the optimal tradeoff between communication rates on the common and private links can be obtained without having to time-share between various operating points (corresponding to different average distortions). The following result presents an explicit characterization of the $D$-admissible rate region $\mathcal{R}(D)[p^{\rho,\delta}_{SUV}]$ for this class of sources.

\begin{lemma}\label{Lem:BinaryExample}
If\footnote{The case where $\delta = 0$ or $1$ will treated in the next section.} $0 < \rho,\delta < 1$ and $D\in[0,\frac{1}{2}]$, then 
\begin{equation}\label{Eqn:Lem:BinaryExample1}
\R(D)[p^{\rho,\delta}_{SUV}] 
=
\left\{ 
(r_{uv},r_v) \in [0,\infty)^2: 
\begin{array}{c}
r_{uv}  
\geq 
h(\rho \ast D) - h(D)\\[2pt]
r_v + r_{uv}  
\geq
h\big((\rho \ast \delta) \ast D\big) - h(D)\\
\end{array}
\right\},
\end{equation}
where $x \ast y := x(1-y)+y(1-x)$ denotes the \emph{binary convolution} operation, and $h$ is the \emph{binary entropy function}. Otherwise if $D \geq 1/2$, then $R(D)[p^{\rho,\delta}_{SUV}] = [0,\infty)^2$.
\end{lemma}
\begin{IEEEproof}
If the distortion $D \geq 1/2$, then we can trivially meet the distortion requirement by setting $\hat{S}^n = \tilde{S}^n = 0^n$. Consider the non-trivial range of distortions $0 \leq D < 1/2$. For $\delta \in (0,1)$, the corresponding joint pmf $p^{\rho,\delta}_{SUV}$ satisfies $S \mkv U \mkv V$ and $\supp(U,V)=\supp(U)\times\supp(V)$. Therefore, Case B of Theorem~\ref{thm-3-cor} applies, and we have
\begin{equation}\label{Eqn:Lem:BinaryExample:1}
\R(D)[p^{\rho,\delta}_{SUV}] = \RQB(D)[p^{\rho,\delta}_{SUV}].
\end{equation}

\begin{figure}[h!]
\begin{center}
\includegraphics[width=0.5\textwidth]{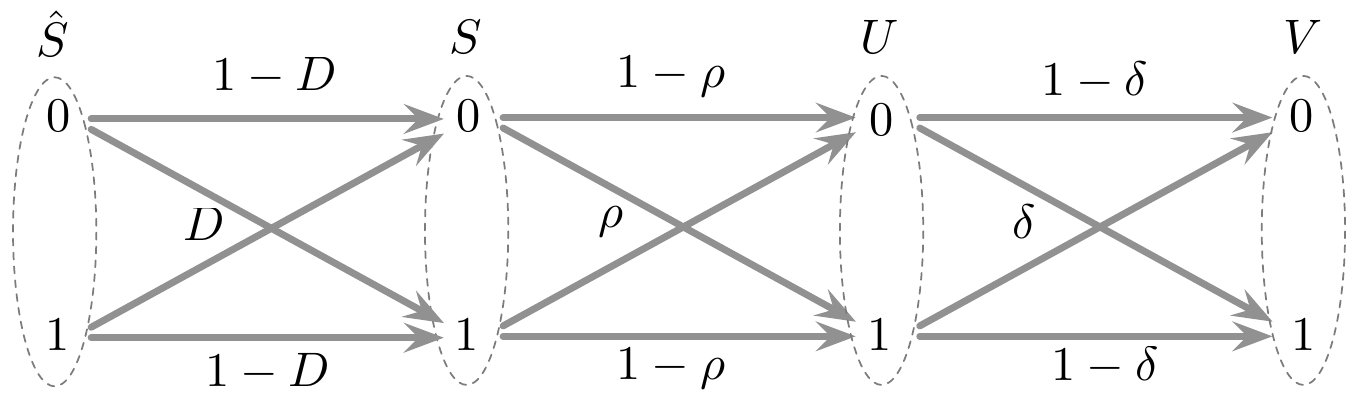}
\caption{Illustration of the quantization $\hat{S}$ used in the proof of Lemma~\ref{Lem:BinaryExample}.}
\label{FIGURE:4}
\end{center}
\end{figure}

We first show that the right hand side of~\eqref{Eqn:Lem:BinaryExample1} is an inner bound for $\RQB(D)[p^{\rho,\delta}_{SUV}]$. Choose $\hat{S}$ as illustrated in Figure~\ref{FIGURE:4} with $\Pr[\hat{S} = 1] = \Pr[\hat{S} = 0] = 1/2$ and $\hat{S} \mkv S \mkv U \mkv V$. Then, $\Exp[d(S,\hat{S})] = D$ and
\begin{subequations}\label{Eqn:Lem:BinaryExample:InnerRate}
\begin{align}
I(S;\hat{S}|U)
&= 
h(D \ast \rho) - h(D),\\
I(S;\hat{S}|V)
&= 
h( (\rho \ast \delta)\ast D) - h(D).
\end{align}
\end{subequations}
Thus,  from the above, we conclude that 
\begin{align}
\left\{(r_{uv}, r_v) \in [0,\infty)^2\,\,\bigg |
\begin{array}{rl} 
r_{uv}\geq & h\big(\rho \ast D \big) - h(D)\\[3pt]
r_v+r_{uv}\geq &h\big((\rho \ast \delta) \ast D \big) - h(D) 
\end{array}
\right\}\subseteq \RQB(D)[p^{\rho,\delta}_{SUV}]\step{\eqref{Eqn:Lem:BinaryExample:1}}{=} \R(D)[p^{\rho,\delta}_{SUV}].
\end{align}

To establish that choosing $\hat S$ according to Fig.~\ref{FIGURE:4} suffices to cover the entire rate region, we need to show that the LHS of the above equation is an outer  bound for $\RQB(D)[p^{\rho,\delta}_{SUV}]$. To that end, we proceed as follows.
\begin{align}
\R(D)[p^{\rho,\delta}_{SUV}]
&\step{(a)}{=}
\bigcup_{
\substack{\hat{S} \mkv S \mkv (U,V)\\\Exp[d(S,\hat{S})] \leq D}}
\left\{(r_{uv}, r_v) \in [0,\infty)^2\,\,\bigg |
\begin{array}{rl} 
r_{uv}\geq & I(S;\hat{S}|U)\\[5pt]
r_v+r_{uv}\geq & I(S;\hat{S}|V)
\end{array}
\right\}\notag\\
&\step{(b)}{\subseteq}
\left\{(r_{uv}, r_v) \in [0,\infty)^2\,\,\Bigg |
\begin{array}{rl} 
r_{uv}\geq & \min\limits_{\substack{\hat{S} \mkv S \mkv U\\\Exp[d(S,\hat{S})] \leq D}} I(S;\hat{S}|U)\\[5pt]
r_v+r_{uv}\geq & \min\limits_{\substack{\hat{S} \mkv S \mkv V\\\Exp[d(S,\hat{S})] \leq D}} I(S;\hat{S}|V)
\end{array}
\right\}\label{eqn-bimini}\\
&\step{(c)}{\subseteq}
\left\{(r_{uv}, r_v) \in [0,\infty)^2\,\,\bigg |
\begin{array}{rl} 
r_{uv}\geq & h\big(\rho \ast D \big) - h(D)\\[3pt]
r_v+r_{uv}\geq &h\big((\rho \ast \delta) \ast D \big) - h(D) 
\end{array}
\right\},
\end{align}
where (a) follows from Case B of Theorem~\ref{thm-3-cor} of Sec.~\ref{Mainresults-rrdefns} and the definition of $\RQB(D)$; (b) creates an outer bound by relaxing the optimizing problem; and (c) follows from using Steinberg's \emph{common-reconstruction function for the binary symmetric source} (see (14) of \cite{SteinbergCR}) to obtain the solutions to the two minimizations in \eqref{eqn-bimini}. The channel $p_{S|\hat S}$ that \emph{simultaneously} minimizes both optimization problems in \eqref{eqn-bimini} is precisely the choice in Fig.~\ref{FIGURE:4}.
\end{IEEEproof}
Notice carefully that the proof of the above result uses one important aspect of Steinberg's characterization of the quantize-and-bin rate region  for the point-to-point rate-distortion problem.
\begin{itemize}
\item When the source and the receiver side-information are related by a binary symmetric channel , the optimal  \emph{reverse test channel} $p_{S|\hat S}$ to be a binary symmetric channel with crossover probability $D$ \emph{independent} of the crossover probability of the channel relating the source and the receiver side information (see (14) of \cite{SteinbergCR}). 
\end{itemize} 
Consequently, the following general observation holds.
\begin{remark}\label{rem:3}
The $D$-admissible rate region for any source $q_{SUV}$ that meets the conditions of Case A or Case B of Theorem~\ref{thm-3-cor} and for which $q_{U|S}$ and $q_{V|S}$ are binary symmetric channels with crossover probabilities $\rho, \rho\ast \delta$, respectively, is given by  \eqref{Eqn:Lem:BinaryExample1}. 
 \end{remark}

\begin{figure}[t!]
\begin{center}
\includegraphics[width=0.6\textwidth]{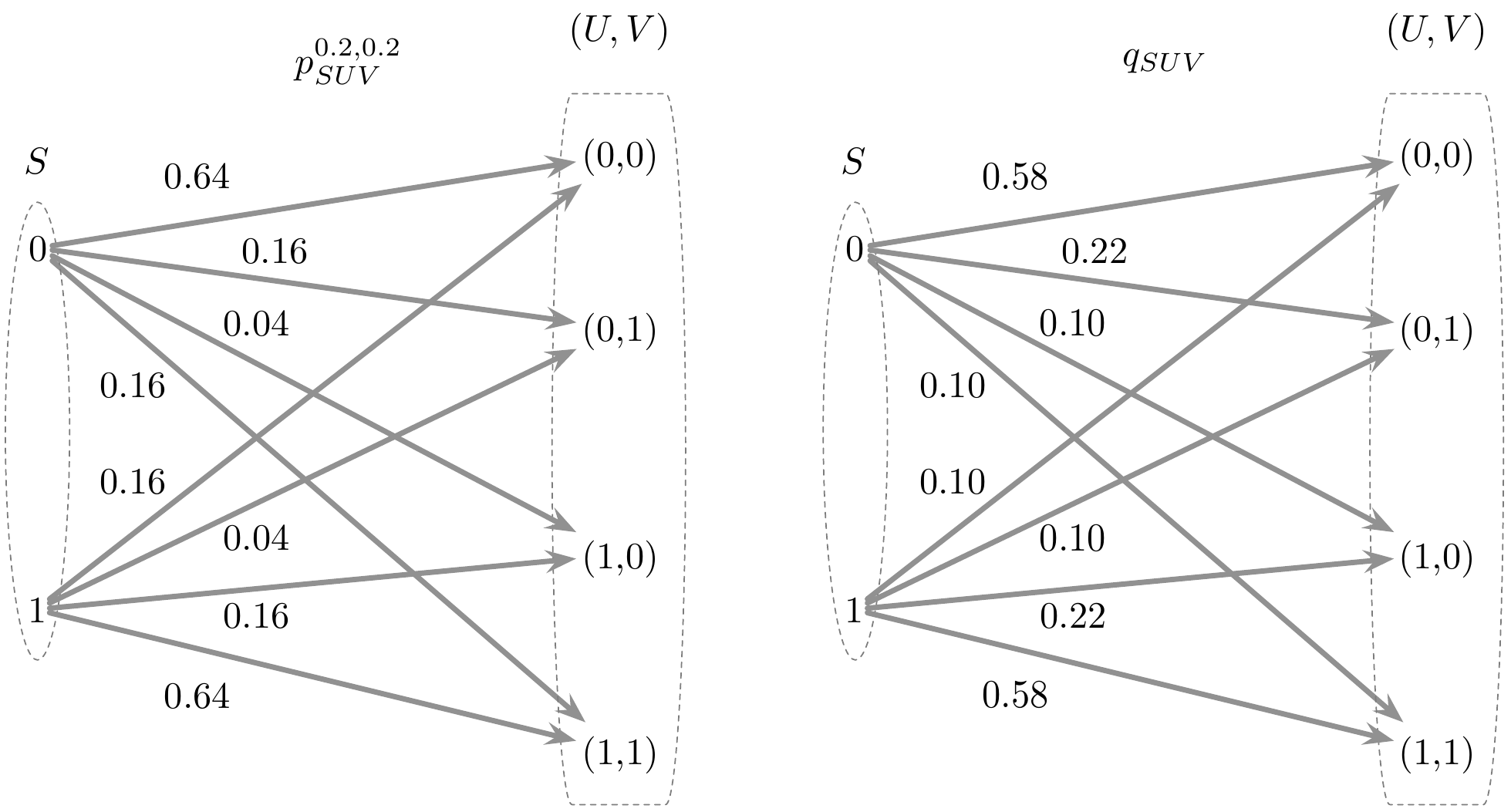}
\caption{Two pmfs that have the same $D$-admissible rate regions. }
\label{FIGURE:5}
\end{center}
\end{figure}

Consider the two pmfs given in Fig.~\ref{FIGURE:5} in support of the remark. A simple computation will yield that when  $(S,U,V)\sim q_{SUV}$ or $(S,U,V)\sim p_{SUV}^{0.2,0.2}$: (a) $\supp(S,U,V) = \eS\times\U\times \V$; (b) $q_{U|S}$ is a binary symmetric channel with crossover probability $0.2$;  and (c) $q_{V|S}$ is a binary symmetric channel with crossover probability $0.32$. Remark~\ref{rem:3} assures that the quantize-and-bin strategy is optimal for both these sources, and that $\R(D)[p^{0.2,0.2}_{SUV}] =\R(D)[q_{SUV}] $.
\begin{figure}[h!]
\begin{center}
\includegraphics[width=0.5\textwidth]{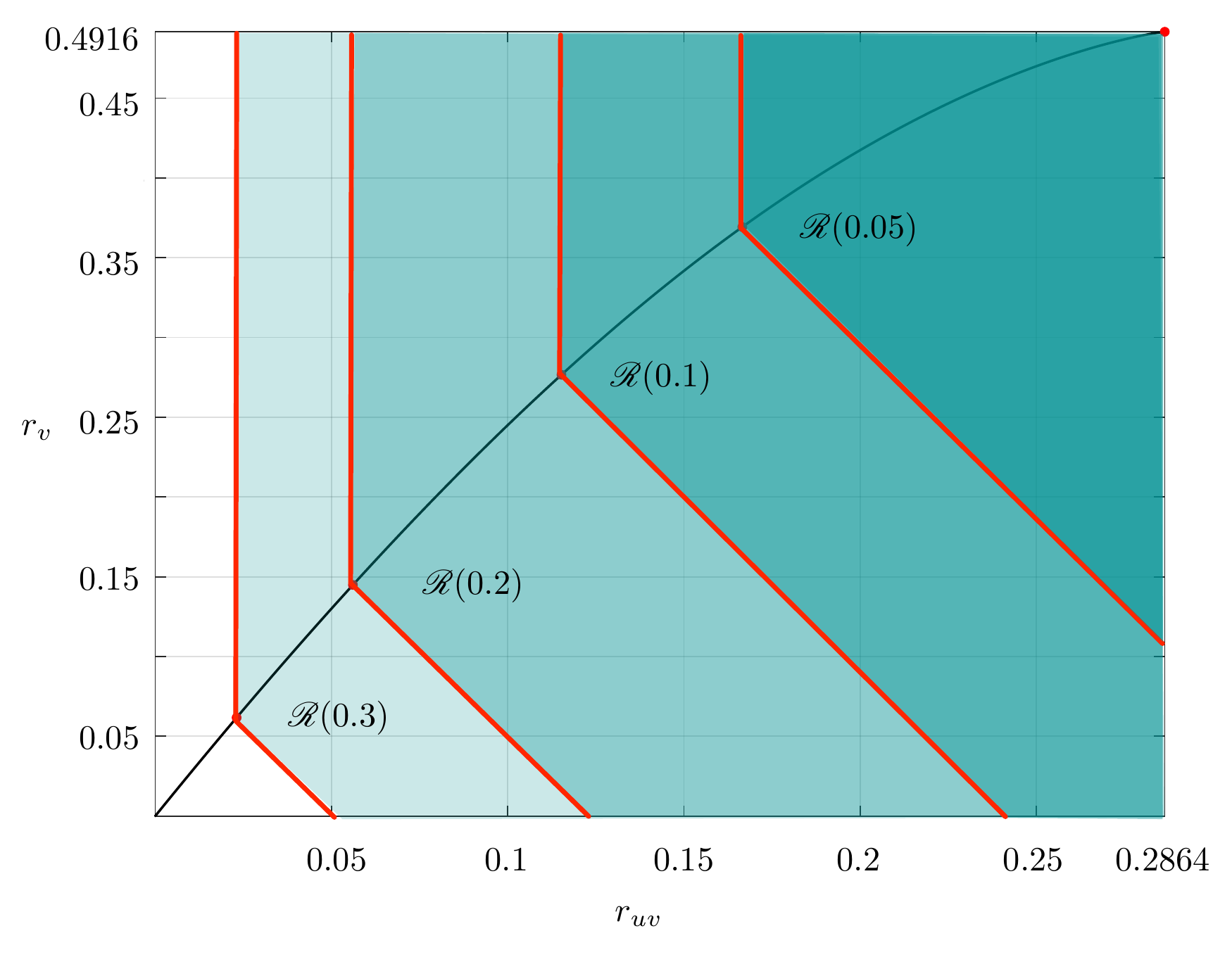}
\caption{The $D$-admissible rate region for $p^{\rho,\delta}_{SUV}$ for $\rho = 0.05$ and $\rho = 0.2$. }
\label{FIGURE:6}
\end{center}
\end{figure}

Figure~\ref{FIGURE:6} presents a simulation of the $D$-admissible rate region for $p^{\rho,\delta}_{SUV}$ for $\rho = 0.05$ and $\rho = 0.2$ as an illustration of the result in Lemma~\ref{Lem:BinaryExample}. The figure presents the rate region for four values of of $D$, namely $D=0.05, 0.1, 0.2$ and $0.3$. The rate region for each of these values for $D$ is bounded by two lines -- one with slope $-1$ corresponding to the message sum rate $r_{uv}+r_v$, and one with infinite slope corresponding to the common message rate $r_{uv}$. When $D=0.5$, no communication is required, and the rate region is the entire non-negative quadrant. As $D$ is made smaller, the $D$-admissible rate region shrinks, and the minimum required communication rate for the common message increases. The admissible rate region shrinks until eventually $D=0$, at which point, the corresponding admissible rate region is given by 
\begin{subequations}
\begin{align}
r_{uv} &\geq H(S|U) = h(\rho) = h(0.05)=0.2864,\\
r_{uv}+r_v& \geq H(S|V) = h(\rho\ast \delta) = h(0.23)= 0.7780.
\end{align}
\end{subequations}
This region is entirely outside this figure except for the vertex, which is located at the top-right corner of the figure. 

\subsection{On the Discontinuity of $\R(D)$}

In source coding problems, the continuity of rate regions with the underlying source distributions allows for small changes in source distributions to translate to small changes in boundary of the rate region. Continuity is therefore essential in practice to allow the communications system engineer to estimate  the source distribution and use the estimate to choose a suitable system operating point. When a single-letter characterization of a source coding rate region is known, it is possible to establish its continuity w.r.t the underlying rate region using  the continuity of Shannon's information measures on finite alphabets~\cite[Chap.~2.3]{Yeung-2008-B}. For example, \cite[Lem.~7.2]{Csiszar-2011-B} considers the continuity of the standard rate-distortion function, and~\cite{Gu-Sep-2008-C} and~\cite{Chen-Jul-2008-C} study the semicontinuity of various source networks.  However, the rate regions of certain source-coding problems are known to be discontinuous in the source distribution especially when they involve zero-error or functional reconstruction constraints~\cite[Ch.~11]{Csiszar-2011-B},~\cite{Gu-Sep-2008-C, Han-Kobayashi-Functional1987}.

Despite the absence of any such reconstruction constraints, it turns out that the $D$-admissible rate region studied here is \emph{discontinuous} in the pmf $p_{SUV}$. The discontinuity arises rather due to the fact that we require the two  reconstructions that are generated at two different locations in the network to agree (albeit with vanishing block error probability). Intuitively, in each of the cases where a single-letter characterization of the $D$-admissible rate region is known, the discontinuity can be attributed to the G\'acs-K\"orner common randomness in the argument of the single-letter reconstruction function; the G\'acs-K\"orner common randomness, and more precisely, its entropy can easily be seen to be discontinuous in the pmf $p_{SUV}$. 

We illustrate this phenomenon by a simple example. Recall the $D$-admissible rate region $\R(D)[p^{\rho,\delta}_{SUV}]$ of the binary  example given above. We now establish the discontinuity of this problem by showing that 
\begin{equation}
\lim_{\delta \downarrow 0}\ \mathcal{R}(D)[p^{\rho,\delta}_{SUV}]
\neq 
\mathcal{R}(D)[p^{\rho,0}_{SUV}], \qquad 0 < \rho < D < 1/2.
\end{equation}
Suppose that $(S,U,V) \sim p^{\rho,0}_{SUV}$ and $1>D>\rho>0$. Since $\delta=0$, we have $U=V$, and consequently, neither Case A nor Case B of Theorem~\ref{thm-3-cor} is no longer applicable in identifying the $D$-admissible rate region.  However, since  $D > \rho$ and $U = V$, we can obviously achieve the distortion $D$ at $(r_{uv},r_v) = (0, 0)$ by simply choosing $\hat{S}^n = U^n=V^n=\tilde{S}^n$. This yields an average distortion of $\rho<D$, since
\begin{equation}
\frac{1}{n} \sum_{i=1}^n \Exp[d(S_i,\hat{S}_i)] 
= \rho.
\end{equation}
Thus, 
\begin{equation}\label{Eqn:Lem:BinaryExample:q}
\R(D)[p^{\rho,0}_{SUV}] = [0,\infty)^2.
\end{equation}

Lemma~\ref{Lem:BinaryExample} determines $\R(D)[p^{\rho,\delta}_{SUV}]$ for $\delta \in (0,1)$. The mapping $\delta \mapsto h( (\rho \ast \delta) \ast D)$ is continuous on $[0,1]$, so 
\begin{equation}\label{Eqn:Lem:BinaryExample2}
\lim_{\delta \downarrow 0}\ \R(D)[p^{\rho,\delta}_{SUV}]
= \left[\lim_{\delta\downarrow 0} h((\rho \ast \delta) \ast D) - h(D), \infty\right)\times \big[0,\infty\big)
=\big[ h(\rho  \ast D) - h(D), \infty\big)\times \big[0,\infty\big).
\end{equation}
From~\eqref{Eqn:Lem:BinaryExample:q} and~\eqref{Eqn:Lem:BinaryExample2}, we see that the $D$-admissible rate region of $p^{\rho,\delta}_{SUV}$ does not approach that of $p^{\rho,0}_{SUV}$ as $\delta \to 0$. 
\begin{remark}
The above argument can also be used to show the discontinuity at $\delta = 1$, i.e.,
\begin{equation}
\lim_{\delta \uparrow 1}\ \R(D)[p^{\rho,\delta}_{SUV}]
\neq 
\mathcal{R}(D)[p^{\rho,1}_{SUV}].
\end{equation}
\end{remark}

\section{Conclusions}\label{Sec:Conclu}

In this work, we look at a variant of the two-receiver successive refinement problem with the common receiver reconstructions requirement. We present general inner and outer bound for this variant. The outer bound is unique in the sense it is the first information-theoretic single-letter characterization where the source reconstruction at the receivers is explicitly achieved via a function of the G\'acs-K\"orner common randomness between the random variables (both auxiliary and side information) available to the two receivers. Using these bounds, we derive a single-letter characterization of the admissible rate region and the optimal coding strategy for several settings of the joint distribution between the source and the receiver side information variables. Using this characterization, we then establish the discontinuity of the admissible rate region  with respect to the underlying source source distribution even though the problem formulation does not involve zero-error or functional reconstruction constraints.

\appendices 

\section{Proof of Lemma~\ref{Lem:Ancillary2}}\label{APP-A}

Suppose that we have $(a_1,u) \xrightleftharpoons{} (a_1',u')$ in $\mathbb{G}^{A_1U,A_2V}[p_{A_1UA_2V}]$. Then, there must exist a positive integer $k\in \mathbb N$ and  
sets $\big\{(a_2^{(j)},v^{(j)}):  1\leq j \leq k\big\} \textrm{ and } \big\{(a_1^{(j)},u^{(j)}): 1\leq j < k\big\}$ such that 
\begin{subequations}
\begin{align}
p_{A_1UA_2V} (a_1,u,a_2^{(1)},v^{(1)})&>0,\\
p_{A_1UA_2V} (a_1^{(j)},u^{(j)},a_2^{(j)},v^{(j)})&>0, \quad 1\leq j<k,\\
p_{A_1UA_2V} (a_1^{(j)},u^{(j)},a_2^{(j+1)},v^{(j+1)})&>0, \quad 1\leq j<k,\\
p_{A_1UA_2V} (a_1',u',a_2^{(k)},v^{(k)})&>0.
\end{align}
\end{subequations}
Since $\supp(A_1,A_2,U,V) = \supp(A_1,A_2) \times \supp(U,V)$, it follows that the above equations hold if and only if
\begin{subequations}
\begin{alignat}{3}
p_{A_1A_2} (a_1,a_2^{(1)})&>0\textrm{ and } p_{UV} (u,v^{(1)})&>0,&\quad \\
p_{A_1A_2} (a_1^{(j)},a_2^{(j)})&>0\textrm{ and } p_{UV} (u^{(j)},v^{(j)})&>0,&\quad 1\leq j<k,\\
p_{A_1A_2} (a_1^{(j)},a_2^{(j+1)})&>0\textrm{ and }  p_{UV} (u^{(j)},v^{(j+1)})&>0, &\quad 1\leq j<k,\\
p_{A_1A_2} (a_1',a_2^{(k)})&>0\textrm{ and }  p_{UV} (u',v^{(k)})&>0. &\quad
\end{alignat}
\end{subequations}
Thus, it follows that 
\begin{align}
(a_1,u) \xrightleftharpoons{} (a_1',u')\textrm{ in }\mathbb{G}^{A_1V,A_2U}[p_{A_1VA_2U}] \Leftrightarrow {\setstretch{1}\left\{\hspace{-3mm}\begin{array}{c} \,\,\,\,\,a_1\xrightleftharpoons{} a_1'\textrm{ in }\mathbb{G}^{A_1,A_2}[p_{A_1A_2}]\\ u\xrightleftharpoons{} u'\textrm{ in }\mathbb{G}^{U,V}[p_{UV}]\end{array}\right.}. \label{eqn-a1a2uvequiv2}
\end{align}
Similarly, it can be shown that 
\begin{align}
(a_2,v) \xrightleftharpoons{} (a_2',v')\textrm{ in }\mathbb{G}^{A_1V,A_2U}[p_{A_1VA_2U}] \Leftrightarrow {\setstretch{1}\left\{\hspace{-3mm}\begin{array}{c} \,\,\,\,\,a_2\xrightleftharpoons{} a_2'\textrm{ in }\mathbb{G}^{A_1,A_2}[p_{A_1A_2}]\\ v\xrightleftharpoons{} v'\textrm{ in }\mathbb{G}^{U,V}[p_{UV}]\end{array}\right.}. \label{eqn-a1a2uvequiv2}
\end{align}
Consequently, from \eqref{Eqn:GKMapping},  it follows that $\GK^{A_1U,A_2V} \equiv(\GK^{A_1,A_2},\GK^{U,V})$.\hfill\qed

\section{Proof of Lemma~\ref{thm-RegionEquivalence}}\label{APP-B}

Let $(r_{uv},r_v)\in \R^\ddagger_{D,k}$. Then, there must exist pmf $p_{ABCSUV}\in\mc P^{  \ddagger}_{D,k}$ and function $f:\A\times\C\times\V\rightarrow \hat\eS$ such that 
\begin{align}
{\Exp [d (S, f(\GK^{  ABU,ACV}))]\leq D},\label{eqn:reconfn}
\end{align}
and 
\begin{subequations}
\begin{align}
r_{uv} &\geq I(S;A,B|U),\label{eqn-appb1}\\
r_v+r_{uv} &\geq I(S;B|A,C,U,V) +I(S;A,C|V). \label{eqn-appb2}
\end{align}
\end{subequations}

For this choice of $p_{ABCSUV}$, let us define
\begin{align}
\eta^* := \mathop{\min}\limits_{\substack{(u,v)\in\supp(U,V)\\ s\in \supp(S)}} \frac{p_{UV|S}(u,v|s)}{p_{UV}(u,v)}.
\end{align}
Since \eqref{eqn:Condition1} holds, we are guaranteed that $\eta^*>0$. Then, for any $(a,b,c)\in\A\times\B\times \C$ and $(u,v)\in\mathcal U\times \V$,
\begin{align}
p_{ABCUV}(a,b,c,u,v)&=\sum_{s\in\supp(S)}p_{ABCS}(a,b,c,s) p_{UV|S}(u,v|s)\notag\\
&\stackrel{}{\geq}\eta^*\sum_{s\in\supp(S)}p_{ABCS}(a,b,c,s) p_{UV}(u,v)\notag\\
&=\eta^*p_{ABC}(a,b,c)p_{UV}(u,v).\\
p_{ABCUV}(a,b,c,u,v)&=\sum_{s\in\supp(S)}p_{ABC|S}(a,b,c,s) p_{UV|S}(u,v|s)\notag\\
&\stackrel{}{\leq}\frac{1}{\mpv_S}\sum_{s\in\supp(S)}p_{ABCS}(a,b,c,s) p_{UVS}(u,v,s)\notag\\
&\leq \frac{p_{UV}(u,v)}{\mpv_S} \sum_{s\in\supp(S)}p_{ABCS}(a,b,c,s)\notag\\
&=\frac{p_{ABC}(a,b,c)p_{UV}(u,v)}{\mpv_S}  .
\end{align}
Hence, we also have $ \supp(A,B,C,U,V) = \supp(A,B,C)\times \supp(U,V).$
An application of Lemma~\ref{Lem:Ancillary2} of Sec.~\ref{Sec:Gacs-Korner} with $A_1=(A,C)$ and $A_2=(A,B)$ then  yields the following conclusion. 
\begin{align}
\GK^{ABU,ACV}\equiv (\GK^{AB,AC}, \GK^{U,V}) \label{eqn:GK-decorr1}.
\end{align}
Now, define random variable $\tilde A:= \GK^{AB,AC}$ over $\tilde A \subseteq \A\times\C$, and let $\tilde p_{\tilde ASUV}$ denote the {pmf} of $(\tilde A,S,U,V)$. Then, by construction, $\tilde A\mkv S \mkv (U,V)$.  Note that both $\tilde A$ and $\GK^{U,V}$ are functions of $\GK^{\tilde AV,\tilde AU}$, i.e.,
\begin{align}
H\big(\tilde A, \GK^{U,V}\,\big|\,\GK^{\tilde AV,\tilde AU}\big)& = 0.
\end{align}
Combining the above with \eqref{eqn:GK-decorr1}, yields $H(\GK^{ABU,ACV}|\GK^{\tilde AU,\tilde AV})=0$. Let $\tilde{f}:\tilde \A \times \V\rightarrow \A\times\C\times\V$ be such that 
\begin{align}
\tilde{f}(\GK^{\tilde AU,\tilde AV})=\GK^{ABU,ACV}.
\end{align}
Then, from \eqref{eqn:reconfn}, ${\Exp [d (S, f(\tilde{f}(\GK^{\tilde AV,\tilde AU})))]\leq D}$. Hence, since $\tilde \A \subseteq \A\times \C$, we definitely have $\tilde p_{\tilde ASUV}\in \mc{P}^*_{D,k^2}$. Lastly, note that 
\begin{align}
r_{uv}\step{\eqref{eqn-appb1}}\geq I(S;A,B|U)\geq  I(S;\GK^{AB,AC}|U)=I(S;\tilde A|U),\\
r_{v}+r_{uv}\step{\eqref{eqn-appb2}}\geq I(S;A,C|V)\geq  I(S;\GK^{AB,AC}|V)=I(S;\tilde A|V).
\end{align}
Hence, it follows that $(r_{uv}, r_v) \in\R^*_{D,k^2}$.\hfill\qed

\section{Proof of Lemma~\ref{thm-RegionEquivalence2}}\label{APP-C}

 We will begin by limiting the size for the auxiliary RV $A$, and then present bounds for the alphabet sizes for $B$ and $C$, respectively. To bound the size of $\A$, we need to preserve: (a) $\{p_S(s):s\in\eS\}$; (b) six information functionals $H(S|A,U)$, $H(S|A,V)$, $H(S|A,B,U)$, $H(S|A,C,V)$, $H(S|A,C,U,V)$, and $H(S|A,B,C,U,V)$; and lastly, (c) the reconstruction constraint \eqref{eqn-Defn1eq3}.

We begin by fixing $\check{s}\in\eS$. Define the following $|\eS|+6$ continuous, real-valued functions on $\mathscr{P}(\B\times\C\times\eS)$, the set of all {pmf}s on $\B\times\C\times\eS$. Let $\pi\in\mathscr{P}(\B\times\C\times\eS)$.
\begin{subequations}
\begin{align}
\psi_{s}(\pi)&=\sum_{(b,c)\in\B\times \C} \pi(b,c,s),\quad s\in\eS\setminus\{\check{s}\}\label{psifns-1},\\
\psi_{1}(\pi)&=-\sum\limits_{s,u} \left(\sum_{b,c} \pi(b,c,s)p_{U|S}(u|s)\right) \log_2 \left(\frac{\sum\limits_{b,c} \pi(b,c,s)p_{U|S}(u|s)}{\sum\limits_{b,c,s'} \pi(b,c,s')p_{U|S}(u|s')}\right),\\
\psi_{2}(\pi)&=-\sum\limits_{s,v} \left(\sum_{b,c} \pi(b,c,s)p_{V|S}(v|s)\right) \log_2 \left(\frac{\sum\limits_{b,c} \pi(b,c,s)p_{V|S}(v|s)}{\sum\limits_{b,c,s'} \pi(b,c,s')p_{V|S}(v|s')}\right),\\
\psi_{3}(\pi)&=-\sum\limits_{b,s,u} \left(\sum_{c} \pi(b,c,s)p_{U|S}(u|s)\right) \log_2 \left(\frac{\sum\limits_c \pi(b,c,s)p_{U|S}(u|s)}{\sum\limits_{c,s'} \pi(b,c,s')p_{U|S}(u|s')}\right),\\
\psi_{4}(\pi)&=-\sum\limits_{c,s,v} \left(\sum_{b} \pi(b,c,s)p_{V|S}(v|s)\right) \log_2 \left(\frac{\sum\limits_b \pi(b,c,s)p_{V|S}(v|s)}{\sum\limits_{b,s'} \pi(b,c,s')p_{V|S}(v|s')}\right),\\
\psi_{5}(\pi)&=-\sum\limits_{b,s,u,v} \left(\sum_{c} \pi(b,c,s)p_{UV|S}(u,v|s)\right) \log_2 \left(\frac{\sum\limits_c \pi(b,c,s)p_{UV|S}(u,v|s)}{\sum\limits_{c,s'} \pi(b,c,s')p_{UV|S}(u,v|s')}\right),\\
\psi_{6}(\pi)&=-\sum\limits_{b,c,s,u,v}  \pi(b,c,s)p_{UV|S}(u,v|s) \log_2 \left(\frac{\pi(b,c,s)p_{UV|S}(u,v|s)}{\sum\limits_{s'} \pi(b,c,s)p_{UV|S}(u,v|s)}\right)\label{psifns-7}.
\end{align}
\end{subequations}
Note that preserving condition \eqref{eqn-Defn1eq3} is not straightforward because of the presence of G\'{a}cs-K\"{o}rner common randomness function. Consequently, this condition has to be split into two parts, which have to be combined together non-trivially. However, this approach requires the application of the Support Lemma~\cite[p.~631]{ElGamalKimBook} infinitely many number of times, along with a suitable limiting argument. To preserve \eqref{eqn-Defn1eq3}, define for each $m\in\mathbb{N}$, a continuous function $\psi_{7,m}: \mathscr{P}(\B\times\C\times\eS)\rightarrow \mathbb{R}$ by
\begin{equation}
\psi_{7,m}(\pi)=  \min_{\substack{g:\B\times\U\rightarrow \hat{\eS}\\f:\C\times\V\rightarrow \hat{\eS}}} \left[\sum\limits_{b,c,s,u,v}\pi(b,c,s)p_{UV|S}(u,v|s)\Big(m\bar{D}\bar{\mathds{1}}\{f (c,v)=g(b,u)\} +d(s,f(c,v))\Big)\right] \label{eqn:Cara1-1}.
\end{equation}
Note that $\psi_{7,m}$ links together the distortion requirement with the probability that the reconstructions are different. Pick any pmf $\mathsf p_{ABCSUV}\in \mathcal{P}^{\ddagger}_{D,k}$. Then, by definition, it follows that there exist functions $f:\A\times\C\times \V\rightarrow \hat \eS$ and $g:\A\times\B\times \V\rightarrow \hat \eS$
\begin{subequations}
\begin{align}
\Pr[f(A,C,V)\neq g(A,B,U)] &= 0,\\
\Exp[d(S,f(A,C,V))]\leq D.
\end{align}
\end{subequations}
Consequently, 
\begin{align}
\sum_{a\in\A}p_A(a)\psi_{7,m}(\mathsf p_{BCS|A=a})&\leq D, \quad \forall\, m\in\mathbb{N}.
\end{align}
Combining the above with \eqref{psifns-1}-\eqref{psifns-7}, we see that
\begin{subequations}
\begin{align}
\sum_{a\in\A}p_A(a)\psi_s(\mathsf p_{BCS|A=a})&=\mathsf p_S(s)=p_S(s),\quad s\in\eS\setminus\{\check{s}\},\\
\sum_{a\in\A}p_A(a)\psi_1(\mathsf p_{BCS|A=a})&=H(S|A,U)[\mathsf p],\\
\sum_{a\in\A}p_A(a)\psi_2(\mathsf p_{BCS|A=a})&=H(S|A,V)[\mathsf p],\\
\sum_{a\in\A}p_A(a)\psi_3(\mathsf p_{BCS|A=a})&=H(S|A,B,U)[\mathsf p],\\
\sum_{a\in\A}p_A(a)\psi_4(\mathsf p_{BCS|A=a})&=H(S|A,C,V)[\mathsf p],\\
\sum_{a\in\A}p_A(a)\psi_5(\mathsf p_{BCS|A=a})&=H(S|A,B,U,V)[\mathsf p],\\
\sum_{a\in\A}p_A(a)\psi_6(\mathsf p_{BCS|A=a})&=H(S|A,B,C,U,V)[\mathsf p],\\
\sum_{a\in\A}p_A(a)\psi_{7,m}(\mathsf p_{BCS|A=a})&\leq D, \quad \forall\, m\in\mathbb{N}.
\end{align}
\end{subequations}
For each $m\in\mathbb{N}$,  apply the Support Lemma with  $|\eS|+6$ functions $(\{\psi_s\}_{s\in\eS\setminus\{\check s\}}, \psi_1,\psi_2 ,  \psi_3,\psi_4, \psi_5, \psi_6, \psi_{7,m})$ to identify a {pmf} $q_{A_mBCSUV}$ with $|{\A}_m|\leq |\eS|+6$ such that $(A_m,B,C)\mkv S \mkv (U,V)$ and 
\begin{subequations}
\begin{align}
\sum_{a\in\A_m}q_{A_m}(a)\psi_s(q_{BCS|A_m=a})&=\mathsf p_S(s)=p_S(s), \quad s\in\eS\setminus\{\check{s}\} \label{eqn-Qm1},\\
\sum_{a\in\A_m}q_{A_m}(a)\psi_1(q_{BCS|A_m=a})&=H(S|A,U)[\mathsf p]\label{eqn-Qm2},\\
\sum_{a\in\A_m}q_{A_m}(a)\psi_2(q_{BCS|A_m=a})&=H(S|A,V)[\mathsf p]\label{eqn-Qm3},\\
\sum_{a\in\A_m}q_{A_m}(a)\psi_3(q_{BCS|A_m=a})&=H(S|A,B,U)[\mathsf p]\label{eqn-Qm4},\\
\sum_{a\in\A_m}q_{A_m}(a)\psi_4(q_{BCS|A_m=a})&=H(S|A,C,V)[\mathsf p]\label{eqn-Qm5},\\
\sum_{a\in\A_m}q_{A_m}(a)\psi_5(q_{BCS|A_m=a})&=H(S|A,B,U,V)[\mathsf p]\label{eqn-Qm6},\\
\sum_{a\in\A_m}q_{A_m}(a)\psi_6(q_{BCS|A_m=a})&=H(S|A,B,C,U,V)[\mathsf p]\label{eqn-Qm7},\\
\sum_{a\in\A_m}q_{A_m}(a)\psi_{7,m}(q_{BCS|A_m=a})&\leq D.\label{eqn-supportbound}
\end{align}
\end{subequations}

After possibly renaming of elements, we may assume that the alphabet of each of the auxiliary RVs $A_m$ is the same, say $\A^*$. Note that the optimal reconstruction functions $f_m,\,g_m$ (see \eqref{eqn:Cara1-1}) for the choice $q_{A_mBCSUV}$ meeting \eqref{eqn-supportbound}  satisfy
\begin{subequations}
\begin{align}
\Pr\big[f_m(A_m,C,V)\neq g_m(A_m,B,U)\big]&\leq \textstyle \dfrac{D}{\bar{D}m}\leq \textstyle\frac{1}{m}\label{eqn-decoder1}, \\
\Exp\big[d(S,f_m(A_m,C,V)\big]&\leq D\label{eqn-decoder2}.
\end{align}
\end{subequations}
Since the number of functions from the set  $\A^*\times\B\times \U$ (or $\A^*\times\C\times\V$) to $\hat{\eS}$ is finite, the sequence $\{(f_m,g_m)\}_{m\in\mathbb N}$ must contain infinitely many repetitions of at least one pair of reconstruction functions. Therefore, let
\begin{align}
\omega_0=\min\big\{ i\in\mathbb{N}:\big|\{ j\in\mathbb{N}:(f_j,g_j)=(f_i,g_i)\}\big|=\infty\big\}.\nonumber
\end{align}
Let $\{q_{A_{m_j}BCSUV}\}_{j\in\mathbb{N}}$ be the subsequence of $\{q_{A_{m}BCSUV}\}_{m\in\mathbb{N}}$ with $(f_{m_j},g_{m_j})=(f_{\omega_0},g_{\omega_0})$. By the Bolzano-Weierstrass theorem~\cite{MathAnalysisApostol}, we can find a subsequence $\{q_{A_{m_{j_l}}SUV}\}_{l\in\mathbb{N}}\subseteq \{q_{A_{m_j}SUV}\}_{j\in\mathbb{N}}$ that converges to, say, $q^*_{A^*BCSUV}$. Since $\{\psi_s\}_{s\in\eS\setminus\{\check{s}\}}, \psi_1$, $\psi_2$, $\psi_3$, $\psi_4$,  $\psi_5$,  $\psi_6$ and $\{\psi_{7,m}\}_{m\in\mathbb N}$ are continuous in their arguments, by taking appropriate limits of \eqref{eqn-Qm1}-\eqref{eqn-Qm5}, we see that
 \begin{subequations}
\begin{align}
\sum_{a\in\A^*}\,\,q_{A^*}(a)\psi_s(q_{BCS|A^*=a})&\stackrel{\eqref{eqn-Qm1}}{=} \mathsf p_S(s),\,\,\, s\in\eS\setminus\{\check s\},\\
\sum_{a\in\A^*}\,\,q_{A^*}(a)\psi_1(q_{BCS|A^*=a})&\stackrel{\eqref{eqn-Qm2}}{=} H(S|A,U)[\mathsf p],\\
\sum_{a\in\A^*}\,\,q_{A^*}(a)\psi_2(q_{BCS|A^*=a})&\stackrel{\eqref{eqn-Qm3}}{=}H(S|A,V)[\mathsf p],\\
\sum_{a\in\A^*}\,\,q_{A^*}(a)\psi_3(q_{BCS|A^*=a})&\stackrel{\eqref{eqn-Qm4}}{=} H(S|A,B,U)[\mathsf p],\\
\sum_{a\in\A^*}\,\,q_{A^*}(a)\psi_4(q_{BCS|A^*=a})&\stackrel{\eqref{eqn-Qm5}}{=}H(S|A,C,V)[\mathsf p],\\
\sum_{a\in\A^*}\,\,q_{A^*}(a)\psi_5(q_{BCS|A^*=a})&\stackrel{\eqref{eqn-Qm6}}{=}H(S|A,B,U,V)[\mathsf p],\\
\sum_{a\in\A^*}\,\,q_{A^*}(a)\psi_6(q_{BCS|A^*=a})&\stackrel{\eqref{eqn-Qm7}}{=}H(S|A,B,C,U,V)[\mathsf p].
\end{align}
 \end{subequations}
Let $\mc{S}:= \{(a,b,c,u,v):f_{\omega_0}(a,c,v)\neq g_{\omega_0}(a,b,u)\}$. Then, we have
\begin{align}
\Pr[f_{\omega_0}(A^*,C,V)\neq g_{\omega_0}(A,B,U)] &=\Pr[(A^*,B,C,U,V)\in\mc{S}]\notag\\ &= \lim_{l\rightarrow\infty} \Pr[(A_{m_{j_l}},B,C,U,V)\in\mc{S}]\notag \\
&\stackrel{}{=} \lim\limits_{l\rightarrow\infty} \Pr[f_{m_{j_l}}(A_{m_{j_l}},C,V)\neq g_{m_{j_l}}(A_{m_{j_l}},B,U)] \stackrel{\eqref{eqn-decoder1}}{=}0.
\end{align}
Note that the above equality holds because $(f_{\omega_0},g_{\omega_0})=(f_{m_{j_l}},g_{m_{j_l}})$ for all $l\in\mathbb{N}$. Similarly, 
\begin{align}
 \Exp[d(S,f_{\omega_0}(A^*,C,V)]=\lim_{j\rightarrow\infty} \Exp[d(S,f_{\omega_0}(A_{m_j},C,V)]\leq D\label{eqn-decoder3}.
 \end{align}
 Thus, we may, without loss of generality, restrict the size of the alphabet of $A$ to be $|\eS|+6$.
 
 Next, to restrict the size of the alphabets of $B$ and $C$, we proceed by first picking $(A,B,C,S,U,V)\in\mathcal P^{\ddagger}_{D,k}$ with $|\A|\leq |\eS|+6$. By definition, it follows that there exist functions $f:\A\times\C\times \V\rightarrow \hat \eS$ and $g:\A\times\B\times \V\rightarrow \hat \eS$
 \begin{subequations}
\begin{align}
\Pr[f(A,C,V)\neq g(A,B,U)] &= 0,\label{eqn-reconconst-alphbnd1}\\
\Exp[d(S,f(A,C,V))]\leq D.\label{eqn-reconconst-alphbnd2}
\end{align}
\end{subequations}
Unfortunately, we cannot limit the the sizes of $\B$ and ${\C}$ by invoking  Carath\'{e}odory's Theorem because of the inability to preserve $I(S; {AB}|U)$, $(S; {AC}|V)$ and the constraint $\Pr[ f( A,  C, V)  \neq g( A,  B,  U) ]=0$ simultaneously. 
 
 However, without loss of generality we can assume that $|\supp(\mathsf{B})| \leq |\hat \eS|^{|\U|}$. To see why that is the case, let $\{u_1,\ldots, u_k\}$ be an enumeration of $\U$. Define auxiliary random variable $\bar{B}$ by 
\begin{align}
\bar{ B} &:= \big({g}( A, B, u_1),\ldots,{g}( A, B, u_k)\big) \in \hat\eS^{|\U|}.
\end{align}
By construction, $\bar B$ is a function of $A$ and $B$, and hence, we are guaranteed that $(A,\bar B, C) \mkv S \mkv (U,V)$, and 
\begin{align}
H({g}(A,B,U) \mid A,\bar{ B}, U  ) & = 0.
\end{align}
Combining the above with  \eqref{eqn-reconconst-alphbnd1} and \eqref{eqn-reconconst-alphbnd2}, we conclude the existence of functions $\bar g: \A\times \bar{\B} \times \U \rightarrow \hat \eS$ such that 
\begin{align}
\Pr[ f(A, C,V)\neq \bar g(A,\bar B,U)] &= 0,\\
\Exp[d(S, f(A, C,V))]\leq D.
\end{align}
Further, since $H(\bar{ B} \mid A,{ B}) = 0$, we are guaranteed that
\begin{subequations}
\begin{align}
I(S; A,B|U) &\geq I(S; A,\bar B|U), \\
I(S; B|A,C,U,V) &\geq I(S; \bar{B}|A,C,U,V).
\end{align}
\end{subequations}
It then follows from Definitions~ \ref{Def:PddagRegionk}  and \ref{Def:RddagRegionk} of Sec.~\ref{Mainresults-rrdefns} that considering random variables $A$ and $B$ with $|\A|> |\eS|+6$ and $|\B|> |\hat \eS|^{|\U|}$ does not enlarge the region, i.e., we can identify different $A^*$ and $\bar B$ using the above argument that operate at the same rate pair. 

We are now only left with bounding the size of $\C$, for which, we can repeat now repeat steps similar to that of $A$. This time, we preserve: (a) the distribution $\mathsf p_{ABS}$; (b) three information functionals $H(S|A,C,V)$, $H(S|A,C,U,V)$ and $H(S|A,B,C,U,V)$; and (c) the reconstruction constraint. Proceeding similarly as in the case of for the random variable $A$, we conclude that $|\C| \leq |\eS|(|\eS|+6) |\hat \eS|^{|\U|}+3$ suffices. Thus, it follows that 
\begin{align}
\R_{k}^\ddagger (D) \subseteq \R_{|\eS|(|\eS|+6) |\hat \eS|^{|\U|}+3}^\ddagger (D)=\R^\ddagger (D), \quad k\in\mathbb{N}.
\end{align}\hfill\qed
 
 \begin{remark}\label{rem-regionequiv2}
 Since $\R^*_k(D)$ has only one auxiliary random variable, the proof of \eqref{eqn-rateregionalphabetbnd1} of Lemma~\ref{thm-RegionEquivalence2} follows closely the portion of the above proof corresponding to the reduction of the size of $A$ alone. For the purposes of Lemma~\ref{thm-RegionEquivalence2}, we only need to preserve two information functionals, namely $H(S|A,U)$, and $H(S|A,V)$. Hence, the proof will only use $\{\psi_s:s\in\eS\setminus\{\check s\}\}$, $\psi_1$, $\psi_2$, and $\psi_{7,m}$ to conclude that $|\A|\leq |\eS|+2$ suffices, and hence
  \begin{align}
\R_{k}^* (D) \subseteq \R_{|\eS|+2}^* (D)=\R^*(D), \quad k\in\mathbb{N}.
\end{align}
 \end{remark}
 
 \begin{remark}\label{rem-regionequiv2b}
For the proof of \eqref{eqn-rateregionalphabetbnd3} of Lemma~\ref{thm-RegionEquivalence2}, we only need to preserve four information functionals, namely $H(S|A,U)$, $H(S|A,V)$, $H(S|A,B,U)$ and $H(S|A,C,V)$. Hence, the proof will only use $\{\psi_s:s\in\eS\setminus\{\check s\}\}$, $\psi_1$, $\psi_2$, $\psi_3$, $\psi_4$ and $\psi_{7,m}$ to conclude that $|\A|\leq |\eS|+4$ suffices. The bound for $\B$ is the same as in the above proof. The final argument for bounding $\C$ requires the preservation of (a) the distribution $\mathsf p_{ABS}$; (b) the information functionals $H(S|A,C,V)$; and (c) the reconstruction constraint.
\begin{align}
\R_{k}^\dagger (D) \subseteq \R_{|\eS|(|\eS|+4) |\hat \eS|^{|\U|}+1}^\dagger (D)=\R^\dagger (D), \quad k\in\mathbb{N}.
\end{align}
 \end{remark}
 
\section{Proof of Lemma~\ref{Lem:RstarConvex}}\label{APP-D}

Since $\R^{*}(D)\subseteq \mathbb{R}^2$, we only need to show that the line segment between any two points in $\R^{*}(D)$ lies completely within $\R^{*}(D)$. To do so, pick $(r_u',r_{uv}'), (r_u',r_{uv}'') \in \R^{*}(D)$ and $ \lambda\in [0,1]$. Then, by definition, we can find pmfs $q_{A'SUV}, q_{A''SUV}\in \mc P_{D, |\eS|+2}^*$ such that
\begin{subequations} 
\begin{align}
r_{uv}' &\geq I(S; A'|U),\\
r_v'+r_{uv}' &\geq I(S; A'|V),\\
r_{uv}'' &\geq I(S; A''|U),\\
r_v''+r_{uv}'' &\geq I(S; A''|V).
\end{align}
\end{subequations}
Further, there must also exist functions $f':\A' \times \U \rightarrow \hat \eS$, $g':\A' \times \V \rightarrow \hat \eS$, $f'':\A'' \times \U \rightarrow \hat \eS$, and $g'':\A'' \times \V \rightarrow \hat \eS$ such that 
\begin{subequations}\label{eqn:convexR*1}
\begin{alignat}{3}
\Exp(S, f'(A',U) ] &=\sum_{a',s,u,v} q_{A' SUV}(a',s,u,v) d(s, f'(a',u)) &\leq D,\\
\Exp(S,f''(A'',U) ] &=\sum_{a'',s,u,v} q_{A'' SUV}(a'',s,u,v) d(s, f''(a'',u))\, &\leq D,
\end{alignat}
\end{subequations}
and 
\begin{subequations}\label{eqn:convexR*4}
\begin{align}
\Pr[f'(A',U) \neq g'(A',V)] &= 0,\\
\Pr[f''(A'',U) \neq g''(A'',V)] &= 0,
\end{align}
\end{subequations}
Without loss of generality, we may assume that the alphabets of $A'$ and $A''$ are disjoint, i.e., $\A' \cap \A'' =\emptyset$. Let us define $\tilde \A \triangleq \A'\cup\A''$. Now, define a joint pmf $q_{T\tilde ASUV}^*$ over $\{0,1\}\times \tilde\A \times \eS \times \U \times \V$ as follows:
\begin{itemize}
\item[1.] Let $T\in \{0,1\}$ be an RV such that $\Pr[T=0] = \lambda$. 
\item[2.] Let 
\begin{align}
q^*_{\tilde ASUV | T=\alpha} = \left\{ \begin{array}{cc} q_{A'SUV} ,& \alpha = 0\\ q_{A''SUV}, & \alpha = 1 \end{array}\right..\label{eqn:convexR*5}
\end{align}
\end{itemize}
Then by definition $T$ is a function of $\tilde A$ (since $T=0$ if and only if $\tilde A\in \A'$) and $T$ is independent of $(S,U,V)$. Let $q_{\tilde A,S,U,V}^*$ be the marginal of $(\tilde A, S,U,V)$ obtained from $q_{T,\tilde A,S,U,V}^*$. Then, the following hold:
\begin{itemize}
\item[(i)] $q_{\tilde A,S,U,V}^* \in \mc P_{D, 2|\eS|+4}^*$. This follows by defining $\tilde f: \tilde \A \times \U \rightarrow \hat \eS$ and $\tilde f: \tilde \A \times \V \rightarrow \hat \eS$ by 
\begin{subequations} \label{eqn-convfuncdefn}
\begin{align}
\tilde f(\tilde A, U) = \left\{ \begin{array}{cc} f'(A',U), & \tilde A\in\A' \\f''(A'',U), & \tilde A\in\A'' \end{array}\right.,\\
\tilde g(\tilde A, V) = \left\{ \begin{array}{cc} g'(A',V) ,& \tilde A\in\A' \\g''(A'',V), & \tilde A\in\A'' \end{array}\right.,
\end{align}
\end{subequations}
and verifying that 
\begin{align}
\Pr[\tilde f(\tilde A,U) \neq \tilde g(\tilde A,V)] &= \sum_{t\in\{0,1\}} q^*_T(t)\Pr[\tilde f(\tilde A,U) \neq \tilde g(\tilde A,V)| T=t] \notag \\
&\step{\eqref{eqn:convexR*5}, \eqref{eqn-convfuncdefn}}{=} \lambda \Pr[f'( A',U) \neq g'( A',V)]+(1-\lambda)\Pr[f''( A'',U) \neq g''( A'',V)] \step{\eqref{eqn:convexR*4}}{=} 0.\\
\Exp(S,\tilde f(\tilde A,U) ] &=\sum_{\tilde a,s,u,v} q_{\tilde ASUV}^*(\tilde a,s,u,v) d(s, \tilde f(\tilde a,u))\\
&=\sum_{\tilde a,s,u,v} q_T^*(t) q_{\tilde ASUV|T}^*(\tilde a,s,u,v|t) d(s, \tilde f(\tilde a,u))\\
&\step{\eqref{eqn:convexR*5}, \eqref{eqn-convfuncdefn}}{=} \lambda\Exp(S,f'(A',U) ]+ (1-\lambda)\Exp(S,f''(A'',U) ]  \step{\eqref{eqn:convexR*1}}{\leq} D.
\end{align}
\item[(ii)] Further, we also have that 
\begin{align}
I(S;\tilde A | U ) &\stackrel{(a)}{=} I(S;\tilde A, T  | U ) \stackrel{(b)}{=} I(S; \tilde A | U, T) = \sum_{t\in\{0,1\}} \Pr[T=t] I (S; \tilde A| U, T=t) \notag\\ &= \lambda I(S; A'|U) + (1-\lambda) I(S; A''|U),
\end{align}
where (a) follows since $T$ is a function of $\tilde A$, and (b) follows by the independence of $T$ and $(A,U)$. Similarly, we can also show that $I(S;\tilde A|V) = \lambda I(S; A'|V) + (1-\lambda) I(S; A''|V)$. 
\end{itemize}
It then follows that 
\begin{subequations}
\begin{align}
\lambda r_{uv}' + (1-\lambda) r_{uv}'' &\geq \lambda I(S; A'|U) + (1-\lambda) I(S; A''|U) = I (S; \tilde A | U),\\
\lambda (r_v+r_{uv}') + (1-\lambda) (r_v''+r_{uv}'') &\geq \lambda I(S; A'|V) + (1-\lambda) I(S; A''|V) = I (S; \tilde A | V).
\end{align}
\end{subequations}
Hence, it follows that any point that is a linear combination of  $(r_u',r_{uv}'), (r_u',r_{uv}'') \in \R^{*}(D)$ is a point in $ \R^{*}_{2|\eS|+6}(D)$, which by Lemma~\ref{thm-RegionEquivalence2}, is identical to $\R^{*}(D)$. Hence, the claim of convexity follows. 

\section{Proof of Lemma~\ref{Lem:RstarCont}}\label{APP-E}

From Definition~\ref{Def:RstarRegionk}, we can find $q_{A_iSUV} \in \mc P_{D_i,k}^*$, and functions $f_i:\A_i\times \V \rightarrow \hat \eS$ and $g_i:\A_i\times \U \rightarrow \hat \eS$ such that 
\begin{subequations}
\begin{align}
r_{uv}^{(i)}        & \geq I(S;A_i|U),\\
r_v^{(i)}+r_{uv}^{(i)} & \geq I(S;A_i|V),
\end{align}
\end{subequations}
and 
\begin{subequations}\label{eqn-APPE3}
\begin{align}
\Pr[ g_i(A_i,U) \neq f_i(A_i,V) ] &=0,\\
 \Exp \big[d \big(S, f_i(A_i,V)\big)\big]&\leq D_i.
\end{align}
\end{subequations}
Perhaps after a round of renaming, we may assume that the alphabets of $\A_i$s are identical, i.e., $\A_i = \A$. Since there are only a finite number of functions from $\A\times \V$ or $\A\times \U$ to $\hat \eS$, the sequence $(f_1,g_1),(f_2,g_2),\ldots$ must contain infinitely many copies of some pair of functions. Let 
\begin{align}
\omega = \min\big\{j\in\mathbb {N}: |\{i\in\mathbb {N}: (f_i,g_i) = (f_j,g_j)\}| =\infty\big\}. 
\end{align}
Let $\{i_j\}_{j\in\mathbb{N}}$ be a subsequence of indices such that $(f_{i_j},g_{i_j})=(f_{\omega},g_\omega)$ for all $j\in\mathbb{N}$. Consider the sequence of pmfs $\{ q_{A_{i_j} SUV}\}_{j\in\mathbb{N}} $. Since $\A\times\eS\times\U\times\V$ is a finite set, by Bolzano-Weierstrass theorem~\cite{MathAnalysisApostol}, a subsequence of pmfs must be convergent. Let $\{{i_j}_k\}_{k\in\mathbb N}$ be one such subsequence, and let $q_{A_{{i_j}_k} SUV } \rightarrow q_{\mathring ASUV}$. By the continuity of the information functional~\cite{Yeung-2008-B}, we see that 
\begin{align}
I(\mathring A; U,V | S) = \lim_{k\rightarrow \infty} I(A_{{i_j}_k}; U,V | S) = 0. \label{eqn-APPE5}
\end{align}
Now, let $\mc F = \{(a,u,v)\in\A\times \U\times \V: f_\omega(a,v) \neq g_\omega(a,u)\}$. Then, 
\begin{align}
\Pr[f_\omega(\mathring A,V) \neq g_\omega (\mathring A,U) ] &= \Pr[(\mathring A,U,V)\in \mc F]\notag\\
&=\sum_{(a,u,v)\in\mc F} q_{\mathring AUV}(a,u,v) \notag\\
& =\sum_{(a,u,v)\in\mc F} \left(\lim_{k\rightarrow \infty} q_{A_{{i_j}_k}UV}(a,u,v)\right)\notag \\
& = \lim_{k\rightarrow \infty} \sum_{(a,u,v)\in\mc F}q_{A_{{i_j}_k}UV}(a,u,v)\notag\\
& = \lim_{k\rightarrow \infty} \Pr[f_\omega(A_{{i_j}_k},V) \neq g_\omega (A_{{i_j}_k},U) ]\step{\eqref{eqn-APPE3}}= 0.\label{eqn-APPE6}
\end{align}
 Further, we see that 
\begin{align}
\Exp \big[d \big(S, f_\omega(\mathring A,V)\big)\big] &= \sum_{a,s,v} q_{\mathring ASV} (a,s,v) d(s, f_\omega(a,v))\notag\\
& = \sum_{a,s,v} \left(\lim_{k\rightarrow \infty} q_{A_{{i_j}_k}UV}(a,u,v)d(s, f_\omega(a,v))\right)\notag\\
& = \lim_{k\rightarrow \infty} \sum_{a,s,v} q_{A_{{i_j}_k}UV}(a,u,v)d(s, f_\omega(a,v))\notag\\
&=  \lim_{k\rightarrow \infty} \Exp \big[d \big(S, f_\omega(A_{{i_j}_k},V)\big)\big]  \step{\eqref{eqn-APPE3}}\leq \lim_{k\rightarrow \infty} D_i = \mathsf D.\label{eqn-APPE7}
\end{align}
Note that in the above two arguments, we have used the fact that $(f_{{i_j}_k},g_{{i_j}_k})=(f_{\omega},g_\omega)$ for $k\in\mathbb N$. Combining \eqref{eqn-APPE5}, \eqref{eqn-APPE6}, and \eqref{eqn-APPE7}, we see that $q_{\mathring ASUV}\in \mc P^*_{D,k}$. Lastly, using the continuity of the information functional~\cite{Yeung-2008-B}, we see that 
\begin{subequations}
\begin{align}
\mathsf{r_{uv}} &= \lim_{k \rightarrow \infty} r_{uv}^{({i_j}_k)} \geq  \lim_{k \rightarrow \infty} I(S; A_{{i_j}_k} | U ) = I(S;\mathring A|U), \\
\mathsf{r_{uv}}+ \mathsf{r_{v}}&= \lim_{k \rightarrow \infty} r_{uv}^{({i_j}_k)}+r_{v}^{({i_j}_k)} \geq  \lim_{k \rightarrow \infty} I(S; A_{{i_j}_k} | V ) = I(S;\mathring A|U). 
\end{align}
\end{subequations}
Hence, it follows that $(\mathsf{r_{uv}}, \mathsf{r_{v}}) \in \R^*_{k}(\mathsf D)$. \hfill\qed

\section{Proof of Theorem~\ref{thm-achieve-thm-2} (Inner Bound)}\label{APP-F}

\begin{figure}[h!]
\begin{center}
\includegraphics[width=0.6\textwidth]{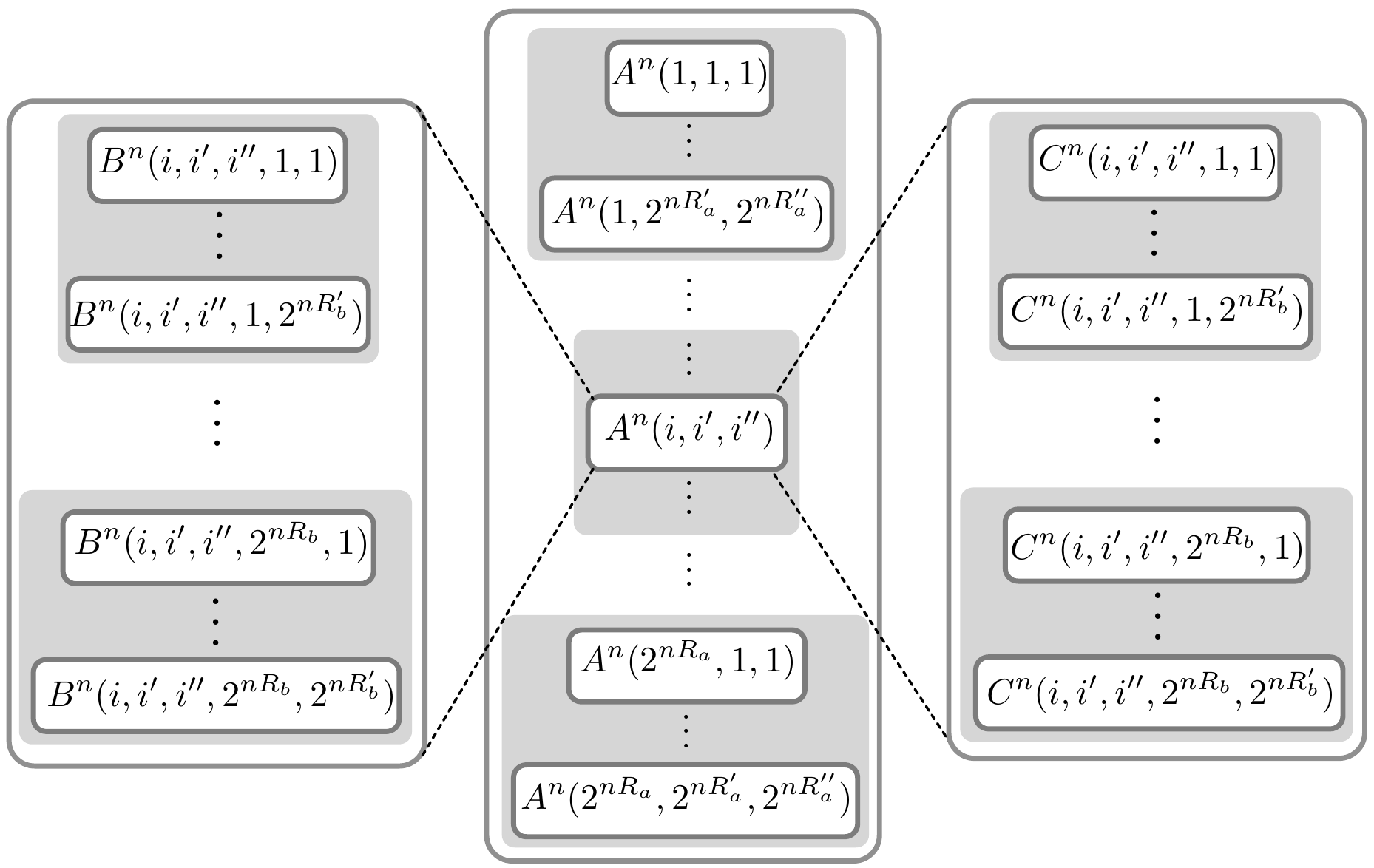}
\caption{A random-coding achievable scheme.}\label{FIGURE:7}
\end{center}
\end{figure}

Pick $p_{ABCSUV}\in \mc{P}_{D,k}^{  \ddagger}$ and $\ve\in(0,\frac{1}{6})$. Let $f:\A\times \B \times \U \rightarrow \A\times\C\times \V$ and $g:\A\times \C \times \V \rightarrow \A\times\C\times \V$ be such that 
$$ f_U(A,B,U) = f_V(A,C,V) = \GK^{ABU,ACV}(A,C,V).$$
Let us denote
\begin{subequations}
\begin{align}
R_a &= I(S;A|U)+2\ve,\\
R_a' &=\max\{I(S;A|V)-I(S;A|U), 0\},\\
R_a''&=\max\{\min\{I(A;U), I(A;V)\}-\ve, 0\},\\
R_b&=I(S;B|A,U)+2\ve,\\
R_b'&=\max\{I(U;B|A) -\ve, 0\},\\
R_c&=I(S;C|A,V)+2\ve,\\
R_c'&=\max\{I(V;C|A) -\ve, 0\}.
\end{align}\label{eqn-InnerBndRates}
\end{subequations}
We build a codebook using the marginals $p_A$, $p_{B|A}$ and $p_{C|A}$ obtained from the chosen joint pmf.  The codebooks for the three auxiliary RVs are constructed as follows using the structure illustrated in Fig.~\ref{FIGURE:7}.
\begin{itemize}
\item For each triple $(i,i',i'')\in\llbracket 1, 2^{nR_a}\rrbracket \times \llbracket 1, 2^{nR_a'}\rrbracket \times \llbracket 1, 2^{nR_a''}\rrbracket$, generate a random codeword $A^n(i,i',i'')\sim p_A^n$ independent of all other codewords. Note that by the choice of rates, the total rate of the $A$-codebook is
\begin{align}
R_a+R_a'+R_a'' \step{\eqref{eqn-InnerBndRates}}{>} I(S;A). \label{eqn-InnerBndArate}
\end{align}
\item For each triple $(i,i',i'')\in\llbracket 1, 2^{nR_a}\rrbracket \times \llbracket 1, 2^{nR_a'}\rrbracket \times \llbracket 1, 2^{nR_a''}\rrbracket$, and pair $(\ell,\ell')\in\llbracket 1, 2^{nR_b}\rrbracket \times \llbracket 1, 2^{nR_b'}\rrbracket $, generate a random codeword $B^n(i,i',i'',\ell,\ell')\sim \prod_{k=1}^n p_{B|A_k(i,i',i'')}$ independent of all other codewords. Note that by the choice of rates, the total rate of the $B$-codebook is
\begin{align}
R_b+R_b'' \step{\eqref{eqn-InnerBndRates}}{>} I(S;B|A).\label{eqn-InnerBndBrate}
\end{align}
\item Similarly, for each triple $(i,i',i'')\in\llbracket 1, 2^{nR_a}\rrbracket \times \llbracket 1, 2^{nR_a'}\rrbracket \times \llbracket 1, 2^{nR_a''}\rrbracket$, and pair $(l,l')\in\llbracket 1, 2^{nR_c}\rrbracket \times \llbracket 1, 2^{nR_c'}\rrbracket $, generate a random codeword $C^n(i,i',i'',l,l')$ randomly using $\prod_{k=1}^n p_{C|A_k(i,i',i'')}$ independent of all other codewords. Note that by the choice of rates, the total rate of the $B$-codebook is
\begin{align}
R_c+R_c'' \step{\eqref{eqn-InnerBndRates}}{>} I(S;C|A). \label{eqn-InnerBndCrate}
\end{align}
\end{itemize}

Upon receiving a realization $s^n$ of $S^n$, the encoder does the following:
\begin{itemize}
\item[1.] It searches for a triple $(i,i',i'')$ such that $(A^n(i,i',i''),s^n)\in  T_\ve^n[p_{AS}]$. 
\item[2.] It then searches for a pair $(\ell,\ell')$ such that $(A^n(i,i',i''),B^n(i,i',i'',\ell,\ell',s^n)\in  T_\ve^n[p_{ABS}]$, and a pair $(l,l')$ such that $(A^n(i,i',i''),C^n(i,i',i'',l,l',s^n)\in  T_\ve^n[p_{ABS}]$. Using \eqref{eqn-InnerBndArate}-\eqref{eqn-InnerBndCrate} and invoking the lossy source coding theorem~\cite[p.~57]{ElGamalKimBook}, and  the Covering Lemma~\cite[p.~62]{ElGamalKimBook}, we see that
\begin{align}
\Pr\left[ \exists\, (i,i',i'',\ell, \ell'): (A^n(i,i',i''), B^n(i,i',i'',\ell,\ell'), C^n(i,i',i'',l,l'),S^n)\in T_\ve^n[p_{ABCS}] \right] \stackrel{n\rightarrow\infty}{\longrightarrow} 0.\label{eqn-codebook1}
\end{align}
\item[3.] The encoder conveys $(i,\ell)$ to both receivers, and $(i',l)$ to the receiver with side information $V$. Note that this strategy corresponds to the following rates 
\begin{subequations}
\begin{align}
r_{uv}&= R_a+R_b=I(S;A,B|U)+4\ve, \\ 
r_{uv}+r_v&= R_a+R_a'+R_b+R_c\notag\\
&=\max\{I(S;A|U), I(S;A|V)\}+ I(S;B|A,U) + I(S;C|A,V)+6\ve.\end{align}
\end{subequations}

\end{itemize}
Further, for any $(i,i',i'',\ell,\ell',l,l')$, by the Markov Lemma~\cite[p.~296] {ElGamalKimBook}, we are guaranteed that 
\begin{align}
\Pr\left[ \begin{array}{c} (A^n(i,i',i''), B^n(i,i',i'',\ell,\ell'),  C^n(i,i',i'',l,l'), S^n,U^n,V^n)\notin T_{2\ve}^n[p_{ABCSUV}]\\\Big |\, (A^n(i,i',i''), B^n(i,i',i'',\ell,\ell'),  C^n(i,i',i'',l,l'), S^n)\in T_\ve^n[p_{ABCS}]\end{array}\right]\stackrel{n\rightarrow\infty}{\longrightarrow} 0.\label{eqn-codebook2}
\end{align}
Thus for sufficiently large $n$, the probability with which we will find a tuple $(i,i',i'',\ell,\ell',l,l')$ such that the corresponding codewords and the source and side information realizations is $\ve$-typical can be made arbitrarily small. Moreover,  as a consequence of the Packing Lemma~\cite[p.~46]{ElGamalKimBook}, we are also guaranteed that
\begin{subequations}
\begin{align}
\Pr\left[ \begin{array}{c}
\exists\, (\tilde i',\tilde i'')\neq (i',i'') \textrm{ s.t. } (A^n(i,i\tilde ',\tilde i''), U^n)\in T_{2\ve}^n[p_{AU}] \\\qquad \Big |\,(A^n(i,i',i''),U^n)\in T_{2\ve}^n[p_{AU}] \end{array}\right]&\stackrel{n\rightarrow\infty}{\longrightarrow} 0,\\
\Pr\left[ \begin{array}{c}
\exists\, (\tilde i',\tilde i'')\neq (i',i'') \textrm{ s.t. } (A^n(i,i\tilde ',\tilde i''), V^n)\in T_{2\ve}^n[p_{AV}] \\\qquad \Big |\,(A^n(i,i',i''),V^n)\in T_{2\ve}^n[p_{AV}] \end{array}\right]&\stackrel{n\rightarrow\infty}{\longrightarrow} 0,\\
\Pr\left[ \begin{array}{c}
\exists\, \tilde \ell' \neq \ell \textrm{ s.t. } (A^n(i,i',i''), B^n(i,i',i'',\ell,\tilde \ell'), U^n)\in T_{2\ve}^n[p_{ABU}] \\\qquad \Big |\,(A^n(i,i',i''), B^n(i,i',i'',\ell,\ell'), U^n)\in T_{2\ve}^n[p_{ABU}] \end{array}\right]&\stackrel{n\rightarrow\infty}{\longrightarrow} 0,\\
\Pr\left[ \begin{array}{c}
\exists\, \tilde l' \neq l \textrm{ s.t. } (A^n(i,i',i''), C^n(i,i',i'',l,l'), V^n)\in T_{2\ve}^n[p_{ACV}] \\\qquad \Big |\,(A^n(i,i',i''), C^n(i,i',i'',l,l'), V^n)\in T_{2\ve}^n[p_{ACV}] \end{array}\right]&\stackrel{n\rightarrow\infty}{\longrightarrow} 0.
\end{align}\label{eqn-codebook4}
\end{subequations}

From~\eqref{eqn-codebook1}, \eqref{eqn-codebook2}, \eqref{eqn-codebook4}, we can choose $n$ large enough such that when averaging over all realizations of the random codebooks, the probability with which all of the following events occur is at least $\ve$. 
\begin{itemize}
\item[(a)] the encoder will be able to identify indices $(i,i',i'',l,l',\ell,\ell')$ such that the corresponding codewords and the realization of $S^n$ are jointly $\ve$-letter typical;
\item[(b)] the identified codewords and the realizations of $(S^n, U^n, V^n)$ are jointly $2\ve$-letter typical;
\item[(c)] the receiver with side information $V$ will identify the indices $(i'',l')$  determined by the encoder; and
\item[(d)] the receiver with side information $U$ will identify the indices $i',i'',\ell'$  determined by the encoder.
\end{itemize}

Then, there must exist a realization of the three codebooks $\{\mathsf{a}^n(i,i',i'')\}$, $\{\mathsf{b}^n(i,i',i'',\ell,\ell')\}$, and $\{\mathsf{c}^n(i,i',i'',l,l')\}$ such that the above four events occur simultaneously with a probability of at least $1-\ve$. For this realization of the codebooks, with probability of at least $1-\ve$, the realizations of the source and side informations $(s^n,u^n,v^n)$, and the selected codewords will be jointly $2\ve$-letter typical, i.e., 
\begin{align} (\mathsf{a}^n(i,i',i''), \mathsf{b}^n(i,i',i'',\ell,\ell'),  \mathsf{c}^n(i,i',i'',l,l'), s^n,u^n,v^n)\in T_{2\ve}^n[p_{ABCSUV}].\label{eqn-codebook5}\end{align}
Note that letter typicality ensures that the support of the empirical distribution $\tilde{q}_{ABCSUV}$ induced by the tuple $ (\mathsf{a}^n(i,i',i''), \mathsf{b}^n(i,i',i'',\ell,\ell'),  \mathsf{c}^n(i,i',i'',l,l'), s^n,u^n,v^n)$ matches that of $q_{ABCSUV}$. Thus, it follows that whenever \eqref{eqn-codebook5} holds,  we have for each $k=1,\ldots, n$, 
\begin{align}
\hat{s}_k &= f_U( \mathsf a_k(i,i',i''), \mathsf b_k(i,i',i'',\ell,\ell'),u_k)= f_V( \mathsf a_k(i,i',i''), \mathsf c_k(i,i',i'',l,l'),v_k) = \tilde{s}_k.
\end{align}
Lastly, since \eqref{eqn-codebook5} holds with a probability of at least $1-\ve$, it is also true that the reconstructions at either receiver will offer an expected distortion of no more than $D(1+2\ve)$ with a probability of at least $1-\ve$. The proof is complete by noting that $\ve$ can be chosen to be arbitrarily small. \hfill\qed

\section{{Proof of Theorem~\ref{thm-achieve-thm-2a} (Outer Bound)}}\label{APP-G}

 Let $(r_{uv},r_v)^T\in {\mathsf{\R}(D)}$. Let  $0<\iota<1$. Choose $0<\eta<\log 2$ such that
 \begin{align}
\max\left\{2|\eS|(e^\eta-1), \sqrt[4]{{2\eta}} \log \frac{|\eS|^2}{ \sqrt[4]{{2}\eta}} +2|\eS|(e^\eta-1)\log |\eS| \right\} <\iota.
 \end{align}
 Let $\ve=\min\left\{\iota, \left(\frac{1}{4}\mpv_{SUV}\mpv_S(1-e^{-\eta})\right)^2\right\}$. 
 By Definition~\ref{defn-CRRRateregion}, there exist  functions $ \mathtt E_{uv}$, $ \mathtt E_v$, $ \mathtt D_u$, $ \mathtt D_v$ satisfying \eqref{eqn-RRDefn1}. Let $M_{uv}:= \mathtt E_{uv}(S^n)$, $M_v :=  \mathtt E_v(S^n)$, $\hat S^n:=  \mathtt D_u(M_{uv},U^n)$, and $\tilde{S}^n :=  \mathtt D_v(M_{uv},M_v,V^n)$. Then,
\begin{align}
r_{uv}+\ve&\geq \frac{1}{n}H(M_{uv})\notag \\
&\geq \frac{1}{n}H(M_{uv}|U^n) \notag\\
&\geq \frac{1}{n}I(S^n; M_{uv}|U^n)\notag\\
&=\frac{1}{n} \sum\limits_{i=1}^nI(S_i; M_{uv}|U^n ,S^{i-1})\notag\\
&=\frac{1}{n} \sum\limits_{i=1}^nI(S_i; M_{uv},U^{n\setminus i},S^{i-1}|U_i) \notag\\
&\step{(a)}{=} \frac{1}{n}\sum\limits_{i=1}^n I(S_i;A_i,B_i|U_i)\notag\\
& \step{(b)}{=} I (S_Q; A_Q,B_Q | U_Q,Q)\notag\\
& \step{(c)}{=} I (S_Q; A_Q,B_Q,Q | U_Q)\notag\\
& \step{(d)}{=}  I(S;\bar{A},\bar{B}|U),\label{eqn-ruvexpn}
\end{align}
where in (a), we let $A_i:= (M_{uv}, U^{i-1})$ and $B_i:= (U_{i+1}^n,S^{i-1})$;  in (b), we introduce the time-sharing random variable $Q$ that is uniform over $\{1,\ldots,n\}$; in (c) we use the fact that $Q$ is independent of $(S_Q,U_Q)$; and in (d), we denote $\bar{A}:= (A_Q,Q)$, and $\bar{B} = (B_Q,Q)$.
Similarly,
\begin{align}
r_{uv}+r_v+2\ve&\geq \frac{1}{n}H(M_{uv},M_v)\notag\\
&\geq \frac{1}{n}I(S^n; M_{uv},M_v|V^n)\notag\\
&=\frac{1}{n}\Big[I(S^n; M_{uv},M_v,U^n|V^n)-I(S^n;U^n|M_{uv},M_v,V^n)\Big]\notag\\
&=\frac{1}{n}\sum\limits_{i=1}^n \Big[ I(S_i; M_{uv},M_v, U^n|V^n ,S^{i-1})-I(S^n; U_i|M_{uv},M_v,V^n,U^{i-1})\Big]\notag\\
&=\frac{1}{n}\sum\limits_{i=1}^n \Big[ I(S_i; M_{uv},M_v,U^n,V^{n\setminus i},S^{i-1}|V_i)- I(S^n; U_i|M_{uv},M_v,V^n,U^{i-1})\Big]\notag\\
&=\frac{1}{n}\sum\limits_{i=1}^n \Big[ I(S_i; A_i,B_i,M_v,V^{n\setminus i},U_i|V_i)- H(U_i|A_i,M_v,V^n) +H(U_i|A_i,M_v,V^n,S^n)\Big]\notag\\
&\step{(a)}{=}\frac{1}{n}\sum\limits_{i=1}^n \Big[ I(S_i; A_i,B_i,C_i,U_i|V_i)- H(U_i|A_i,C_i,V_i)+ H(U_i|A_i,C_i,V_i,S_i)\Big]\notag\\
&=\frac{1}{n} \sum\limits_{i=1}^n \Big[ I(S_i; A_i,B_i,C_i,U_i|V_i) -I(S_i;U_i|A_i,C_i,V_i)\Big]\notag\\
&= \frac{1}{n}\sum\limits_{i=1}^n \Big[ I(S_i; A_i,C_i|V_i)+I(S_i; B_i|A_i,C_i,U_i,V_i)\Big]\notag\\
&\step{(b)}{=} I(S_Q; A_Q, C_Q|V_Q, Q) + I (S_Q; B_Q | A_Q, C_Q ,U_Q ,V_Q, Q) \notag\\
&\step{(c)}{=}  I(S; \bar{A},\bar{C}|V)+I(S; \bar{B}|\bar A,\bar{C},U,V),\label{eqn-ruv+rvexpn}
\end{align}
where in (a), we have denoted $C_i:= (M_v,V^{n\setminus i})$ and used the chain 
\begin{align}
U_i\mkv (M_{uv},M_v,V^n,U^{i-1},S_i) \mkv (M_{uv},M_v,V^n,U^{i-1},S^n)\label{eqn-OB-MkvChn}; \end{align} in (b) we make use of the uniform time-sharing random variable $Q$; and in (c), we use the independence of $Q$ and $(S_Q,U_Q,V_Q)$ and define $\bar{A}:= (A_Q,Q)$, $\bar{B} = (B_Q,Q)$, and $\bar{C} = (C_Q,Q)$. Note that the following holds.
\begin{align}
I (U,V ; \bar {A}, \bar B, \bar C | S) &= I( U_Q, V_Q; A_Q ,B_Q ,C_Q | S_Q, Q)
= \sum_{i=1}^n  \frac{1}{n }I( U_i,V_i; A_i ,B_i, C_i | S_i) \stackrel{\eqref{eqn-OB-MkvChn}}{=} 0.
\end{align}

Now, note that $H(M_{uv},M_v,V^n|A_i,C_i,V_i)=0$ and  $H(M_{uv},U^n|A_i,B_i,U_i)=0$. Hence, $\hat S_i$ and $\tilde S_i$ are functions of $(A_i,B_i, U_i)$ and $(A_i,C_i, V_i)$, resp. Let $\hat{S}_i:= g_i(A_i,B_i,U_i)$ and $\tilde S_i := f_i(A_i,C_i,V_i)$.  Let $\bar{f}: \bar{\mathcal{A}} \times \bar{\mathcal{C}} \times \V \rightarrow \hat{\eS}$ and $\bar{g}: \bar{\mathcal{A}} \times \bar{\mathcal{B}} \times \U \rightarrow \hat{\eS}$ be defined as follows. For each $i\in\{1,\ldots,n\}$ and  $(a,b,c,u,v)\in\A_i\times\B_i\times \C_i\times \U\times\V$,
\begin{subequations}
\begin{align}
\bar{f}((a,i),(c,i),v) &:= f_i(a,c,v),\\
\bar{g}((a,i),(b,i),u) &:= g_i(a,b,u).
\end{align}
\end{subequations}
In other words, 
\begin{subequations}
\begin{align}
 \tilde S_Q & =\bar{f}((A_Q,Q),(C_Q,Q), V_Q),\\
\hat S_Q & =  \bar{g}((A_Q,Q),(B_Q,Q), U_Q). 
\end{align}
\end{subequations}
Using the above notation, we can then verify that
\begin{align}
\Pr[ \bar{f}(\bar A, \bar C, V)  \neq \bar{g}(\bar A, \bar B, U) ] &=\sum_{i=1}^n \frac{1}{n}\Pr[ \bar{f}(\bar A, \bar C, V )  \neq \bar{g}(\bar A, \bar B, U) | Q=i ]\notag\\&= \sum_{i=1}^n \frac{\Pr[ \tilde{S}_i \neq \hat S_ i ]}{n} \leq \ve\leq \iota, \label{eqn:InterAux1}\\
\Exp[d(S, \bar f(\bar A,\bar C, V)]& = \sum_{i=1}^n\frac{1}{n}\Exp[d({S}_i,  f_i(A_i,C_i,V_i) | Q=i ] \notag\\&= \sum_{i=1}^n\frac{1}{n}\Exp[d({S}_i, \tilde{S}_i]  \stackrel{(c)}{\leq}  D+\ve\leq D+\iota,\label{eqn:InterAux2}\\
\Exp[d(S, \bar g(\bar A,\bar B, U)]& = \sum_{i=1}^n\frac{1}{n}\Exp[d({S}_i,  g_i(A_i,B_i,U_i) | Q=i ] \notag\\&= \sum_{i=1}^n\frac{1}{n}\Exp[d({S}_i, \hat{S}_i]  \stackrel{(c)}{\leq}  D+\ve\leq D+\iota.\label{eqn:InterAux3}
\end{align}
Now, we have to establish the existence of auxiliary RVs such that the RHS of \eqref{eqn:InterAux1} is, in fact, zero. To do so, we make use of the two \emph{pruning theorems} in Appendix~\ref{APP-H}. The first step is to only allow realizations of auxiliary random variables for which the reconstructions agree most of the time, and prune out the rest. To this end, define
\begin{align}
\mc{E} := \left\{(\bar a,\bar b, \bar c)\in \supp(\bar A, \bar B, \bar C): \Pr\Big[ \bar{f}(\bar A, \bar C, V)  \neq \bar{g}(\bar A, \bar B, U)\,\big|\, (\bar A,\bar B, \bar C)=(\bar a,\bar b, \bar c) \Big]\leq \sqrt{\ve} \right\}. \label{eqn:mcEDefn}
\end{align}
By a simple application of Markov's inequality, one can argue that
 \begin{align}
 \Pr[ (\bar A, \bar B, \bar U)\in\mc E] \geq 1-\sqrt{\ve}.
 \end{align}
  Define auxiliary RVs $\mathring{A}\in\bar\A$, $\mathring{B}\in\bar\B$ and $\mathring{C}\in\bar\C$ with $(\mathring A, \mathring B, \mathring C) \mkv \bar S \mkv (\bar U,\bar V)$ by
\begin{align}
p_{\mathring A \mathring B \mathring C S}(a,b,c,s) =\left\{\begin{array}{ll} \dfrac{p_{ABCS}(a,b,c, s)}{\Pr[(A,B,C)\in \E|S=s]},& (a,b,c)\in\E \textrm{ and } s\in\supp(S)\\ 0, &\textrm{otherwise}\end{array}\right..
\end{align}
The above pmf $p_{\mathring A \mathring B \mathring C S}$ is precisely the pmf $p_{ABCS}$ obtained by an application of Pruning Method A  defined in \eqref{eqn-Prune1Def} of Appendix~\ref{APP-H} applied with $\delta =\sqrt{\ve}< m_S(1-e^{-\eta})$. Hence, the properties of Theorem~\ref{Thm:Ancillary3} of Appendix~\ref{APP-H} can be applied to $p_{\mathring A \mathring B \mathring C S}$. First, by an application of Property (d) of Theorem~\ref{Thm:Ancillary3} of Appendix~\ref{APP-H}, we see that
\begin{align}
| I(S; AB | U) - I(S; \mathring A \mathring B | U)| & = | H(S| ABU) - H(S| \mathring A \mathring B U)|\leq \iota.\label{eqn:AfterPrune1}\\
| I(S; AC | V) - I(S; \mathring A \mathring C | V)| & = | H(S| ACV) - H(S| \mathring A \mathring C V)|\leq \iota.
\end{align}
Similarly, using Property (e) of Theorem~\ref{Thm:Ancillary3} of Appendix~\ref{APP-H}, we see that
\begin{align}
| I(S; B | ACUV) - I(S; \mathring B |  \mathring A \mathring C UV)| & = \left(\begin{array}{l} | H(S| ACUV) - H(S| \mathring A \mathring C UV)|\\ \quad + | H(S| ABCUV) - H(S| \mathring A \mathring B \mathring C UV)| \end{array}\right)\leq 2\iota.
\end{align}
From \eqref{eqn:InterAux2}, \eqref{eqn:InterAux3} and Property (a) of Theorem~\ref{Thm:Ancillary3} of Appendix~\ref{APP-H}, we see that
\begin{subequations}
\begin{align}
\Exp[d({S}, \bar f(\mathring A,\mathring C,  V)]&\leq \Exp[d({S}, \bar f(\bar A,\bar C,  V)]+ \lVert p_{ACSV}-p_{\mathring A \mathring C SV}\rVert\overline{D}\\
& \leq D + \iota + \overline D \iota\\
\Exp[d({S}, \bar g(\mathring A,\mathring B,  U)]& \leq \Exp[d({S}, \bar g(\bar A,\bar B,  U)]+ \lVert p_{ABSU}-p_{\mathring A \mathring B SU}\rVert\overline{D}\\
&\leq D+\iota+\overline D \iota. 
\end{align}\label{eqn:AfterPrune2}
\end{subequations}
Since $\supp(\mathring A, \mathring B, \mathring C) = \mc E$, for any $(\bar a,\bar b, \bar c)\in\mc E$, we can invoke Property (b) of Theorem~\ref{Thm:Ancillary3} of Appendix~\ref{APP-H} with $\mathcal{F}:= \{(u,v): \bar{f}(\bar a, \bar c, v) \neq \bar g(\bar a, \bar b, u)\}$ to infer that
\begin{align}
\Pr\Big[ \bar{f}(\mathring A, \mathring C, V)  \neq \bar{g}(\mathring A, \mathring B,  U)\,\big|\, (\mathring A,\mathring B, \mathring C)=(\bar a,\bar b, \bar c) \Big]\leq e^\eta\sqrt{\ve} = 2\sqrt{\ve}, \label{eqn:mce'estim}
\end{align}

Now, let's proceed by pruning $p_{\mathring A\mathring B \mathring C S}$ further by using Pruning Method B defined in \eqref{eqn-Anc4-1} of Appendix~\ref{APP-H} with $\delta =\frac{2\sqrt{\ve}}{\mpv_{UVS}}$. Let us define
\begin{align}
\mc{E}' = \left\{ (a,b,c,s): p_{S|\mathring A \mathring B \mathring C}(s|\bar a,\bar b, \bar c) > \frac{2\sqrt{\ve}}{\mpv_{UVS}}\right\},
\end{align}
and auxiliary RVs $\mathsf{A}\in\bar\A$, $\mathsf{B}\in\bar\B$, and $\mathsf{C}\in\bar\C$ with $(\mathsf A, \mathsf B, \mathsf C) \mkv S \mkv (U,V)$ by
\begin{align}
p_{\mathsf{ABC}S}(a,b,c,s) =\left\{\begin{array}{ll} \dfrac{p_{\mathring{A}\mathring B \mathring CS}(a,b,c,s)}{\Pr[(\mathring A,\mathring B, \mathring C, S)\in \mc E' |S=s]},& (a,b,c,s)\in\mc E' \\ 0, &\textrm{otherwise}\end{array}\right..
\end{align}
Since $p_{\mathsf{ABC}S}$ is obtained from $p_{\mathring A \mathring B \mathring C S}$ by Pruning Method B,  the properties of Theorem~\ref{Thm:Ancillary4} of Appendix~\ref{APP-H} can be applied to $p_{\mathring A \mathring B \mathring C S}$. Combining \eqref{eqn:mce'estim} with Property (b) of Theorem~\ref{Thm:Ancillary4} of Appendix~\ref{APP-H}, we see that
\begin{align}
\Pr\big[ \bar{f}(\mathsf A, \mathsf C, V)  \neq \bar{g}(\mathsf A, \mathsf B,  U)\,\big|\, (\mathsf A,\mathsf B, \mathsf C)=(\bar a,\bar b, \bar c) \big]=0, \quad (\bar a,\bar b, \bar c)\in\mc E.
\end{align}
Since by construction, $\supp(\mathsf{A},\mathsf{B},\mathsf{C}) \subseteq \supp(\mathring A,\mathring B, \mathring C) \subseteq \mc E$, it follows that 
\begin{align}
\Pr[ \bar{f}(\mathsf A, \mathsf C, V)  \neq \bar{g}(\mathsf A, \mathsf B,  U) ] = 0.
\end{align}
Invoking Property (a) of Theorem~\ref{Thm:Ancillary4} of Appendix~\ref{APP-H}, it follows that 
\begin{subequations} 
\begin{align}
\Exp[d({S}, \bar f(\mathsf A,\mathsf C, V)]&\leq \Exp[d({S}, \bar f(\mathring A,\mathring C,  V)]+ \overline{D} \lVert p_{\mathsf{AC}SV}-p_{\mathring A \mathring C SV}\rVert\\
& \leq D + \iota + 2\overline D \iota\\
\Exp[d({S}, \bar g(\mathsf A,\mathsf B,  U)]& \leq\Exp[d({S}, \bar f(\mathring A,\mathring B,  U)]+ \overline{D} \lVert p_{\mathsf{AB}SU}-p_{\mathring A \mathring B SU}\rVert\\
& \leq D + \iota + 2\overline D \iota.
\end{align}\label{eqn:AfterPrune3}
\end{subequations}
By an application of Property (e) of Theorem~\ref{Thm:Ancillary4} of Appendix~\ref{APP-H}, we see that
\begin{subequations} 
\begin{align}
| I(S; \mathsf{AB} | U) - I(S; \mathring A \mathring B | U)| & = | H(S| \mathsf{AB}U) - H(S| \mathring A \mathring B U)|\leq \iota,\\
| I(S; \mathsf{AC} | V) - I(S; \mathring A \mathring C | V)| & = | H(S| \mathsf{AC}V) - H(S| \mathring A \mathring C V)|\leq \iota.
\end{align}
\end{subequations}
Similarly, an application of Property (f) of Theorem~\ref{Thm:Ancillary4} of Appendix~\ref{APP-H} yields
\begin{align}
| I(S; \mathsf B | \mathsf{AC}UV) - I(S; \mathring B |  \mathring A \mathring C UV)| & = \left(\begin{array}{l} | H(S| \mathsf{AC}UV) - H(S| \mathring A \mathring C UV)|\\ \quad + | H(S| \mathsf{ABC}UV) - H(S| \mathring A \mathring B \mathring C UV)| \end{array}\right)\leq 2\iota\label{eqn:AfterPrune4}.
\end{align}
Combining \eqref{eqn:AfterPrune1}-\eqref{eqn:AfterPrune2} with \eqref{eqn:AfterPrune3}-\eqref{eqn:AfterPrune4}, we have auxiliary RVs $(\mathsf A, \mathsf B, \mathsf C) \mkv S \mkv (U,V)$ such that
\begin{subequations} 
\begin{align}
r_{uv} + 2\iota &\geq I(S; \mathsf{AB}|U),\\
r_v+ r_{uv} + 2\iota &\geq I(S; \mathsf{AC}|V)+I(S; \mathsf{B} | \mathsf{AC}UV),\\
\Pr[ \bar{f}(\mathsf A, \mathsf C, V)  \neq \bar{g}(\mathsf A, \mathsf B,  U) ] &= 0,\\
\Exp[d({S}, \bar f(\mathsf A,\mathsf C, V)]&\leq D+\iota+ 2\overline{D}\iota,\\
\Exp[d({S}, \bar g(\mathsf A,\mathsf B,  U)]& \leq D+\iota+2\overline{D}\iota.
\end{align}
\end{subequations}
Thus, from  \eqref{eqn-rateregionalphabetbnd2} of Lemma~\ref{thm-RegionEquivalence2} of Sec.~\ref{Mainresults-rrdefns}, it follows that $(r_{uv} + \iota, r_{v} + \iota) \in \R^\ddagger (D+\iota+2\overline{D}\iota)$. Finally, by  constructing an appropriate sequence of infinitesimals and invoking Lemma~\ref{Lem:RstarCont} of Sec.~\ref{Mainresults-rrdefns}, we conclude that $(r_{uv} , r_{v}) \in \R^\ddagger (D)$. \hfill\qed

\section{Pruning Theorems}\label{APP-H}

We now present two \emph{pruning theorems} that concern any five random variables $(A_1,A_2,S,B_1,B_2)$ where $(A_1,A_2)\mkv S \mkv (B_1,B_2)$ forms a Markov chain. The theorems will be applied to the CRR successive refinement problem, where $S$ will be identified as the source, $A_1, A_2$ will be associated with auxiliary random variables, and $B_1$ and $B_2$ are the side information random variables available at each of the receivers. The pruning theorems will help us to understand how small alterations to the marginal pmf of $(A_1,A_2,S)$ will change
\begin{itemize}
\item the joint pmf of $(A_1, A_2, S, B_1, B_2)$ with respect to variational distance, and
\item information functionals such as $I(S; A_1 | B_1)$, $I(S; A_2| B_2)$ and $I(S; A_1 A_2 | B_1 B_2)$. 
\end{itemize}

\begin{figure*}[t!]
\begin{center}
\includegraphics[width=0.7\textwidth]{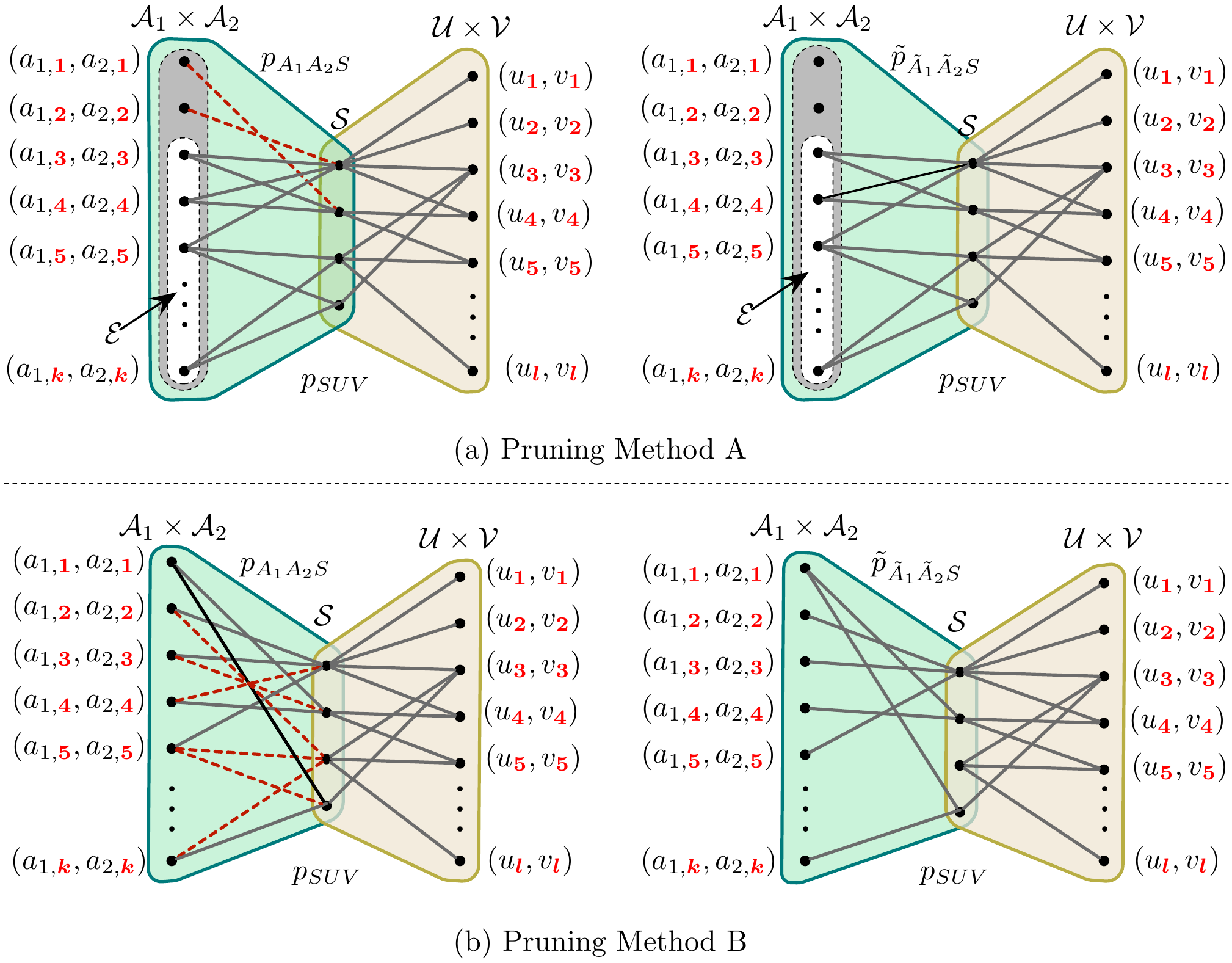}
\caption{Illustration of the two kinds of pruning.}\label{FIGURE:8}
\end{center}
\end{figure*}
Figure~\ref{FIGURE:8} illustrates the two kinds of pruning. In the Pruning Method A, we have $(A_1,A_2,S) \sim p_{A_1A_2S}$ as shown in Figure~\ref{FIGURE:8}(a). We select an appropriate threshold $0 < \delta < 1$, and consider any subset $\mathcal{E} \subseteq \mathcal{A}_1 \times \mathcal{A}_2$ with $\Pr[(A_1,A_2) \in \mathcal{E}] \geq 1-\delta$. We then take the joint pmf $p_{A_1A_2S}$ and construct a new joint pmf $\tilde{p}_{\tilde{A}_1 \tilde{A}_2 S}$ whose support set satisfies $\supp(\tilde A_1,\tilde A_2, S) \subseteq \mc E \times \eS$. In other words, the edges belonging to $\mathcal{E}^c \times \eS$ in the bipartite graph of $p_{A_1 A_2 S}$ indicated by red dashed lines in Figure~\ref{FIGURE:8}(a) are removed (and the remaining edges are scaled appropriately) to define $\tilde{p}_{\tilde{A}_1 \tilde{A}_2 S}$. 

In the Pruning Method B that is illustrated in Figure.~\ref{FIGURE:8}(b), we first select an appropriate threshold $0 < \delta < 1$, and consider any subset of edges $\mathcal{E} \subseteq \mathcal{A}_1 \times \mathcal{A}_2 \times \eS$ with
\begin{equation}
p_{S|A_1,A_2}(s| a_1,a_2)> \delta,
\quad \forall\ (a_1,a_2,s) \in \mc{E}
\end{equation}
Edges that are not in $\mc E$ (shown by red dashed lines in Fig.~\ref{FIGURE:8}(b)) are removed, and the probability mass of the rest of the edges are scaled appropriately to  construct a new $\tilde{p}_{\tilde{A}_1 \tilde{A}_2  S}$ such that $\supp(\tilde A_1,\tilde A_2, S) \subseteq \mc E$. The precise details of the two kinds of pruning, and the required results pertaining to them are elaborated below.

\subsection{Pruning Method A}

Suppose that we have a random tuple $(A_1,A_2,S,B_1,B_2)$ over $\mathcal{A}_1\times\mathcal{A}_2\times \mathcal{S} \times \mathcal{B}_1 \times \mathcal{B}_2$ distributed according to $p_{A_1A_2S B_1B_2}$  such that $(A_1,A_2) \mkv S \mkv (B_1,B_2)$ and $\supp(S)=\eS$. Pick any $\delta,\eta > 0$ satisfying  
\begin{equation}\label{Eqn:Proof:PruningThm:Delta1}
\delta\leq\mpv_{S}(1-e^{-\eta}),
\end{equation}
and consider any subset $\E\subseteq \A_1\times\A_2$ satisfying $\Pr[(A_1,A_2)\in \E]\geq 1-\delta$. Let
\begin{equation}
\tilde{p}_{\tilde{A}_1\tilde{A}_2 S}(a_1,a_2,s) 
:=
\left\{
\begin{array}{ll}
\dfrac{p_{A_1A_2S}(a_1,a_2,s)}{\Pr[(A_1,A_2)\in \E|S=s]}
& \text{ if } (a_1,a_2)\in\E\\
0 
& \text{ if } (a_1,a_2)\in\E^c
\end{array},
\right. \label{eqn-Prune1Def}
\end{equation}
and $\tilde{p}_{\tilde{A}_1 \tilde{A}_2 S B_1 B_2} :=\tilde{p}_{\tilde{A}_1\tilde{A}_2S}p_{B_1B_2|S}$. Define $\Xi: [0, \frac{\mpv_{SUV}}{|\eS|}) \rightarrow \mathbb R$ by 
\begin{equation}
\Xi(x):= \frac{|\eS|x}{\mpv_{SUV}-|\eS|x}\log \frac{|\eS| ^2(\mpv_{SUV}-|\eS|x)}{x}. \label{eqn-XiDefn}
\end{equation}

\begin{theorem}\label{Thm:Ancillary3}
$\tilde{p}_{\tilde{A}_1 \tilde{A}_2 S B_1 B_2}$ defines a valid joint pmf with $(\tilde{A}_1,\tilde{A}_2) \mkv S \mkv (B_1,B_2)$ for which the following holds:
\begin{enumerate}[(a)]
\item
$
\big\lVert\tilde{p}_{\tilde A_1\tilde A_2 S B_1B_2} -p_{A_1A_2S B_1B_2}\big\rVert_1 
\leq \frac{2\delta}{\mpv_S -\delta}
\leq 2(e^\eta-1).\vspace{1mm}
$

\item For any event $\mathcal{F}\subseteq \B_1\times \B_2$ and $(a_1,a_2) \in\mathcal{E}$, 
\begin{equation}
\Pr[(B_1,B_2)\in \mathcal{F}| (\tilde A_1,\tilde A_2)=(a_1,a_2)]\\
\leq  
e^\eta\ \Pr[(B_1,B_2)\in \mathcal{F}| (A_1, A_2)=(a_1,a_2)].
\end{equation}

\item $D_{KL}(\tilde{p}_{\tilde{A}_1\tilde{A}_2 S B_1 B_2} \lVert\,  p_{A_1 A_2 S B_1 B_2} ) \leq \eta.
$

\item 
$
\big| H(S|\tilde{A}_1 B_1) - H(S|A_1 B_1) \big |
\leq \sqrt[4]{{2\eta}} \log \dfrac{|\eS|^2}{ \sqrt[4]{{2}\eta}} +2(e^\eta-1)\log |\eS|.
$

\item 
$
\big|H(S|\tilde A_1 \tilde A_2B_1B_2)-H(S|A_1A_2B_1B_2)\big| 
\leq \sqrt[4]{{2\eta}} \log \dfrac{|\eS|^2}{ \sqrt[4]{{2}\eta}} +2(e^\eta-1)\log |\eS|.
$
\end{enumerate}
\end{theorem}


\begin{IEEEproof}
For brevity, we will omit the subscripts on the joint distributions $p_{A_1 A_2 S B_1 B_2}$ and $\tilde{p}_{\tilde{A}_1 \tilde{A}_2 S B_1 B_2}$ (and their marginals) throughout the proof; for example, $p(a_1,s) = p_{A_1 S}(a_1,s)$ and $\tilde{p}(a_1,s) = \tilde{p}_{A_1 S}(a_1,s)$.

\subsubsection*{Assertion (a)}
For any $(a_1,a_2,s,b_1,b_2)$ with $(a_1,a_2) \in \E$ and $p(a_1,a_2,s,b_1,b_2)>0$, we have
\begin{equation}\label{eqn:P1-1}
1 \leq 
\frac{\tilde{p}(a_1,a_2,s,b_1,b_2)}{p(a_1,a_2,s,b_1,b_2)} 
=\frac{\tilde{p}(a_1,a_2,s)}{p(a_1,a_2,s)}  
=
\frac{p(s)}{\Pr[(A_1,A_2)\in\E, S=s]} 
\step{(a)}{\leq }
\frac{\mpv_S}{\mpv_S-{\delta}},
\end{equation}
which (a) follows because
\begin{equation}
1 
\geq 
\frac{\Pr[(A_1,A_2)\in\E, S=s]}{p(s)}
\geq 
\frac{p(s) - \Pr[(A_1,A_2)\in\E]}{p(s)}\\
\geq 1- \frac{\Pr[(A_1,A_2)\in\E]}{p(s)}
\geq 
\frac{\mpv_S - \delta}{\mpv_S}.
\end{equation}

Thus, from~\eqref{eqn:P1-1} we now have 
\begin{align}
\notag
\sum_{(a_1,a_2,s,b_1,b_2) \in \E \times \eS \times \B_1 \times \B_2} 
&\Big| p(a_1,a_2,s,b_1,b_2)  - \tilde{p}(a_1,a_2,s,b_1,b_2) \Big|\\
\notag
&\leq
\sum_{(a_1,a_2,s,b_1,b_2) \in \E \times \eS \times \B_1 \times \B_2}
\left(\frac{\mpv_S}{\mpv_S - \delta} - 1\right) p(a_1,a_2,s,b_1,b_2)
\\
\label{Eqn:Proof:Thm:3:AssA:1}
&= 
\left(\frac{\delta}{\mpv_S-{\delta}}\right)\Pr\big[(A_1,A_2)\in\mathcal{E}\big].
\end{align}
Next, we can bound the contribution for $\mathcal{E}^c \times \mathcal{S}\times\mathcal{B}_1\times\mathcal{B}_2$ by 
\begin{equation}\label{Eqn:Proof:Thm:3:AssA:2}
\sum_{(a_1,a_2,s,b_1,b_2) \in \E^c \times \eS \times \B_1 \times \B_2}
\big| p(a_1,a_2,s,b_1,b_2) - \tilde{p}(a_1,a_2,s,b_1,b_2) \big| 
= \Pr\big[(A_1,A_2) \in \E^c\big] \leq \delta.
\end{equation}
Combining~\eqref{Eqn:Proof:Thm:3:AssA:1} and~\eqref{Eqn:Proof:Thm:3:AssA:2} yields Assertion (b), since $\delta+ \frac{\delta}{\mpv_S -\delta}<\frac{2\delta}{\mpv_S -\delta}=2(e^\eta-1)$.

\subsubsection*{Assertion (b)}

It is sufficient to prove the assertion for any singleton $\mathcal{F} = \{(b_1,b_2)\}$ with $(b_1,b_2)\in\B_1\times \B_2$. Let $(a_1,a_2)\in \mc E$. Without loss of generality, we may assume that $p(b_1,b_2|a_1,a_2) > 0$, since if not, the claim is trivially true. Thus,
\begin{align}
\frac{\tilde{p}(b_1,b_2|a_1,a_2)}{p(b_1,b_2|a_1,a_2)} 
&= 
\frac{\sum\limits_{s \in \supp(S)} \tilde{p}(a_1,a_2,s)\ p(b_1, b_2|s)}
{\tilde{p}(a_1,a_1)\ p(b_1,b_2|a_1,a_2)}
\step{{\eqref{eqn:P1-1}}}{\leq} 
\left({\frac{\mpv_S}{\mpv_S-{\delta}}}\right) 
\frac{\sum\limits_{s \in \supp(S)} p(a_1,a_2,s)\ p(b_1, b_2|s)}
{\tilde{p}(a_1,a_2)\ p(b_1,b_2|a_1,a_2)}\notag\\
&= 
\left({\frac{\mpv_S}{\mpv_S-{\delta}}}\right)
\frac{p(a_1,a_2,b_1, b_2)}
{\tilde{p}(a_1,a_2)\ p(b_1,b_2|a_1,a_2)}\notag\\
&= 
\left({\frac{\mpv_S}{\mpv_S-{\delta}}}\right) 
\frac{ p(a_1,a_2)}{\tilde{p}(a_1,a_2)} \step{(a)} {\leq}
\frac{\mpv_S}{\mpv_S-{\delta}} \step{\eqref{Eqn:Proof:PruningThm:Delta1}}{\leq}
e^\eta,
\end{align}
where (a) follows by rearranging \eqref{eqn:P1-1}, and summing over elements of $\eS \times \B_1\times \B_2$. 

\subsubsection*{Assertion (c)}
 \begin{align}
     D_{KL}(\tilde{p}_{\tilde{A}_1\tilde{A}_2 S B_1 B_2} \lVert\,  p_{A_1 A_2 S B_1 B_2} ) &=D_{KL}(\tilde{p}_{\tilde{A}_1\tilde{A}_2 S} \lVert\,  p_{A_1 A_2 S} )\notag\\
     &= \sum_{(a_1,a_2,s)\in\supp(\tilde{A}_1\tilde{A}_2 S)} \tilde{p}_{\tilde{A}_1\tilde{A}_2 S} (a_1,a_2,s) \log \frac{ \tilde{p}_{\tilde{A}_1\tilde{A}_2 S} (a_1,a_2,s) }{ {p}_{{A}_1{A}_2 S} (a_1,a_2,s) }\notag\\
     &\stackrel{\eqref{eqn:P1-1}}{\leq}  \sum_{(a_1,a_2,s)\in\supp(\tilde{A}_1\tilde{A}_2 S)} \tilde{p}_{\tilde{A}_1\tilde{A}_2 S} (a_1,a_2,s)  \log \frac{\mpv_S}{\mpv_S-\delta}\leq \eta. \label{eqn:P3-1}
     \end{align}
\subsubsection*{Assertion (d)}
     \begin{align}
    D_{KL}(\tilde{p}_{S|\tilde{A_1}B_1} \lVert\, p_{S|A_1 B_1}) &:= \sum_{a_1 b_1} \tilde{p}_{\tilde{A}_1B_1}(a_1,b_1) D_{KL} \big( \tilde{p}_{S\mid \tilde A_1 B_1} ( \cdot | a_1, b_1) | {p}_{S\mid  A_1 B_1} ( \cdot | a_1, b_1)\big)\notag \\
    &\leq D_{KL}(\tilde{p}_{\tilde{A}_1\tilde{A}_2 S B_1 B_2} \lVert\,  p_{A_1 A_2 S B_1 B_2} )\stackrel{\eqref{eqn:P3-1}}{\leq} \log \frac{\mpv_S}{\delta -\mpv_S}=\eta.
     \end{align}
     An application of Pinsker's followed by Jensen's inequality yields the following.
     \begin{align}
    \sum_{a_1 b_1} \tilde{p}_{\tilde{A}_1B_1}(a_1,b_1) \big\lVert\tilde{p}_{S\mid \tilde A_1 B_1} ( \cdot | a_1, b_1) - {p}_{S\mid  A_1 B_1} ( \cdot | a_1, b_1)\big\rVert_1 \leq \sqrt{{2}\eta}.
     \end{align}
     Let $\mathcal{D} := \left\{(a_1,b_1): \big\lVert\tilde{p}_{S\mid \tilde A_1 B_1} ( \cdot | a_1, b_1) - {p}_{S\mid  A_1 B_1} ( \cdot | a_1, b_1)\big\rVert_1 \leq \sqrt[4]{{2}\eta}.\right\}$. Then, by Markov's inequality, 
     \begin{align}
     \Pr[ (\tilde A_1, B_1) \in \mathcal{D} ] &\geq 1- \sqrt[4]{{2}\eta}.
     \end{align}
     Further for every $(a_1,b_1)\in \mc D$, Lemma~2.5 of \cite{Csiszar-2011-B} guarantees that 
     \begin{align}
     \Delta_{a_1,b_1} &:= \Big| H(S\,|\,(\tilde A_1 ,B_1)=(a_1,b_1)) - H(S\,|\,(A_1 ,B_1)=(a_1,b_1))\Big|\leq    \sqrt[4]{{2}\eta} \log \frac{|\eS|}{ \sqrt[4]{{2}\eta}}.   \end{align}
     Finally, 
     \begin{align}
     \big|H(S|\tilde A_1 B_1) -H(S| A_1 B_1) \big| &\leq \sum_{a_1,b_1} \tilde{p}_{\tilde A_1,B_1} (a_1,b_1)  \Delta_{a_1,b_1} + \big\lVert \tilde{p}_{\tilde A_1,B_1}- {p}_{A_1,B_1}\big\rVert_1 \cdot \log |\eS|\notag\\
     &\leq \sum_{(a_1,b_1)\in \mc D} \tilde{p}_{\tilde A_1,B_1} (a_1,b_1)  \Delta_{a_1,b_1}+\Pr[(\tilde A_1,B_1)\notin \mc D] \log |\eS|+ \frac{2\delta\log |\eS|}{\mpv_S -\delta}\notag\\
     &\leq  \sqrt[4]{{2}\eta} \log \frac{|\eS|^2}{ \sqrt[4]{{2}\eta}} +2(e^\eta-1) \log |\eS|. 
     \end{align}
     \subsubsection*{Assertion (e)}
     The proof of (e) is identical to that of (d), with the only difference being that the commencement of the argument is as follows.
       \begin{align}
        D_{KL}(\tilde{p}_{S|\tilde{A_1}\tilde{A}_2 B_1 B_2} \lVert\, p_{S|A_1 A_2 B_1 B_1})\leq  D_{KL}(\tilde{p}_{\tilde{A}_1\tilde{A}_2 S B_1 B_2} \lVert\,  p_{A_1 A_2 S B_1 B_2} )\stackrel{\eqref{eqn:P3-1}}{\leq} \log \frac{\mpv_S}{\delta -\mpv_S}=\eta.
     \end{align}
 The remaining steps are identical to those in (d), with the exception that all four variables (either $(\tilde A_1, \tilde A_2, B_1, B_2)$ or $(A_1, A_2, B_1, B_2) $) appear in the conditioning.    
\end{IEEEproof}

\subsection{Pruning Method B}

Let {pmf} $p_{A_1A_2S B_1B_2}$ over $\mathcal{A}_1\times\mathcal{A}_2\times \mathcal{S} \times ,\mathcal{B}_1\times\mathcal{B}_2$ be given such that $(A_1,A_2)\mkv S \mkv (B_1,B_2)$ and $\supp(S)=\eS$.
 Let $\delta\leq\mpv_{S}(1-e^{-\eta})$ for some $0<\eta<1$, and let
 \begin{subequations}
 \begin{align}
 \E&:= \left\{ (a_1,a_2,s)\in\supp(A_1,A_2,S): p_{S|A_1A_2}(s|a_1,a_2)> \delta\right\},\\
\mathcal{E}_s &:= \big\{(a_1,a_2): (a_1,a_2,s)\in \mc E\big\},\,\,\quad s\in\eS,\\
\mathcal{E}_{a_1,a_2} &:= \big\{s: (a_1,a_2,s)\in \mc E\big\},\qquad\qquad s\in\eS.
 \end{align}
 \end{subequations}
 Define {pmf} $\tilde{p}_{\tilde{A}_1\tilde{A}_2 S }$ by 
\begin{align}
\tilde p_{A_1A_2S}(a_1,a_2,s) =\left\{\begin{array}{ll} {\displaystyle \frac{p_{A_1A_2S}(a_1,a_2,s)}{\Pr[(A_1,A_2)\in \E_s|S=s]}},& (a_1,a_2,s)\in\E\\ 0, &\textrm{otherwise}\end{array}\right., \label{eqn-Anc4-1}
\end{align}
and extend it to {pmf} $\tilde p_{\tilde A_1 \tilde A_2 S B_1 B_2}$ using the Markov chain $(\tilde A_1,\tilde A_2)\mkv S \mkv (B_1,B_2)$ with  $\tilde{p}_{SB_1 B_2}:=p_{SB_1B_2}$. Let $
\Xi(\cdot)$ be as defined in \eqref{eqn-XiDefn}. The following properties hold for the pmf $\tilde{p}_{\tilde{A}_1 \tilde{A}_2 S B_1 B_2}$ defined above.

\begin{theorem}\label{Thm:Ancillary4}
\begin{itemize}
\item[(a)]$ \lVert{\tilde{p}_{\tilde A_1\tilde A_2 S B_1B_2} -p_{A_1A_2S B_1B_2}}\rVert_1 \leq  \frac{2\delta|\eS|}{\mpv_S -\delta}=2|\eS|(e^\eta-1).$\vspace{1mm}
\item[(b)] Given $\mathcal{F}\subseteq \B_1\times \B_2$ and $(a_1,a_2)\in\supp(A_1,A_2)$ with $\Pr\big[(B_1,B_2)\in\mc F \big|(A_1,A_2)=(a_1,a_2)\big]\leq \delta\: \mpv_{B_1B_2|S}$, 
\begin{align}\Pr\big[(B_1,B_2)\in\mc F \big|(\tilde A_1,\tilde A_2)=(a_1,a_2)\big]\leq \delta \mpv_{B_1B_2|S} &= 0.\end{align}
\item[(c)]  $D_{KL}(\tilde{p}_{\tilde{A}_1\tilde{A}_2 S B_1 B_2} \lVert\,  p_{A_1 A_2 S B_1 B_2} ) \leq \log\dfrac{\mpv_S}{\mpv_S-\delta}=\eta$.\vspace{1mm}
\item[(d)] $\big| H(S|\tilde{A}_1 B_1) - H(S|A_1 B_1) \big | \leq \sqrt[4]{{2}\eta} \log \dfrac{|\eS|^2}{ \sqrt[4]{{2}\eta}} +2|\eS|(e^\eta-1)\log |\eS|$.\vspace{0mm}
\item[(e)] $\big|H(S|\tilde A_1 \tilde A_2B_1B_2)-H(S|A_1A_2B_1B_2)\big|\leq \sqrt[4]{{2}\eta} \log \dfrac{|\eS|^2}{ \sqrt[4]{{2}\eta}} +2|\eS|(e^\eta-1)\log |\eS|.$
\end{itemize}
\end{theorem}

\begin{IEEEproof}
The proof follows on the same steps as Theorem~\ref{Thm:Ancillary3} and the difference is in the evaluation of the probabilities of the normalization term in \eqref{eqn-Anc4-1}. 

\subsubsection*{Assertion (a)} Since for $(a_1,a_2,s)\in \mc E$, we have $p(s|a_1,a_2) \leq \delta$, we can argue that
\begin{align}
\Pr[(A_1,A_2,S)\in\mathcal{E}^c] = \sum_{(a_1,a_2,s)\in\E^c} p_{A_1A_2 S}(a_1,a_2,s) \leq \delta \sum_{a_1,a_2,s} p_{A_1A_2 }(a_1,a_2) \leq \delta|\eS|. \label{eqn-Anc4-3}
\end{align}
Further, for any $s\in \supp(S)$,
\begin{align}
\Pr[(A_1,A_2)\in\E_s|\, S=s] &= \sum_{(a_1,a_2)\in \E_s} \frac{p_{A_1A_2S}(a_1,a_2,s) }{p_S(s)} = 1- \sum_{(a_1,a_2'\notin \E_s} \frac{p_{A_1A_2S}(a_1,a_2,s) }{p_S(s)} \notag\\
&\geq 1-\delta\sum_{a_1,a_2} \frac{ p_{A_1A_2}(a_1,a_2)}{p_S(s)} \geq 1-\frac{\delta}{\mpv_S}. \label{eqn-Anc4-4}
\end{align}
To prove (a), first note that for any $(a_1,a_2,b_1,b_2)$ and $s\in \mathcal{S}$, 
\begin{align}
p_{A_1A_2SB_1B_2}(a_1,a_2,s,b_1,b_2)=0 \Rightarrow \tilde p_{\tilde A_1\tilde A_2S B_1 B_2}(a_1,a_2,s,b_1,b_2)=0,
\end{align}
and if $(a_1,a_2,s)\in\E$ and $p_{A_1A_2SB_1B_2}(a_1,a_2,s,b_1,b_2)>0$, then
\begin{align}
1\leq \frac{\tilde p_{\tilde A_1\tilde A_2SB_1B_2}(a_1,a_2,s,b_1,b_2)}{p_{A_1A_2SB_1B_2}(a_1,a_2,s,b_1,b_2)} &= \frac{1}{\Pr[(A_1,A_2)\in\E_s\mid S=s]} \stackrel{\eqref{eqn-Anc4-4}}{\leq} \frac{\mpv_S}{\mpv_S-\delta} \label{eqn-Anc4-5}.
\end{align}
Thus,
\begin{align}
\sum_{\substack{b_1,b_2\\(a_1,a_2,s)\in\mathcal{E}}} \big| p_{A_1A_2SB_1B_2}(a_1,a_2,s,b_1,b_2) - \tilde p_{\tilde A_1\tilde A_2SB_1B_2}(a_1,a_2,s,b_1,b_2) \big| \leq \frac{\delta\,\Pr[(A_1,A_2,S)\in\mathcal{E}]}{\mpv_S-{\delta}}.\end{align}
Next, we can bound the contribution for $\mathcal{E}^c \times\mathcal{B}_1\times\mathcal{B}_2$ by 
\begin{align}
\sum_{\substack{b_1,b_2\\(a_1,a_2,s)\notin\mathcal{E}}} \big| p_{A_1A_2SB_1B_2}(a_1,a_2,s,b_1,b_2) - \tilde p_{\tilde A_1\tilde A_2SB_1B_2}(a_1,a_2,s,b_1,b_2) \big|&
= \Pr[(A_1,A_2,S)\notin\mathcal{E}] \stackrel{\eqref{eqn-Anc4-3}}{\leq} \delta|\eS|.
\end{align}
Combining the above two equations together with the fact that $\delta|\eS|+ \frac{\delta}{\mpv_S -\delta}<\frac{2\delta|\eS|}{\mpv_S -\delta}$ establishes (a). 

\subsubsection*{Assertion (b)} We may assume that $\mc F\neq \emptyset$, or else the claim is trivial. Now, let $(a_1,a_2)\in\supp(A_1,A_2)$. Then, $(a_1,a_2)\in\supp(\tilde A_1,\tilde A_2)$ because
\begin{align}
\tilde{p}_{\tilde A_1, \tilde A_2}(a_1,a_2)\cdot \Pr[(A_1,A_2)\in \E_s|S=s] &=  {\sum\limits_{s\in \mc{E}_{a_1,a_2}} p_{A_1A_2S}(a_1,a_2,s)}{}\notag\\
&= {p_{A_1,A_2}(a_1,a_2) -\sum\limits_{s\notin \mc{E}_{a_1,a_2}} p_{A_1A_2S}(a_1,a_2,s)}\notag\\
&\geq {p_{A_1,A_2}(a_1,a_2) -  \sum\limits_{s\notin \mc{E}_{a_1,a_2}} \delta\, p_{A_1A_2}(a_1,a_2)}\notag\\
&= (1-\delta|\eS|)p_{A_1,A_2}(a_1,a_2) >0.
\end{align}
 Now, by hypothesis, 
 \begin{align}
 \Pr[(B_1,B_2)\in \mathcal{F} \mid (A_1,A_2)=(a_1, a_2)] = \sum_{(s',b_1,b_2)\in\,\supp(S)\times \mathcal F} p_{S|A_1A_2}(s'|a_1,a_2) p_{B_1B_2|S}(b_1,b_2|s')\leq \delta m_{B_1B_2|S}. \label{eqn-AssBFcond}
 \end{align}
Thus, for any $(b_1,b_2)\in \mc F$ and $s\in \eS$ such that $p_{B_1B_2|S}(b_1,b_2|s)>0$, it must be true that
\begin{align}
p_{S|A_1A_2}(s|a_1,a_2) &\step{}{\leq} \sum\limits_{(s,b_1,b_2)\in\,\supp(S)\times \mathcal F}  \frac{p_{S|A_1A_2}(s'|a_1,a_2) p_{B_1B_2|S}(b_1,b_2|s')}{p_{B_1B_2|S}(b_1,b_2|s)}\step{\eqref{eqn-AssBFcond}}{\leq} \frac{\delta m_{B_1B_2|S}}{p_{B_1B_2|S}(b_1,b_2|s)} \leq\delta,
\end{align}
Hence, $(a_1,a_2,s)\notin \mc{E}$. Thus, for any $(b_1,b_2)\in \mc F$ and $s\in \eS$ such that $p_{B_1B_2|S}(b_1,b_2|s)>0$, $\tilde{p}_{S|\tilde{A}_1\tilde A_2} (s|a_1,a_2)=0$, and therefore, (b) follows. 

  \subsubsection*{Assertion (c)} To prove (c), we proceed as follows.
     \begin{align}
     D_{KL}(\tilde{p}_{\tilde{A}_1\tilde{A}_2 S B_1 B_2} \lVert\,  p_{A_1 A_2 S B_1 B_2} ) &=D_{KL}(\tilde{p}_{\tilde{A}_1\tilde{A}_2 S} \lVert\,  p_{A_1 A_2 S} )\notag\\
     &= \sum_{a_1,a_2,s} \tilde{p}_{\tilde{A}_1\tilde{A}_2 S} (a_1,a_2,s) \log \frac{ \tilde{p}_{\tilde{A}_1\tilde{A}_2 S} (a_1,a_2,s) }{ {p}_{{A}_1{A}_2 S} (a_1,a_2,s) }\notag\\
     &= \sum_{s\in\supp(S), (a_1,a_2)\in\mc E_s} \tilde{p}_{\tilde{A}_1\tilde{A}_2 S} (a_1,a_2,s) \log \frac{1}{ \Pr[ (A_1,A_2)\in \mc E_s|S=s]}\notag\\
     &\stackrel{\eqref{eqn-Anc4-4}}{\leq}  \sum_{\substack{s\in\supp(S)\\ (a_1,a_2)\in\mc E_s}} \tilde{p}_{\tilde{A}_1\tilde{A}_2 S} (a_1,a_2,s) \log \frac{\mpv_S}{\mpv_S-\delta}=\eta. \label{eqn-Anc4-12}
     \end{align}
    \subsubsection*{Assertion (d)}  Consider the following.
     \begin{align}
    \sum_{a_1 b_1} \tilde{p}_{\tilde{A}_1B_1}(a_1,b_1) D_{KL} ( \tilde{p}_{S\mid \tilde A_1 B_1} ( \cdot | a_1, b_1) | {p}_{S\mid  A_1 B_1} ( \cdot | a_1, b_1) &\leq  D_{KL}(\tilde{p}_{\tilde{A}_1\tilde{A}_2 S B_1 B_2} \lVert\,  p_{A_1 A_2 S B_1 B_2} )\stackrel{\eqref{eqn-Anc4-12}}{\leq} \eta.
     \end{align}
     An application of Pinsker's followed by Jensen's inequality yields the following.
     \begin{align}
    \sum_{a_1 b_1} \tilde{p}_{\tilde{A}_1B_1}(a_1,b_1) \big\lVert\tilde{p}_{S\mid \tilde A_1 B_1} ( \cdot | a_1, b_1) - {p}_{S\mid  A_1 B_1} ( \cdot | a_1, b_1)\big\rVert_1 \leq \sqrt{{2}\eta}.
     \end{align}
     Let $\mathcal{D} := \left\{(a_1,b_1): \big\lVert\tilde{p}_{S\mid \tilde A_1 B_1} ( \cdot | a_1, b_1) - {p}_{S\mid  A_1 B_1} ( \cdot | a_1, b_1)\big\rVert_1 \leq \sqrt[4]{{2}\eta}.\right\}$. Then, by Markov's inequality, 
     \begin{align}
     \Pr[ (\tilde A_1, B_1) \in \mathcal{D} ] &\geq 1- \sqrt[4]{{2}\eta}.
     \end{align}
     Further for every $(a_1,b_1)\in \mc D$, by Lemma~2.5 of \cite{Csiszar-2011-B}, we have
     \begin{align}
     \Delta_{a_1,b_1} &:= \Big| H(S| (\tilde A_1 ,B_1)=(a_1,b_1)) - H(S|(A_1 ,B_1)=(a_1,b_1))\Big| \leq    \sqrt[4]{2\eta} \log \frac{|\eS|}{ \sqrt[4]{2\eta}}.   \end{align}
     Finally, 
     \begin{align}
     \big|H(S|\tilde A_1 B_1) -H(S| A_1 B_1) \big| &\leq \sum_{a_1,b_1} \tilde{p}_{\tilde A_1,B_1} (a_1,b_1)  \Delta_{a_1,b_1} + \big\lVert \tilde{p}_{\tilde A_1,B_1}- {p}_{A_1,B_1}\big\rVert_1\log |\eS|\notag\\
     &\leq \sum_{(a_1,b_1)\in \mc D} \tilde{p}_{\tilde A_1,B_1} (a_1,b_1)  \Delta_{a_1,b_1}+ \Pr[(\tilde A_1,B_1)\notin \mc D]\log |\eS| + \frac{2\delta|\eS|\log |\eS|}{\mpv_S -\delta}\notag\\
     &\leq  \sqrt[4]{2\eta} \log \frac{|\eS|^2}{ \sqrt[4]{2\eta}}+2|\eS|(e^{\eta}-1)\log |\eS|.
     \end{align}
     
       \subsubsection*{Assertion (e)} The proof of (e) is identical to that of (d), with the only difference being that the commencement of the argument is as follows.
       \begin{align}
    \sum_{a_1 a_2,b_1,b_2} \tilde{p}_{\tilde{A}_1\tilde{A}_2,B_1,B_2}(a_1 a_2,b_1,b_2)& D_{KL} ( \tilde{p}_{S\mid \tilde A_1 \tilde A_2 B_1 B_2} ( \cdot | a_1 a_2,b_1,b_2) | {p}_{S\mid  A_1 A_2 B_1 B_2} ( \cdot | a_1 a_2,b_1,b_2)\notag\\
     &\leq  D_{KL}(\tilde{p}_{\tilde{A}_1\tilde{A}_2 S B_1 B_2} \lVert\,  p_{A_1 A_2 S B_1 B_2} )\stackrel{\eqref{eqn-Anc4-12}}{\leq} \eta.
     \end{align}
 The remaining steps are identical to those in (d), with the exception that all four variables, i.e., either $(\tilde A_1, \tilde A_2, B_1, B_2)$ or $(A_1, A_2, B_1, B_2) $, appear in the conditioning.    
\end{IEEEproof}

\section{{Proof of Theorem~\ref{thm-3}}}\label{APP-I}
In each case, the overall approach is to show that a rate pair $(r_{uv},r_v)$ that is included in the outer bound rate region $\R^\ddagger (D)$ is also included in $\R^D$. To establish that, in each case, we pick a rate pair $(r_{uv},r_v) \in {\R}_{ |\eS|(|\eS|+6)|\hat\eS|^{|\U|}+4}^{ \ddagger}(D)$ and establish explicitly that the rate pair is also an element of ${\R}_{k}^{ *}(D)$ for some $k$. The proof is then complete by invoking the achievability of $\R_k^*(D)$ proved in Theorem~\ref{thm-achieve-thm-2} of Sec.~\ref{Sec-SLChars}.

\subsubsection*{{Case A}}\label{proof-thm3-A}

The proof follows from the following series of arguments:
\begin{align}
{\R}^{*}(D)\step{(a)}{\subseteq}  {\mathsf{\R}(D)} &\step{(b)}{\subseteq } \R^\ddagger(D)\step{(c)}= {\R}^\ddagger_{ |\eS|(|\eS|+6)|\hat\eS|^{|\U|}+4}(D)\stackrel{(d)}{\subseteq } \R^{*}_{( |\eS|(|\eS|+6)|\hat\eS|^{|\U|}+4)^2}(D),
\end{align}
where (a) and (b) follow from Theorems~\ref{thm-achieve-thm-2} and \ref{thm-achieve-thm-2a} of Sec.~\ref{Sec-SLChars};  (c) from Lemma~\ref{thm-RegionEquivalence2} of Sec.~\ref{Mainresults-rrdefns}; and (d) from Lemma~\ref{thm-RegionEquivalence} of Sec.~\ref{Mainresults-rrdefns}, where the full-support condition specific to this case is incorporated.

\subsubsection*{{Case B}}\label{proof-thm3-B}

Since $S\mkv V \mkv U$, from the outer bound in Theorem~\ref{thm-achieve-thm-2a} can be rewritten as follows.  
\begin{align}
\R(D)\subseteq {{\R}_{ |\eS|(|\eS|+6)|\hat\eS|^{|\U|}+4}^{ \ddagger}(D):= \bigcup_{q\in\mc{P}^{\ddagger}_{D, |\eS|(|\eS|+6)|\hat\eS|^{|\U|}+4}} \left\{(r_{uv}, r_v):{{\setstretch{1.25}\begin{array}{rl} 
r_{uv}\geq &\textstyle I(S;A,B|U)\\
r_v+r_{uv}\geq &I(S;A,B,C|V)
\end{array}}}\right\}}. \label{eqn-CaseBRR}
\end{align}
Note that the sum rate constraint incorporates the chain $S\mkv V \mkv U$ specific to this case.
Now, {let us} fix a pmf $q_{ABCSUV} \in \mc{P}_{D, |\eS|(|\eS|+6)|\hat\eS|^{|\U|}+4}^{  \ddagger}$. Let  
$ f_U:\A\times\B\times \U \rightarrow \A\times\C\times \V$ and 
$f_V:\A\times\C\times \U \rightarrow \A\times\C\times \V$
be such that 
\begin{align}
f_U(A,B,U) = f_V(A,C,V)= \GK^{ABU, ACV}.
\end{align}
Now,  $\Pr[ f_U(A,B,U)\neq f_V(A,C,V)]=0$ can be expanded using $ C \mkv (A,B,V) \mkv U$ in the following manner. 
\begin{align}
\sum_{a,b,v\in \supp(A,B,V)} q_{ABV}(a,b,v)\sum_{u,c} q_{C|ABV}(c|a,b,v) q_{U|V}(u|v)\bar{ \mathds 1}\left\{f_U(a,b,u) = f_V(a,c,v) \right\} = 0.
\end{align}
Hence, for any $(a,b,v)\in \supp(A,B,V)$, 
\begin{align}
\sum_{u,c} q_{C|ABV}(c|a,b,v) q_{U|V}(u|v) \bar{\mathds 1}\left\{f_U(a,b,u)= f_V(a,c,v)\right\} = 0.
\end{align}
Then, by Lemma~\ref{lem-identical} of Appendix~\ref{APP-L}, it follows that $(a,b,v)\in \supp(A,B,V)$, there exists $\eta(a,b,v)$ such that
\begin{subequations}
\begin{align}
\Pr[f_U(A,B,U)= \eta(a,b,v)\mid (A,B,V)=(a,b,v) ] &=1,\\  \Pr[f_V(A,C,V)= \eta(a,b,v)\mid (A,B,V)=(a,b,v) ] &=1.
\end{align}
\end{subequations}
Hence, we are guaranteed of the existence of a function $g_V:\A\times\B\times \V\rightarrow\A\times\C\times \V$ such that 
\begin{align}
g_V(A,B,V) = f_U(A,B,U) = f_V(A,C,V). \label{eqn-CaseBCondition1Met}
\end{align}
Thus, $\tilde{A} := g_V(A,B,V)$ is a function of $(A,B,U)$ as well as a function of $(A,B,V)$. Then, notice that 
\begin{subequations}
\begin{alignat}{3}
r_{uv}&\step{\eqref{eqn-CaseBRR}}{\geq} I(S;A,B|U)&\geq I(S; \tilde A|U),\\
r_v+r_{uv}&\step{\eqref{eqn-CaseBRR}}{\geq} I(S;A,B,C|V)\,\,&\geq I(S;\tilde A| V),
\end{alignat}
\end{subequations}
and further $\tilde A \mkv S \mkv (U,V)$. Lastly, we see that the alphabet of $\tilde{A} \subseteq \A\times\C\times \V$, and hence 
\begin{align}
|\tilde A| \leq \big( |\eS|(|\eS|+6)|\hat\eS|^{|\U|}+4\big)^2 |\U|. \label{eqn-CaseBCondition4Met}
\end{align}
From \eqref{eqn-CaseBCondition1Met}-\eqref{eqn-CaseBCondition4Met}, we conclude that $q_{\tilde ASUV} \in \mc P^*_{D,\big( |\eS|(|\eS|+6)|\hat\eS|^{|\U|}+4\big)^2 |\U|}$ and $(r_{uv}, r_v)\in \R^*_{\big( |\eS|(|\eS|+6)|\hat\eS|^{|\U|}+4\big)^2 |\U|}(D)$. 
Since $q_{ABCSUV}$ is any arbitrary pmf in $ \mc{P}_{D, |\eS|(|\eS|+6)|\hat\eS|^{|\U|}+4}^{  \ddagger}$, it follows that 
\begin{align}
{\textstyle {\R}_{ |\eS|(|\eS|+6)|\hat\eS|^{|\U|}+4}^{  \ddagger}(D)} \subseteq\R^*_{( |\eS|(|\eS|+6)|\hat\eS|^{|\U|}+4)^2 |\U|}(D).
\end{align}

\subsubsection*{{Case C}}\label{proof-thm3-C}
Let  $(r_{uv}, r_v)\in\R_{ |\eS|(|\eS|+6)|\hat\eS|^{|\U|}+4}^{  \ddagger}(D)$. Then, there must exist   $q_{ABCSUV}\in \mc{P}_{D, |\eS|(|\eS|+6)|\hat\eS|^{|\U|}+4}^{  \ddagger}$, and a function $f:\A\times\C\times\V \rightarrow \hat\eS$ such that $\Exp[d(S, f(\GK^{ABU,ACV}))] \leq D$, and
\begin{subequations} \label{eqn-conditions-Case-C}
\begin{align}
r_{uv}&\geq \textstyle I(S;A,B|U),\\
r_v+r_{uv}&\geq I(S;A,C|V)+I(S;B|A,C,U).
\end{align}
 \end{subequations}

Since $(A,B,C) \mkv S \mkv (U,V)$, we have for any $(a,b,c,u,v)\in \supp(A,B,C,U,V)$,
\begin{align}
q_{ABCUV} (a,b,c,u,v)  &= \sum_{s\in\supp(S)} q_{ABCSUV} (a,b,c,s,u,v) \notag\\
& = \sum_{s\in\supp(S)} q_{ABCS} (a,b,c,s) q_{UV|S} (u,v|s)\notag\\
& \leq \frac{1}{\mpv_S} \sum_{s\in\supp(S)} q_{ABCS} (a,b,c,s) q_{SUV}(s,u,v)\notag\\
& \leq \frac{1}{\mpv_S} q_{UV}(u,v)\sum_{s\in\supp(S)} q_{ABCS} (a,b,c,s)\notag\\
& \leq \frac{1}{\mpv_S} q_{ABC} (a,b,c) q_{UV}(u,v).
\end{align}
Further, since $\supp(S,U)=\supp(S)\times\supp(U)$, it follows that for any $(a,b,c,u,v)\in \supp(A,B,C,U,V)$,
\begin{align}
q_{ABCUV} (a,b,c,u,v) &= \sum_{s\in\eS} q_{ABCS}(a,b,c,s) q_{U|S}(u|s) q_{V|U}(v|u) \notag\\
&\geq \mpv_{U|S} \sum_{s\in\eS} q_{ABCS}(a,b,c,s) q_{V|U}(v|u)\notag \\
&=  \mpv_{U|S}  q_{ABC}(a,b,c) q_{V|U}(v|u)\notag\\
&\geq  \mpv_{U|S}  q_{ABC}(a,b,c) q_{UV}(u,v).
\end{align}
Hence,
\begin{align}
\supp(A,B,C,U,V) = \supp (A,B,C)\times \supp(U,V).
\end{align}
Then, from Lemma~\ref{Lem:Ancillary2} of Sec.~\ref{Sec:Gacs-Korner}, it follows that 
\begin{align}
\GK^{ABU, ACV}  \equiv (\GK^{AB, AC} , \GK^{U, V}).\label{eqn:GK-decorr}
\end{align} 
Now, define $\tilde A:= \GK^{AB,AC}$, and let $ p_{\tilde ASUV}$ denote the {pmf} of $(\tilde A,S,U,V)$. Then, by construction the Markov chain $\tilde A\mkv S \mkv (U,V)$ holds. Further,
\begin{align}
H\big(\tilde A, \GK^{U,V}\,\big|\,\GK^{\tilde AV,\tilde AU}\big)& = 0.
\end{align}
Then, by \eqref{eqn:GK-decorr}, there must exist a function $\tilde{f}:\tilde \A \times \V\rightarrow \A\times\C\times\V$ such that 
\begin{align}
\tilde{f}(\GK^{\tilde AU,\tilde AV})\equiv \GK^{ABU,ACV}.
\end{align}
Then, {from~\eqref{eqn-conditions-Case-C}}, we see that 
\begin{equation}
{\Exp \big[d (S, f(\tilde{f}(\GK^{\tilde AV,\tilde AU})))\big]\leq D}.
\end{equation}
Since $\tilde \A\subseteq \A\times \C$, we are guaranteed that $|\tilde \A|\leq \big( |\eS|(|\eS|+6)|\hat\eS|^{|\U|}+4\big)^2 $, and hence, $ p_{\tilde ASUV}\in \mc{P}^*_{D,( |\eS|(|\eS|+6)|\hat\eS|^{|\U|}+4)^2}$. Further,
\begin{subequations}
\begin{align}
r_{uv}           &\geq I(S;A,B|U)\geq  I(S;\GK^{AC,AB}|U)=I(S;\tilde A|U),\\
r_{v}+r_{uv}  &\geq I(S;A,C|V)\geq  I(S;\GK^{AC,AB}|V)=I(S;\tilde A|V).
\end{align} 
\end{subequations}
Consequently, $(r_{uv}, r_v) \in \mathscr R^*_{( |\eS|(|\eS|+6)|\hat\eS|^{|\U|}+4)^2}(D)$. Since the rate pair was chosen arbitrarily, it follows that
\begin{equation}
\R_{ |\eS|(|\eS|+6)|\hat\eS|^{|\U|}+4}^{  \ddagger}(D) \subseteq \R_{ (|\eS|(|\eS|+6)|\hat\eS|^{|\U|}+4)^2}^{*}(D).
\end{equation}

\subsubsection*{{Case D}}\label{proof-thm3-D}

In this case, since $H(S|U)=0$, the receiver with side information $U$ does not require the encoder to communicate any message.  Hence, we see that if $(r_{uv}, r_v)\in\R(D)$, then any $(r_{uv}', r_v')\in \R(D)$ provided $r_{uv}'+r_v' \geq r_{uv}+ r_v$, i.e., only the sum-rate constraint is relevant. 

Let $(r_{uv}, r_v)\in  {\R}_{ |\eS|(|\eS|+6)|\hat\eS|^{|\U|}+4}^{  \ddagger}(D)$. Then, there must exist {a joint pmf}  $q_{ABCSUV}\in \mc{P}_{D, |\eS|(|\eS|+6)|\hat\eS|^{|\U|}+4}^{  \ddagger}$ and a function $f:\A\times\C\times\V \rightarrow \hat\eS$ such that $\Exp[d(S, f(\GK^{ABU,ACV}))] \leq D$, and
\begin{subequations}
\begin{align}
r_{uv}\geq &\textstyle I(S;A,B|U)= 0, \label{eqn-caseDrates1}\\
r_v+r_{uv}\geq &I(S;A,C|V)+I(S;B|A,C,U,V)= I(S;A,C|V). \label{eqn-caseDrates2}
\end{align}
\end{subequations}
Let functions $f_U, g_V$ be defined such that $f_U(A,B,U) =f_V(A,C,V)=f(\GK^{ABU,ACV})$. 
Now, since $H(S|U)=0$, it follows that $B\mkv (A,C,U) \mkv V$. Then, we have
\begin{align}
\sum_{a,c,u} q_{ACU}(a,c,u) \sum_{b,v} q_{B|ACU}(b|a,c,u) p_{V|U}(v|u) \bar{\mathds 1} \left\{f_U(a,b,u) = f_V(a,c,v)\right\}= 0.
\end{align}
Then, by Lemma~\ref{lem-identical} of Appendix~\ref{APP-L}, for each $(a,c,u)\in \supp(A,C,U)$, there must exist an $\eta(a,c,u)$ such that 
\begin{subequations}
\begin{align}
\Pr\big[f_U(A,B,U) = \eta(a,c,u) \mid (A,C,U)=(a,c,u)\big] &= 1,\\
\Pr\big[f_V(A,C,V) = \eta(a,c,u) \mid(A,C,U)=(a,c,u)\big] &= 1.
\end{align}
\end{subequations}
Hence, there must exist a function $\tilde f_U: \A\times\C\times \U \rightarrow \A\times\C\times \V$ such that 
\begin{align}
\Pr\big[\tilde f_U (A,C,U) \neq f_V(A,C,V)\big]  = 0. \label{eqn-existencefacu1}
\end{align}
Further, we also have
\begin{align}
\Exp[d(S, \tilde f_U (A,C,U) )]=\Exp[d(S,  f_V(A,C,V))]=\Exp[d(S, f(\GK^{ABU,ACV}))] \leq D\label{eqn-existencefacu2}
\end{align}
Now, set $\tilde{A} = (A,C)$, and let $q_{\tilde A, S,U,V}$ be the joint {pmf} of $(\tilde A, S,U,V)$. From \eqref{eqn-existencefacu1}, \eqref{eqn-existencefacu2}, and from the fact that $|\tilde \A| \leq |\A||\C|$, it follows that $q_{\tilde A, S,U,V}\in \mc{P}^*_{D,\left( |\eS|(|\eS|+6)|\hat\eS|^{|\U|}+4\right)^2}$. Further, from \eqref{eqn-caseDrates1} and \eqref{eqn-caseDrates2}, it follows that 
\begin{subequations}
\begin{align}
r_{uv}&\geq  I(S;A,B|U)=I(S;\tilde A|U)=0,\\
r_v+r_{uv}&\geq I(S;A,C|V) = I (S; \tilde A|V).
\end{align}
\end{subequations}
Hence, $(r_{uv}, r_v)\in\R_{( |\eS|(|\eS|+6)|\hat\eS|^{|\U|}+4)^2}^{*}(D)$. Since the rate pair was chosen arbitrarily, it follows that
\begin{equation}
\R_{ |\eS|(|\eS|+6)|\hat\eS|^{|\U|}+4}^{  \ddagger}(D) \subseteq \R_{ (|\eS|(|\eS|+6)|\hat\eS|^{|\U|}+4)^2}^{*}(D).
\end{equation}
\subsubsection*{Case E}\label{proof-thm3-E}

Pick  $(r_{uv}, r_v)\in  {\R}_{ |\eS|(|\eS|+6)|\hat\eS|^{|\U|}+4}^{  \ddagger}(D)$. Then, there must exist   $q_{ABCSUV}\in \mc{P}_{D, |\eS|(|\eS|+6)|\hat\eS|^{|\U|}+4}^{  \ddagger}$ and a function $f:\A\times\C\times\V \rightarrow \hat\eS$ such that $\Exp[d(S, f(\GK^{ABU,ACV}))] \leq D$, and
\begin{subequations}
\begin{align}
r_{uv}&\geq  I(S;A,B|U),\\
r_v+r_{uv}&\geq I(S;A,C|V)+I(S;B|A,C,U),
\end{align}
\end{subequations}
where in the sum rate we have incorporated side information degradedness. Since $(A,B,C) \mkv S \mkv U \mkv V$, and $\supp(U,V)=\supp(U)\times\supp(V)$, it follows that
\begin{align}
\supp(A,B,C,U,V) = \supp(A,B,C,U) \times \supp(V).
\end{align}
Then, an invocation of Lemma~\ref{lem-drop1var} of Sec.~\ref{Sec:Gacs-Korner} yields
\begin{align}
\GK^{ABU,ACV}  \equiv \GK^{ABU,AC}. \label{eqn-CaseEequivGK}
\end{align}
Hence, there must exist a function $\tilde f: \A\times \C \rightarrow \A \times \C \times \V$ such that $\tilde f (\GK^{ABU,AC}) = \GK^{ABU,ACV}$. Using this function, let us now define
\begin{align}
 \tilde{A} :=  \tilde f(\GK^{ABU,AC}). \end{align} 
 Then, $\hat S = f(\tilde A)$, and
\begin{align}
\Exp[d(S,\hat S )]=\Exp[d(S,f(\tilde A))]= \Exp[d(S, f(\tilde f (\GK^{ABU,AC}))] = \Exp[d(S, f(\GK^{ABU,ACV}))] \leq D. \label{eqn-CaseERes1}
\end{align}
 Since $\tilde A$ is both a function of $(A,B,U)$ and $(A,C)$, we see that $|\tilde \A| \leq |\A||\C|= ( |\eS|(|\eS|+6)|\hat\eS|^{|\U|}+4)^2$, and 
\begin{subequations}
\begin{alignat}{3}
r_{uv}&\geq I(S;A,B|U)=I(S;A,B,\tilde{A}|U)&\geq I(S;\tilde A|U),\\
r_v+r_{uv}&\geq I(S;A,C,\tilde{A}|V)+I(S;B|A,C,U)\,&\geq I(S;\tilde A|V).
\end{alignat}\label{eqn-CaseERes2}
\end{subequations}
Further, $\tilde A \mkv S \mkv U \mkv V$. Then, from \eqref{eqn-CaseERes1} and \eqref{eqn-CaseERes2}, we conclude that $q_{\tilde A SUV} \in \mc P^*_{D,( |\eS|(|\eS|+6)|\hat\eS|^{|\U|}+4)^2}$ and $(r_{uv}, r_v)\in\R_{ (|\eS|(|\eS|+6)|\hat\eS|^{|\U|}+4)^2}^{*}(D)$. Since $(r_{uv}, r_v)$ was chosen arbitrarily, it follows that
\begin{equation}
\R_{ |\eS|(|\eS|+6)|\hat\eS|^{|\U|}+4}^{  \ddagger}(D) \subseteq \R_{ \left(|\eS|(|\eS|+6)|\hat\eS|^{|\U|}+4\right)^2}^{*}(D).
\end{equation}

\subsubsection*{{Case F}}\label{proof-thm3-F}

Pick  $(r_{uv}, r_v)\in  {\R}_{ |\eS|(|\eS|+6)|\hat\eS|^{|\U|}+4}^{  \ddagger}(D)$. Then, there must exist  pmf $q_{ABCSUV}\in \mc{P}_{D, |\eS|(|\eS|+6)|\hat\eS|^{|\U|}+4}^{  \ddagger}$ and a reconstruction function $f:\A\times\V \rightarrow \hat \eS$ such that $\Exp[d(S, f(\GK^{ABU,ACV}))] \leq D$, and 
\begin{subequations}
\begin{align}   
r_{uv}&\geq \textstyle I(S;A,B|U),\\
r_v+r_{uv}&\geq I(S;A,C|V)+I(S;B|A,C,U,V).
\end{align} \label{eqn-conditions}
\end{subequations}

Suppose that $H(V|S) = 0$. Since $H(A|\GK^{ABU,ACV})=0$, we have 
\begin{align}
I\big(\GK^{ABU,ACV};U,V|S\big)&\leq  
I\big(\GK^{ABU,ACV},A,C;U,V|S\big)\notag\\ 
&= 
I\big(A,C;U,V|S\big) 
+ I\big(\GK^{ABU,ACV};U,V|A,C,S\big)\notag\\
&\step{(a)}= I(\GK^{ABU,ACV};U|A,C,S,V\big) \step{(b)} = 0,
\end{align}
where in (a), we use $(A,C)\mkv S\mkv (U,V)$ and that $V$ is a function of $S$; and in (b), we use the fact that $\GK^{ABU,ACV}$ is a function of $(A,V)$.  Hence, 
\begin{align}
\GK^{ABU,ACV} \mkv S \mkv (U,V).\end{align} 
Similarly, when $H(U|S)=0$, we arrive at the same conclusion by reversing the roles of $U$ and $V$. By defining $\tilde A := \GK^{ABU,ACV}$, we see that $\tilde A \mkv S \mkv (U,V)$ with $|\A| \leq \big( |\eS|(|\eS|+3)|\hat\eS|^{|\U|}+3\big)^3$, and that $\Exp[d(S, f(\tilde A)] \leq D$. Further,
\begin{subequations}
\begin{align}
r_{uv}&\geq \textstyle I(S;A,B|U) \geq I(S;\tilde A|U)\\
r_v+r_{uv}&\geq I(S;A,C|V)+I(S;B|A,C,U,V)\geq I(S;\tilde A|V)
\end{align}
\end{subequations}
Hence, $(r_{uv}, r_v)\in\R_{( |\eS|(|\eS|+6)|\hat\eS|^{|\U|}+4)^3}^{*}(D)$. Since the rate pair was chosen arbitrarily, it follows that
\begin{equation}
\R_{ |\eS|(|\eS|+6)|\hat\eS|^{|\U|}+4}^{  \ddagger}(D) \subseteq \R_{ \left(|\eS|(|\eS|+6)|\hat\eS|^{|\U|}+4\right)^2}^{*}(D).
\end{equation}
\section{Proof of Theorem~\ref{thm-3-cor}}\label{APP-J}

Clearly we have $\RQB(D) \subseteq \R^*(D)$. Since Theorem~\ref{thm-3} applies to Cases A, B and C, we need only show that $\RQB(D) \supseteq \R^*(D)$ in each case. Let $(r_{uv}, r_v)\in\R^*(D)$. Then, there must exist   $q_{ASUV}\in \mc{P}_{D, |\eS|+2}^{  \ddagger}$, and a function $f:\A\times\V \rightarrow \hat \eS$ such that $\Exp[d(S, f(\GK^{AU,AV}))] \leq D$, and 
\begin{subequations}
\begin{align}
r_{uv}&\geq \textstyle I(S;A|U),\\
r_v+r_{uv}&\geq I(S;A|V).
\end{align}
\end{subequations}
The rest of the proof for each case is as follows. 
\subsubsection*{Case A}
Since $\supp(S,U,V) = \supp(S) \times \supp(U,V)$, it follows from that Lemma~\ref{Lem:Ancillary2} of Sec.~\ref{Sec:Gacs-Korner} that
\begin{equation}
\GK^{AU,AV} = \big(A,\GK^{U,V}\big).
\end{equation}
Since $H(\GK^{U,V}(V)) = 0$, we conclude from the above equation that $\GK^{AU,AV}(A,V) \equiv A$. 

Consequently,  there must exist a function $\tilde f: \A\rightarrow \hat \eS$ such that reconstruction $ \tilde{f}(A)=f(\GK^{AU,AV})$. Define random variable $\hat S =\tilde{f}(A) =f(\GK^{AU,AV})$. Then,
$(U,V) \mkv S \mkv A \mkv \hat{S}$ and 
\begin{subequations}
\begin{align}
r_{uv}&\geq \textstyle I(S;A|U)\geq I(S;\hat{S}|U),\\
r_v+r_{uv}&\geq I(S;A|V)\geq I(S;\hat{S}|V).
\end{align}\label{eqn-APPI1}
\end{subequations}
 Further, 
\begin{align}
D\geq  \Exp\big[d\big(S,f(\GK^{AU,AV})\big)\big] = 
\Exp\big[d\big(S,\hat{S}\big)\big], \label{eqn-APPI2}
\end{align} 
Hence, it follows that $(r_{uv}, r_v)\in\RQB(D)$.

\subsubsection*{Case B}

In this setting, the Markov chain $A \mkv S \mkv U \mkv V$ and the support condition $\supp(U,V) = \supp(U) \times \supp(V)$ together imply $\supp(A,U,V) = \supp(A,U) \times \supp(V)$. Thus, from Lemma~\ref{lem-drop1var} of Sec.~\ref{Sec:Gacs-Korner}, we see that
\begin{equation}
\GK^{AU,AV} \equiv \GK^{AU,A} \equiv A.
\end{equation}
The rest of the proof then follows by setting $\hat S = f(A)=f(\GK^{AU,AV})$ and is identical to that of Case A above. 

\subsubsection*{Case C} 
Repeating the steps of Case F of Theorem~\ref{thm-3}, we see in this case that $\GK^{AU,AV} \mkv S \mkv (U,V)$. The proof is then complete by choosing $\hat{S} = f\big(\GK^{AU,AV}\big)$, and verifying that $\hat S \mkv S \mkv (U,V)$, \eqref{eqn-APPI1} and \eqref{eqn-APPI2} hold.


\section{A Lemma on Functions of Independent Random Variables}\label{APP-L}

\begin{lemma}\label{lem-identical}
Let $X \sim p$  and $Y \sim q$ be independent random variables. Suppose that we are given a finite set $\Z$, and functions $f:\X \rightarrow \Z$ and $g:\Y\rightarrow \Z$ satisfying
\begin{align}
\Pr[f(X)\neq g(Y)] \leq \delta <\frac{1}{25}.
\end{align}
There exists an $z^*\in \Z$ such that 
\begin{subequations}
\begin{align}
\label{eqn:techlemma2-1}
\Pr[f(X)=z^*]  
&\geq  \frac{1+\sqrt{1-\delta-\sqrt{\delta}}}{2},\\
\label{eqn:techlemma2-2}
\Pr[g(Y)=z^*] 
&\geq  \frac{1+\sqrt{1-\delta-\sqrt{\delta}}}{2}.
\end{align}
\end{subequations}
\end{lemma}

\begin{IEEEproof}
Let for $z\in \Z$, $p_z := \Pr[f(X)=z]$ and $q_z := \Pr[g(Y)=z]$. Then for any $z\in \Z$,
\begin{align}
 |p_z-q_z| &= \sqrt{ (p_z-q_z)^2} \leq  \sqrt{p_z(1-q_z)+(1-p_z)q_z}\leq \sqrt{\Pr[f(X)\neq g(Y)]} \leq\sqrt{\delta}.\label{eqn-SimpLem1}
\end{align}
Then, the following holds
\begin{align}
 \sum_{z} p_z^2 & \geq  \sum_{z} p_z( p_z+q_z-q_z) \geq  \sum_z p_z q_z - \sum_z p_z |q_z-p_z| \notag\\& \geq \Pr[f(X)=g(Y)] -\sqrt{\delta} 
 \geq 1-\delta-\sqrt \delta > \frac{19}{25}. \label{eqn-SimpLem2}
\end{align}
Now, let $M_p = \max\limits_z \,p_z$, $M_q= \max\limits_z \,q_z$, $z_p=\arg\max\limits_z\, p_z$ and $z_q=\arg\max\limits_z \,q_z$. Then, $M_p>\frac{19}{25}$, since
\begin{align}
M_p = \sum_z M_p \,p_z \geq \sum_z p_z^2 \geq 1-\delta-\sqrt \delta > \frac{19}{25}. \label{eqn-SimpLem3}
\end{align}
Also, $z_p$ and $z_q$, have to be identical, because $q_z$ can strictly exceed $\frac{1}{2}$ for only one  $z\in\Z$, and
\begin{align}
q_{z_p} \geq p_{z_p}-|p_{z_p}-q_{z_p}| \stackrel{\eqref{eqn-SimpLem1}}{\geq}  M_p -\sqrt{\delta} > \frac{19}{25}-\frac{1}{5}>\frac{1}{2}.
\end{align}
Since the problem is symmetric, we see that $M_q = q_{z_q} > 1-\delta -\sqrt{\delta}$ as well. At this point we are done if the RHS of \eqref{eqn:techlemma2-1} and \eqref{eqn:techlemma2-2} were $1-\delta-\sqrt\delta$. To improve this estimate,  consider for $\gamma\in(0,1)$, the following optimization problem and its solution.
\begin{align}
\max_{\substack{ \forall z\in \Z, r_z \in [0,\gamma]\\ \sum_z r_z =1}} \sum_{z} r_z^2 = \left\lfloor\frac{1}{\gamma} \right\rfloor \gamma^2 + \left(1 -   \left\lfloor\frac{1}{\gamma} \right\rfloor \gamma\right)^2 \leq \gamma^2 + (1-\gamma)^2.\label{eqn-SimpLem4}
\end{align}
Since $M_p$ is the maximum positive value taken by $\{p_z:z\in \Z\}$, it follows that 
\begin{align}
1-\delta-\sqrt{\delta} \leq  \sum_z p_z^2 \stackrel{\eqref{eqn-SimpLem1}}{\leq} M_p^2 +(1-M_p)^2,
\end{align}
which necessitates that 
\begin{align}
\Pr[f(X)=z_p] =M_p \in \left[ 0,  \frac{1-\sqrt{1-\delta-\sqrt{\delta}}}{2}\right] \bigcup \left[ \frac{1+\sqrt{1-\delta-\sqrt{\delta}}}{2} ,1\right] .\end{align}
Selecting the choice meeting \eqref{eqn-SimpLem3} eliminates the first interval. Finally, reversing the roles of $p$ and $q$ along with the fact that $z_p=z_q$ concludes the proof. 
\end{IEEEproof}


\end{document}